\documentclass{article}
\usepackage{graphicx}
\usepackage[font=small,skip=0pt]{caption}
\usepackage{titlesec}
\titlespacing*{\section}{0pt}{1.1\baselineskip}{\baselineskip}
\usepackage{epstopdf, epsfig}
\usepackage[euler]{textgreek}
\usepackage{fixltx2e}
\usepackage{todonotes}
\usepackage{caption}
\usepackage{subfig}
\usepackage{tabularx}
\usepackage{wrapfig}
\usepackage{comment}
\usepackage{amsmath} 
\usepackage{hyperref}
\usepackage{amssymb}
\usepackage{amsfonts}
\usepackage{mathabx}
\usepackage{MnSymbol}
\usepackage{wasysym}

\usepackage{hhline}
\usepackage{multirow}
\usepackage{physics}
\usepackage{tabularx}
\usepackage{array}
\usepackage{subfig}
\usepackage{mathtools}
\usepackage{siunitx}
\usepackage{geometry}

\newcolumntype{M}[1]{>{\centering\arraybackslash}m{#1}}
\usepackage{pdfpages}
\usepackage[running]{lineno}
\usepackage{float}

\usepackage[superscript]{cite}

\newcommand{\ddtt}[1]{\frac{\text{d}#1(t)}{\text{d}t}}

\usepackage{xr}

\makeatletter
\newcommand*{\addFileDependency}[1]{
  \typeout{(#1)}
  \@addtofilelist{#1}
  \IfFileExists{#1}{}{\typeout{No file #1.}}
}
\makeatother

\newcommand*{\myexternaldocument}[1]{%
    \externaldocument{#1}%
    \addFileDependency{#1.tex}%
    \addFileDependency{#1.aux}%
}

\myexternaldocument{Figures}

\usepackage{authblk}

\title{A machine learning aided global diagnostic and comparative tool to assess effect of quarantine control in Covid-19 spread}

\author[1]{Raj Dandekar}
\author[2]{Chris Rackauckas}
\author[3,4]{George Barbastathis}
\affil[1]{Department of Civil and Environmental Engineering, Massachusetts Institute of Technology, Cambridge, MA 02139, USA}
\affil[2]{Department of Applied Mathematics, Massachusetts Institute of Technology, Cambridge, MA 02139, USA}
\affil[3]{Department of Mechanical Engineering, Massachusetts Institute of Technology, Cambridge, MA 02139, USA}
\affil[4]{Singapore-MIT Alliance for Research and Technology (SMART) Centre, Singapore 138602}

\begin{document}

\maketitle
\noindent
\textbf{Article Summary Line:} Data-driven epidemiological model to quantify and compare quarantine control policies in controlling COVID-19 spread in Europe, North America, South America and Asia. \newline \newline
\textbf{Running Title:} Machine Learning aided quarantine model - Covid19. \newline\newline

\section{SUMMARY}
We have developed a globally applicable diagnostic Covid-19 model by augmenting the classical SIR epidemiological model with a neural network module. Our model does not rely upon previous epidemics like SARS/MERS and all parameters are optimized via machine learning algorithms employed on publicly available Covid-19 data. The model decomposes the contributions to the infection timeseries to analyze and compare the role of quarantine control policies employed in highly affected regions of Europe, North America, South America and Asia in controlling the spread of the virus. For all continents considered, our results show a generally strong correlation between strengthening of the quarantine controls as learnt by the model and actions taken by the regions' respective governments. Finally, we have hosted our quarantine diagnosis results for the top $70$ affected countries worldwide, on a public platform, which can be used for informed decision making by public health officials and researchers alike. 

\section{INTRODUCTION}
The Coronavirus respiratory disease 2019 originating from the virus ``SARS-CoV-2" \cite{chan2020familial, cdc} has led to a global pandemic, leading to $12, 552, 765$ confirmed global cases in more than $200$ countries as of July 12, 2020 \cite{worldcoronavirus}. 
As the disease began to spread beyond its apparent origin in Wuhan, the responses of local and national governments varied considerably. The evolution of infections has been similarly diverse, in some cases appearing to be contained and in others reaching catastrophic proportions. In Hubei province itself, starting at the end of January, more than $10$~million residents were quarantined by shutting down public transport systems, train and airport stations, and imposing police controls on pedestrian traffic. Subsequently, similar policies were applied nation-wide in China. By the end of March, the rate of infections was reportedly receding \cite{cyranoski2020china}. \newline

By the end of February 2020, the virus began to spread in Europe, with Italy employing extraordinary quarantine measures starting $11$ March 2020. France enforced a lockdown beginning $17$ March followed later by UK on $23$ March; whereas no lockdown was enforced in Sweden \cite{NatureNews}. South Korea, Iran and Spain experienced acute initial increases, but then adopted drastic generalized quarantine. In the United States, the first infections were detected in Washington State  as early as $20^{\textrm{th}}$ January 2020 \cite{holshue2020first}  and now it is being reported that the virus had been circulating undetected in New York City as early as mid-February \cite{Hidden}. Federal and state government responses were comparatively delayed and variable, with most states having stay at home orders \cite{NatureNews} declared by the end of March. In South America, Brazil, Chile and Peru are the highest affected countries as of $12$ July and they employed differing quarantine policies \cite{SouthAmericaquarantine}. Brazil's first case was reported in the last week of February and the country went into a state of partial quarantine on $24$ March. Chile declared a state of catastrophe for $90$ days in the first week of March, and the military was deployed to enforce quarantine measures. In Peru, a nationwide curfew was employed much later, on March $19$. Thus, affected countries around the world enforced differing quarantine strategies in an effort to mitigate the virus spread. \newline

Given the available Covid-19 data for the infected case count by country and world-wide, it is seen that the infection growth curve also showed significantly diverse behaviour globally. In some countries, the infected case count peaked within a month and showed a subsequent decline, while in certain other countries, it was seen to increase for much longer before plateauing. In some of the highly affected countries, the infected count has not yet reached a plateau and the daily active cases continue to increase or remain stagnant as of $12$ July 2020. \newline 

Given the observed spatially and temporally diverse government responses and outcomes, the role played by the varying quarantine measures in different countries in shaping the infection growth curve is still not clear. With publicly available Covid-19 data by country and world-wide by now widely available, there is an urgent need to use data-driven approaches to bridge this gap, quantitatively estimate and compare the role of the quarantine policy measures implemented in several countries in curtailing spread of the disease. \newline
As of this writing, more than a 100 papers have been made available \cite{Bert1}, mostly in preprint form. Existing models have one or more of the following limitations:
\begin{itemize}
    \item Lack of independent estimation: Using parameters based on prior knowledge of SARS/MERS coronavirus epidemiology and not derived independently from the Covid-19 data \cite{chinazzi2020effect} or parameters like rate of detection, nature of government response fixed prior to running the model \cite{Bert2}.
    \item Lack of global applicability: Not implemented on a global scale \cite{kraemer2020effect}.
    \item Lack of interpretibility: Using several free/fitting parameters making it a cumbersome, complicated model to reciprocate and use by policy makers.\cite{ferguson2020impact}.
\end{itemize}
In this paper, we propose a globally scalable, interpretable model with completely independent parameter estimation through a novel approach: augmenting a first principles-derived epidemiological model with a data-driven module, implemented as a neural network. We leverage this model to quantify the quarantine strengths and analyze and compare the role of quarantine control policies employed to control the virus effective reproduction number \cite{imai2020report, read2020novel, tang2020estimation, li2020early, wu2020nowcasting, kucharski2020early, ferguson2020impact} in the European, North American, South American and Asian continents. In a classical and commonly used model, known as SEIR \cite{SEIR1, SEIR2, SEIR3}, the population is divided into the susceptible $S$, exposed $E$, infected $I$ and recovered $R$ groups, and their relative growths and competition are represented as a set of coupled ordinary differential equations. The simpler SIR model does not account for the exposed population $E$. These models cannot capture the large-scale effects of more granular interactions, such as the population's response to social distancing and quarantine policies. However, a major assumption of these models is that the rate of transitions between population states is fixed. In our approach, we relax this assumption by estimating the time-dependent quarantine effect on virus exposure as a neural network informs the infected variable $I$ in the SIR model. This trained model thus decomposes the effects and the neural network encodes information about the quarantine strength function in the locale where the model is trained. \newline

In general, neural networks with arbitrary activation functions are universal approximators\ \cite{nn:Cybenko89-universal,nn:Hornik91-universal,nn:Sonoda-universal-17}. Unbounded activation functions, in particular, such as the rectified linear unit (ReLU) has been known to be effective in approximating nonlinear functions with a finite set of parameters\ \cite{nn:glorot11,nn:Goodfellow13-maxout,nn:Dahl13-reLU-dropout}. Thus, a neural network solution is attractive to approximate quarantine effects in combination with analytical epidemiological models. The downside is that the internal workings of a neural network are difficult to interpret. The recently emerging field of Scientific Machine Learning \cite{baker2019workshop} exploits conservation principles within a universal differential equation \cite{Rackauckas20}, SIR in our case, to mitigate overfitting and other related machine learning risks. \newline

In the present work, the neural network is trained from publicly available infection and population data for Covid-19 for a specific region under study; details are in the Experimental Procedures section. Thus, our proposed model is globally applicable and interpretable with parameters learned from the current Covid-19 data, and does not rely upon data from previous epidemics like SARS/MERS.

\section{RESULTS}
\begin{figure}
\centering
\begin{tabular}{c}
\subfloat[]{\includegraphics[width=0.6\textwidth]{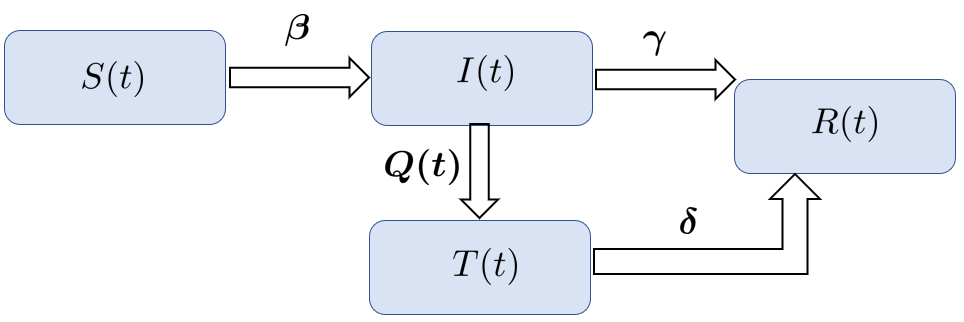}}\\
\subfloat[]{\includegraphics[width=0.4\textwidth]{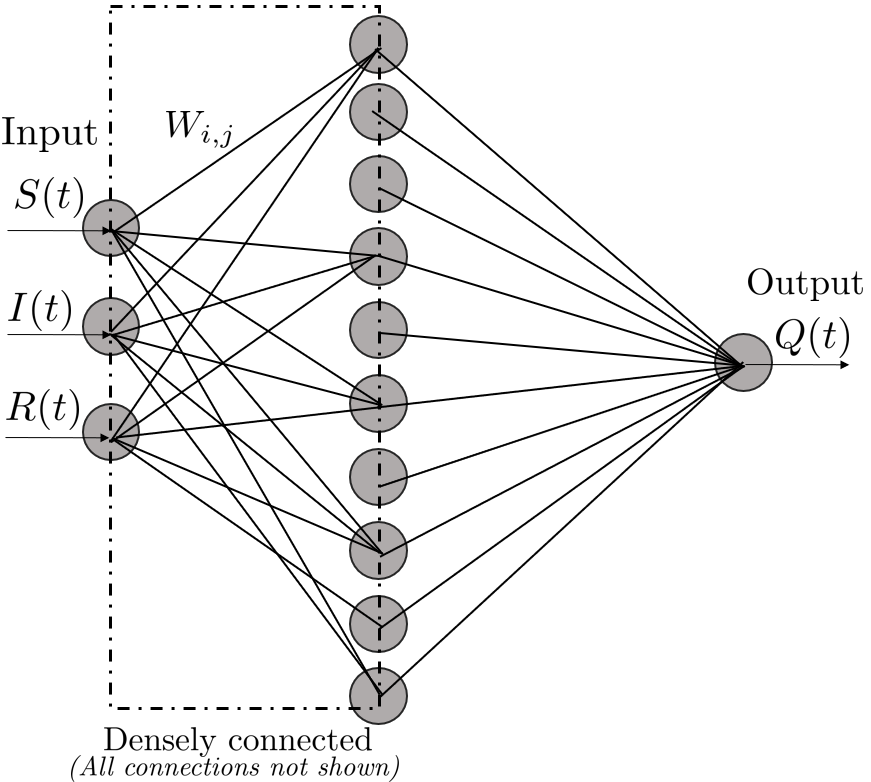}}\\
\end{tabular}
\caption{(a) Schematic of the augmented QSIR model considered in the present study. (b) Schematic of the neural network architecture used to learn the quarantine strength function $Q(t)$.}\label{schem}
\end{figure}

\subsection{Standard SIR model}
The classic SIR epidemiological model is a standard tool for basic analysis concerning the outbreak of epidemics. In this model, the entire population is divided into three sub-populations:  susceptible $S$; infected $I$; and  recovered $R$. The sub-populations' evolution is governed by the following system of three coupled nonlinear ordinary differential equations
\begin{align}
    \ddtt{S} & = -\frac{\beta \: S(t) \: I(t)}{N} \\
    \rule{0cm}{4ex} \ddtt{I} & = \frac{\beta \: S(t) \: I(t)}{N} - \gamma I(t) \\
    \rule{0cm}{4ex} \ddtt{R} & = \gamma I(t).
\end{align}
Here, $\beta$ is the infection rate and $\gamma$ is the recovery rates, respectively, and are assumed to be constant in time. The total population $N=S(t)+ I(t)+R(t)$ is seen to remain constant as well; that is, births and deaths are neglected. The recovered population is to be interpreted as those who can no longer infect others; so it also includes  individuals deceased due to the infection. The possibility of recovered individuals to become reinfected is accounted for by SEIS models \cite{mukhopadhyay2008analysis}, but we do not use this model here, as the reinfection rate for Covid-19 survivors is considered to be negligible as of now. The reproduction number $R_{t}$ in the SEIR and SIR models is defined as
\begin{equation}\label{Rt-SIR}
    R_{t} = \cfrac{\beta}{\gamma}.
\end{equation}
An important assumption of the  SIR  models is homogeneous mixing among the subpopulations. Therefore, this model cannot account for social distancing or or social network effects. Additionally the model  assumes uniform susceptibility and disease progress for every individual; and that no spreading occurs through animals or other non-human means. Alternatively, the  SIR model may be interpreted as quantifying the statistical expectations on the respective mean populations, while deviations from the model's assumptions contribute to statistical fluctuations around the mean.
\subsection{Augmented QSIR model}
To study the effect of quarantine control globally, we start with the SIR epidemiological model. Figure \ref{schem}a shows the schematic of the modified SIR model, the QSIR model, which we consider. We augment the SIR model by introducing a time varying quarantine strength rate term $Q(t)$ and a quarantined population $T(t)$, which is prevented from having any further contact with the susceptible population. Thus, the term $I(t)$ denotes the infected population still having contact with the susceptibles, as done in the standard SIR model; while the term $T(t)$ denotes the infected population who are effectively quarantined and isolated. Thus, we can write an expression for the quarantined infected population $T(t)$ as
\begin{equation} \label{T}
    T(t) = Q(t) \times I(t)
\end{equation}
Further we introduce an additional recovery rate $\delta$ which quantifies the rate of recovery of the quarantined population. Based on the modified model, we define a Covid spread parameter in a similar way to the reproduction number defined in the SIR model (\ref{Rt-SIR}) as
\begin{equation} \label{Cp}
    C_{p}(t) = \cfrac{\beta}{\gamma + \delta +  Q(t)}.
\end{equation}
$C_{p} >1$ indicates that infections are being introduced into the population at a higher rate than they are being removed, leading to rapid spread of the disease. On the other hand, $C_{p} <1$ indicates that the Covid spread has been brought under control in the region of consideration. \newline
Since $Q(t)$ does not follow from first principles and is highly dependent on local quarantine policies, we devised a neural network-based approach to approximate it. \newline
Recently, it has been shown that neural networks can be used as function approximators to recover unknown constitutive relationships in a system of coupled ordinary differential equations \cite{Rackauckas20, Rack2}. Following this principle, we represent $Q(t)$ as a $n$ layer-deep neural network with weights $W_{1}, W_{2} \ldots W_{n}$, activation function $r$ and the input vector $U = (S(t), I(t), R(t))$ as
\begin{equation}\label{NN-1}
   Q(t) = r\left( W_{n} r\left( W_{n-1} \ldots r\left( W_{1} U\right)\right)\right) \equiv \text{NN}(W, U)
\end{equation}

For the implementation, we choose a $n=2$-layer densely connected neural network with $10$ units in the hidden layer and the $\textrm{ReLU}$ activation function. This choice was because we found sigmoidal activation functions to stagnate. The final model was described by $54$ tunable parameters. The neural network architecture schematic is shown in figure \ref{schem}b. The governing coupled ordinary differential equations for the QSIR model are 
\begin{align}\label{model_augment}
    \ddtt{S} & = -\frac{\beta \: S(t) \: I(t)}{N} \\
    \rule{0cm}{4ex} \ddtt{I} & = \frac{\beta \: S(t) \: I(t)}{N} - \left(\rule[-0.5ex]{0cm}{2.5ex} \gamma+Q(t)\right) I(t) =
    \nonumber \\
    & = \frac{\beta \: S(t) \: I(t)}{N} - \left(\rule[-0.5ex]{0cm}{2.5ex} \gamma+\text{NN}(W, U)\right) I(t) \\
    \rule{0cm}{4ex} \ddtt{R} & = \gamma I(t) + \delta T(t) \\
    \rule{0cm}{4ex} \ddtt{T} & = Q(t)\: I(t) = \text{NN}(W, U)\: I(t) - \delta T(t).
\end{align}
\newline
More details about the model initialization and parameter estimation methods is given in the Experimental Procedures section.\newline
In all cases considered below, we trained the model using data starting from the dates when the $500^{\text{th}}$ infection was recorded in each region and up to June $1$ 2020. In each subsequent case study, $Q(t)$ denotes the rate at which infected persons are effectively quarantined and isolated from the remaining population, and thus gives composite information about (a) the effective testing rate of the infected population as the disease progressed and (b) the intensity of the enforced quarantine as a function of time. To understand the nature of evolution of $Q(t)$, we look at the time point when $Q(t)$ approximately shows an inflection point,  or a ramp up point. An inflection point in $Q(t)$ indicates the time when the rate of increase of $Q(t)$ i.e $dQ(t)/dt$ was at its peak while a ramp up point corresponds to a sudden intensification of quarantine policies employed in the region under consideration.\newline

We define the quarantine efficiency, $Q_{\textrm{eff}}$ as the increase in $Q(t)$ within a month following the detection of the $500^{\textrm{th}}$ infected case in the region under consideration. Thus

\begin{equation}\label{Qeff}
    Q_{\textrm{eff}} = Q(30) - Q(1)
\end{equation}

The magnitude of $Q_{\textrm{eff}}$ shows how rapidly the infected individuals were prevented from coming into contact with the susceptibles in the first month following the detection of the $500^{\textrm{th}}$ infected case; and thus contains composite information about the quarantine and lockdown strength; and the testing and tracing protocols to identify and isolate infected individuals.

\subsection{Europe}
\begin{figure}
\centering
\begin{tabular}{cc}
\subfloat[][Russia]{\includegraphics[width=0.5\textwidth]{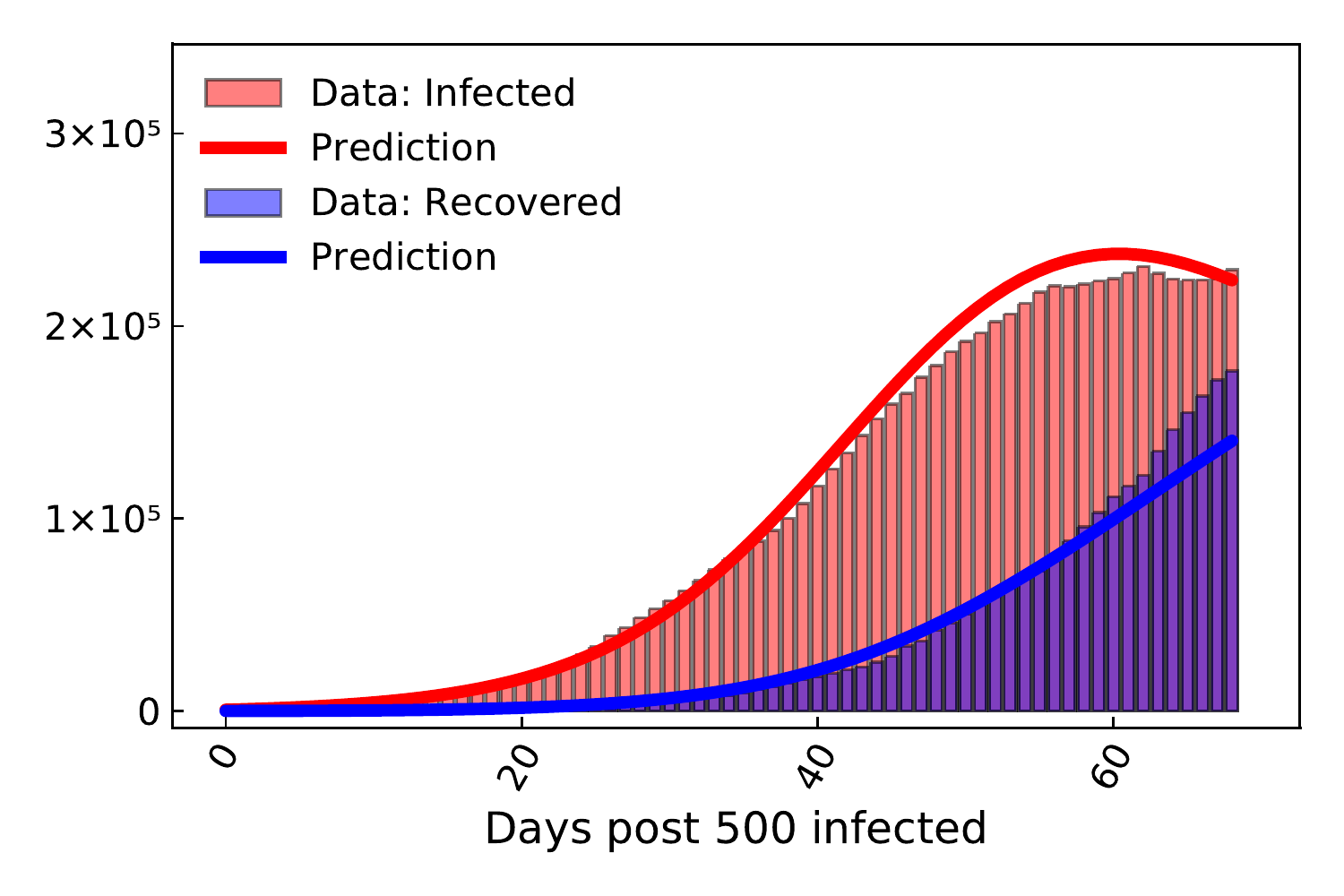}}
\subfloat[][UK]{\includegraphics[width=0.5\textwidth]{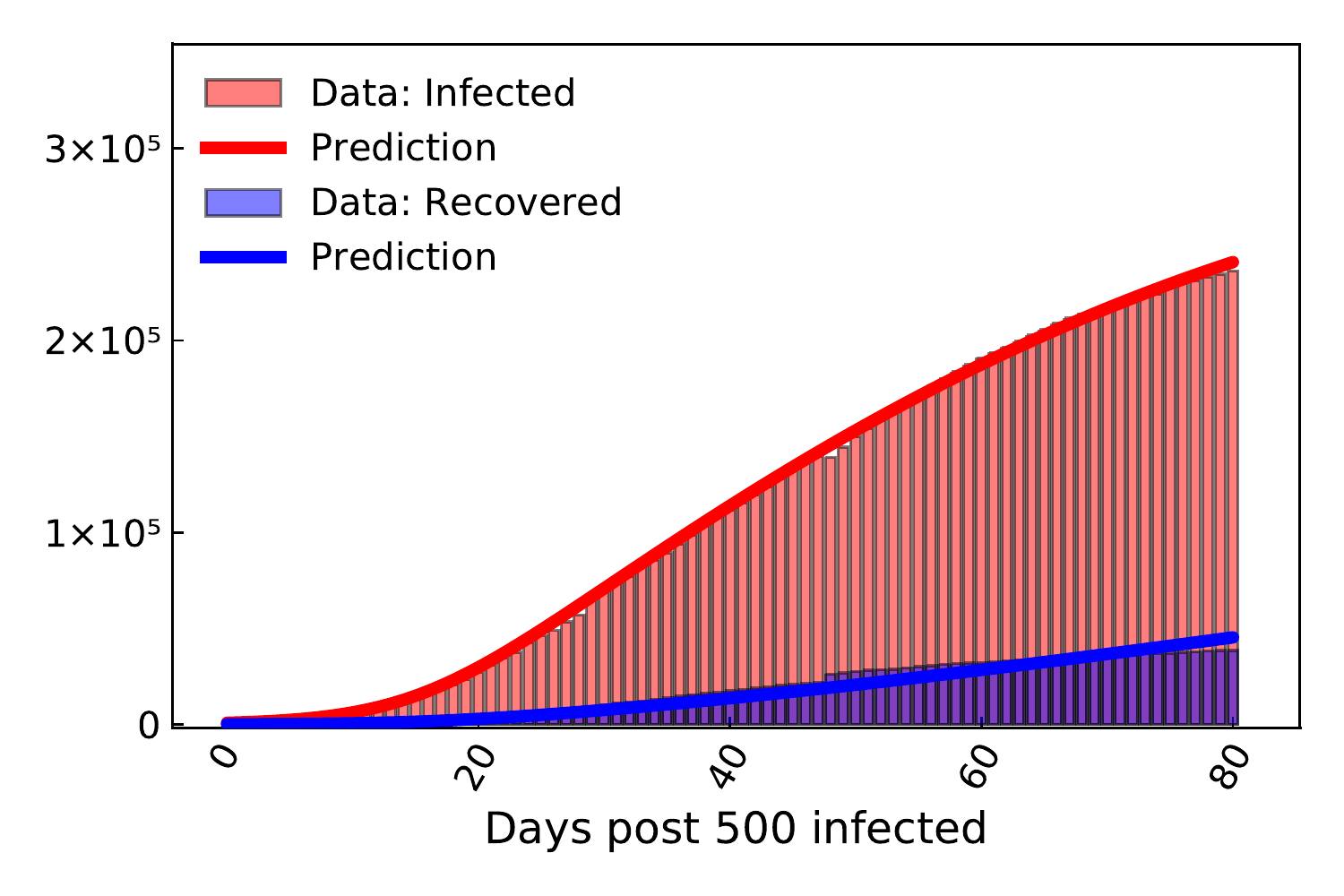}}\\
\subfloat[][Spain]{\includegraphics[width=0.5\textwidth]{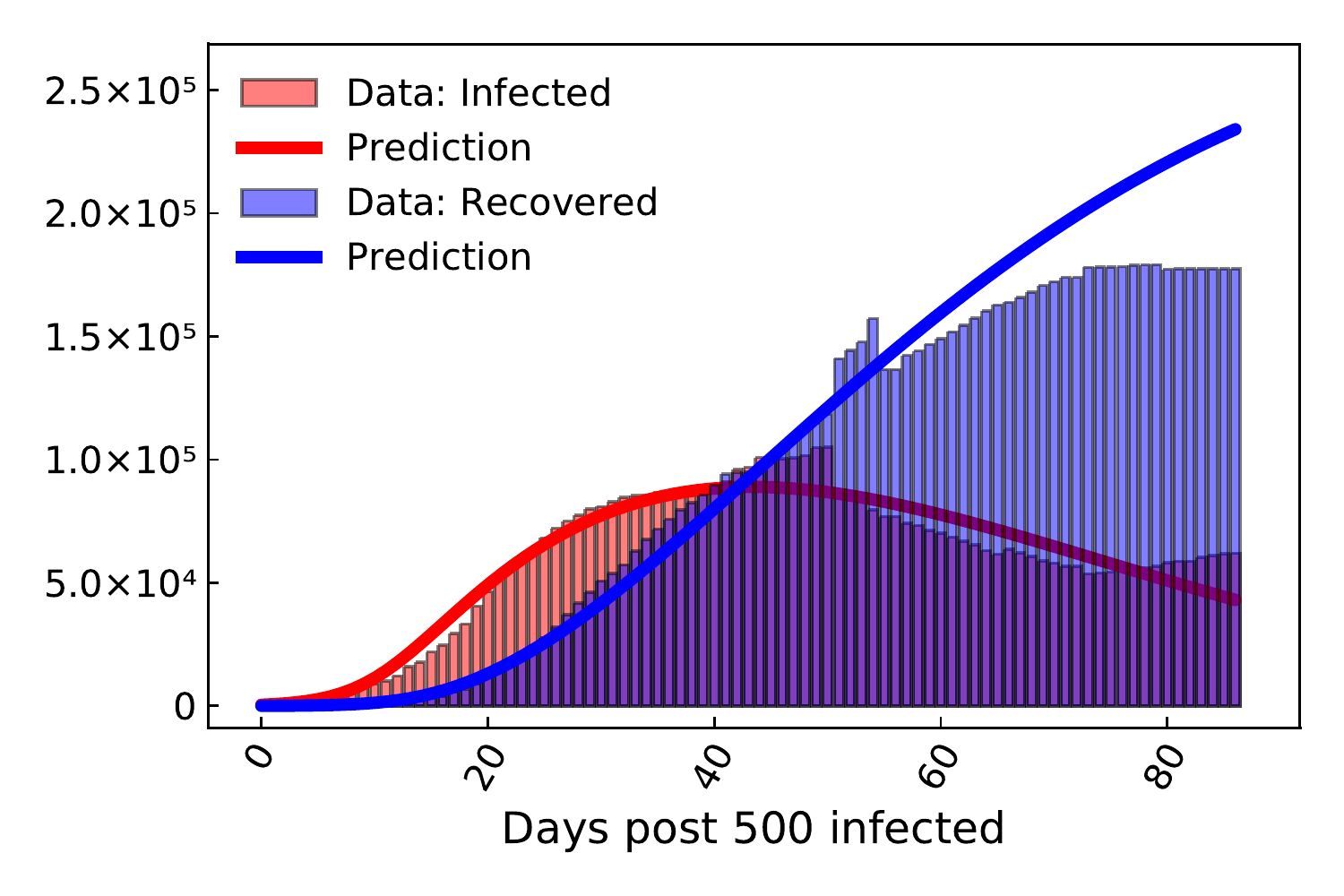}}
\subfloat[][Italy]{\includegraphics[width=0.5\textwidth]{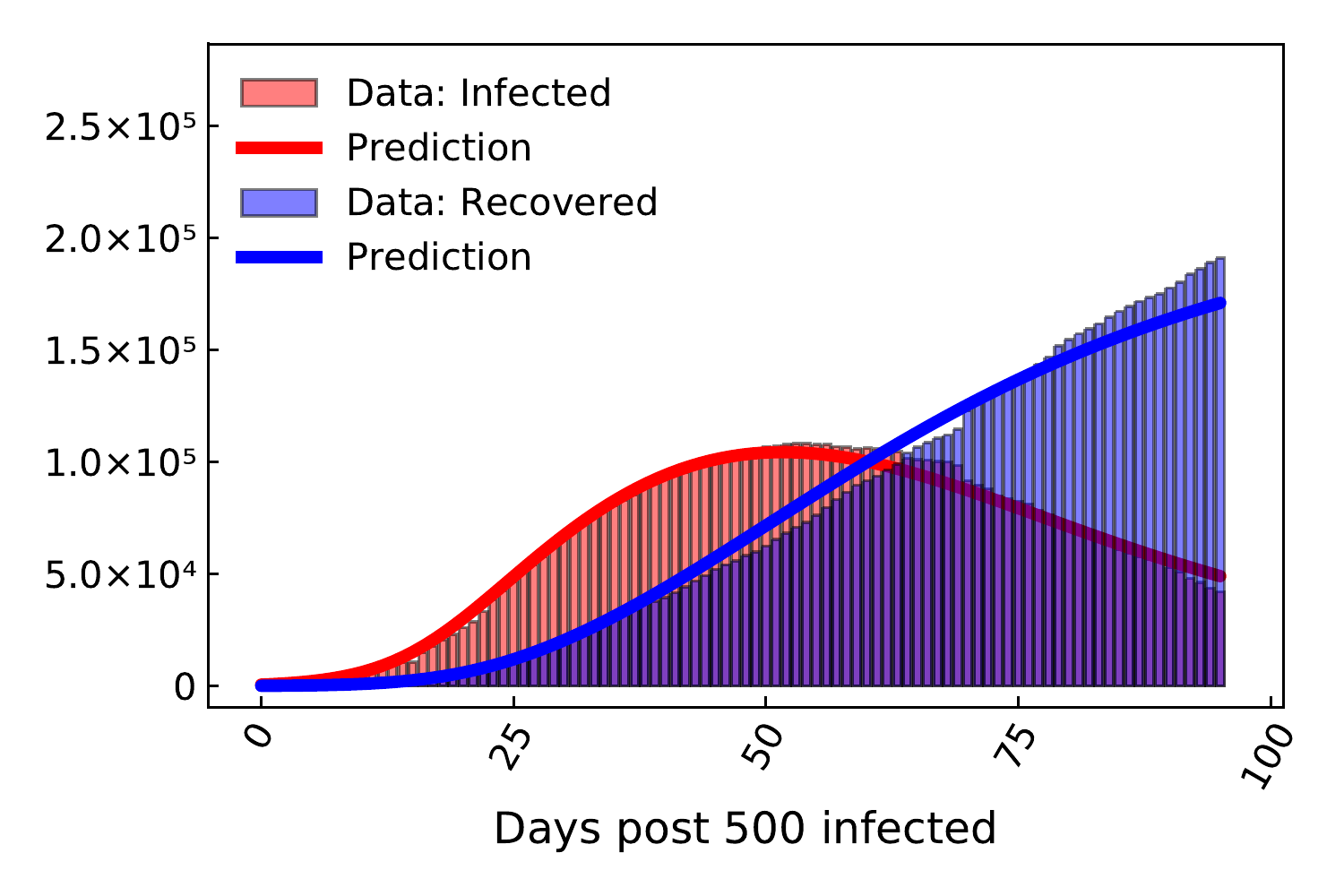}}\\
\end{tabular}
\caption{COVID-19 infected and recovered evolution compared with our neural network augmented model prediction in the highest affected European countries as of June 1, 2020.}\label{Europe1}
\end{figure}

\begin{figure}
\centering
\begin{tabular}{cc}
\subfloat[][Russia]{\includegraphics[width=0.33\textwidth]{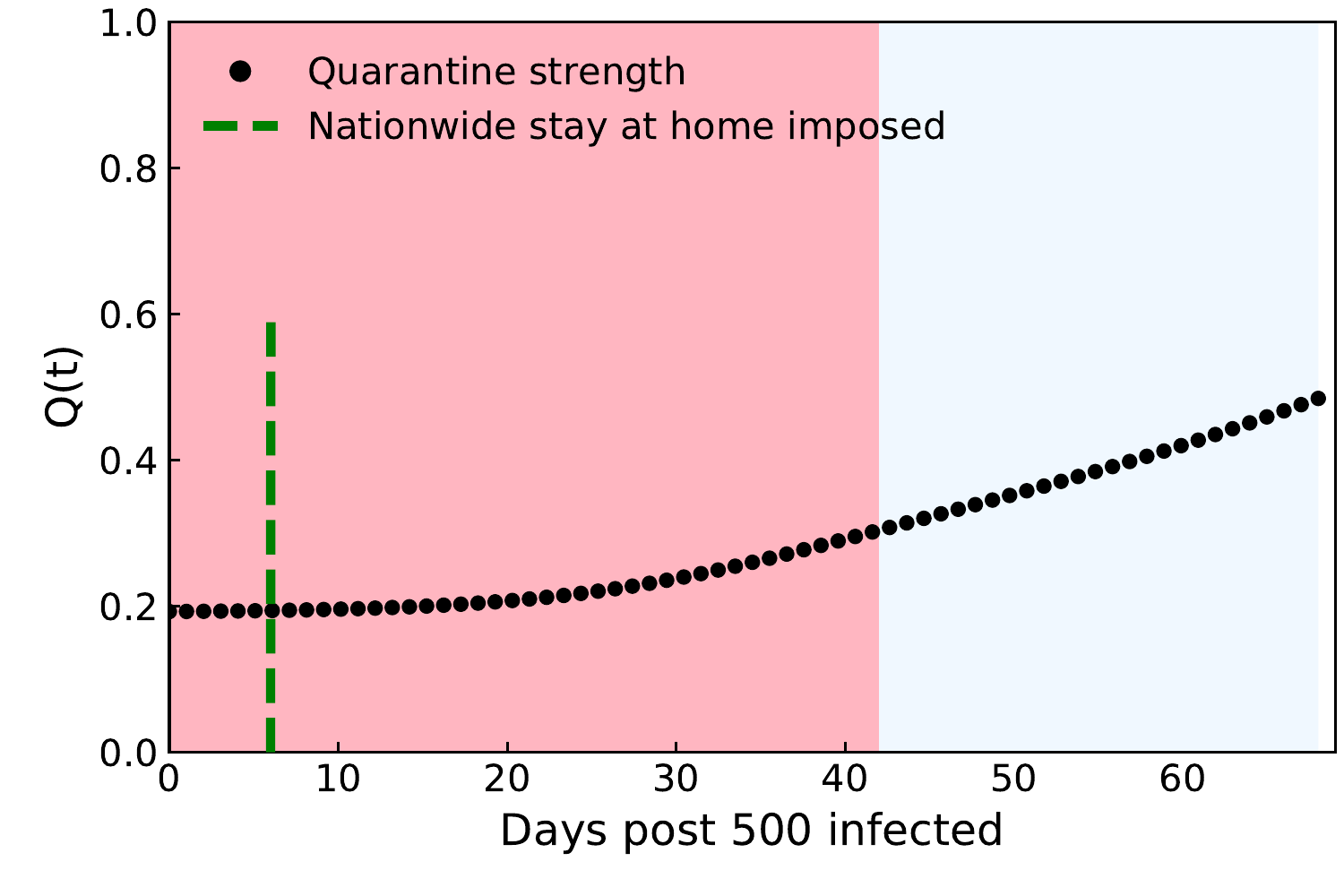}}
\subfloat[][UK]{\includegraphics[width=0.33\textwidth]{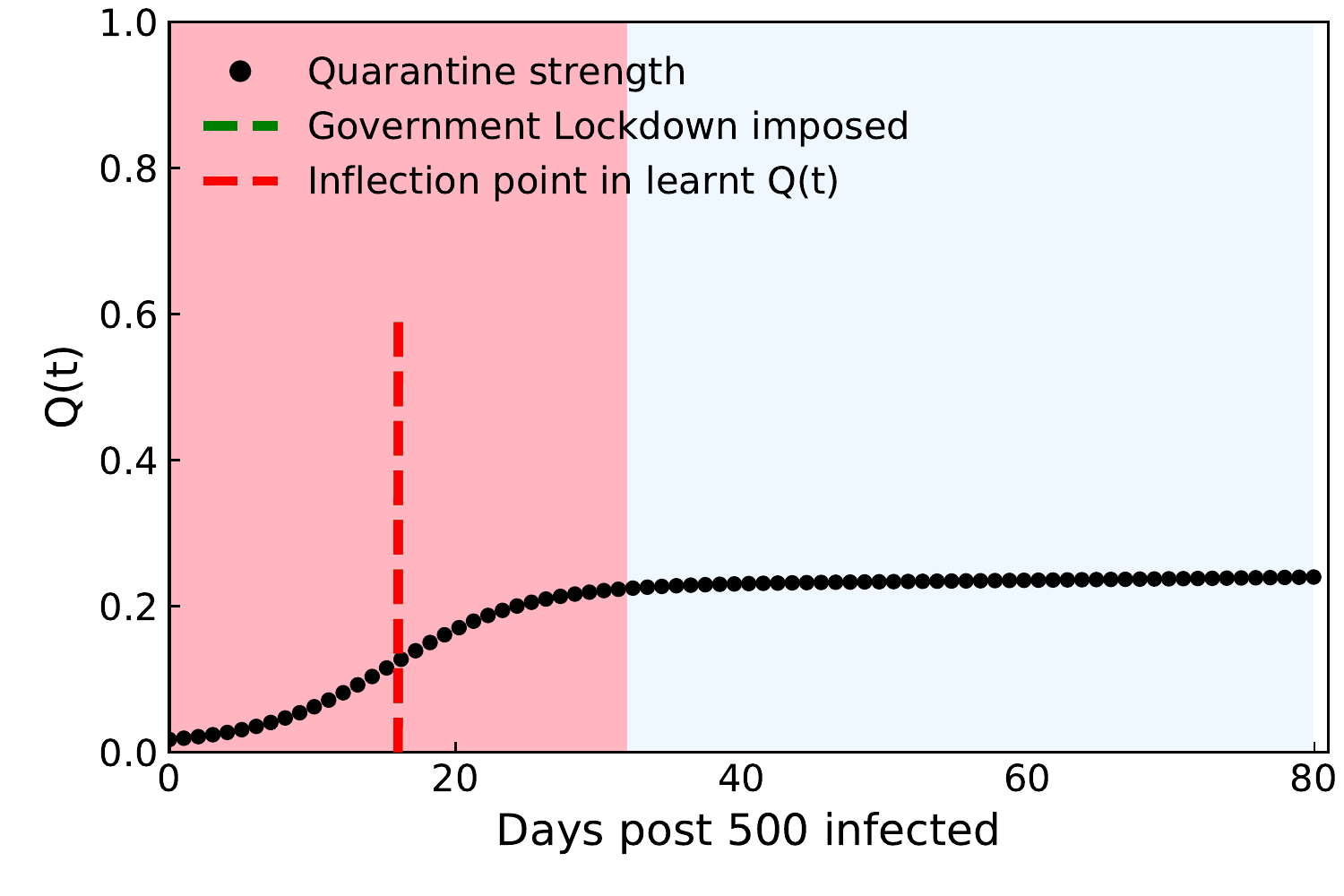}}\\
\subfloat[][Spain]{\includegraphics[width=0.33\textwidth]{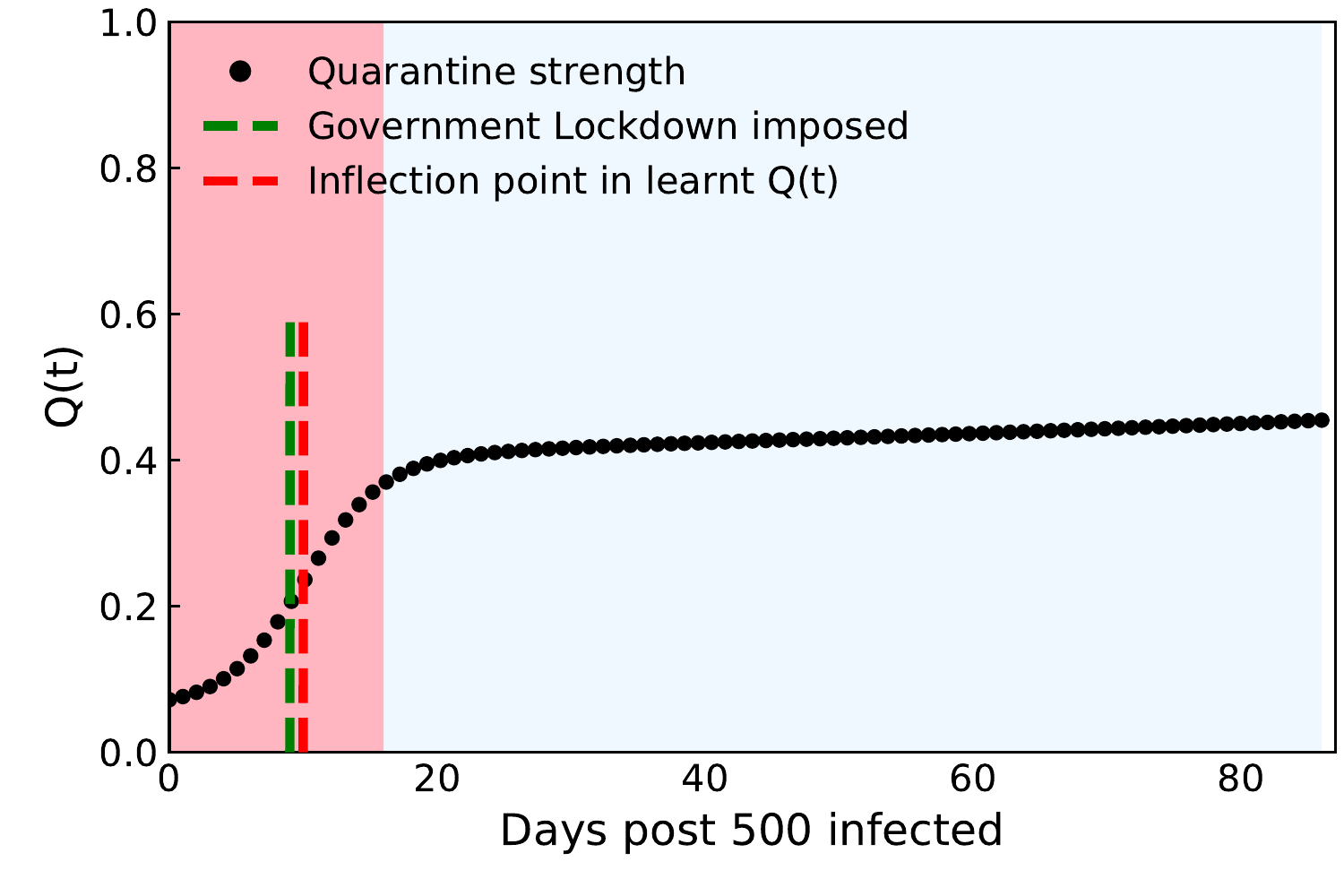}}
\subfloat[][Italy]{\includegraphics[width=0.33\textwidth]{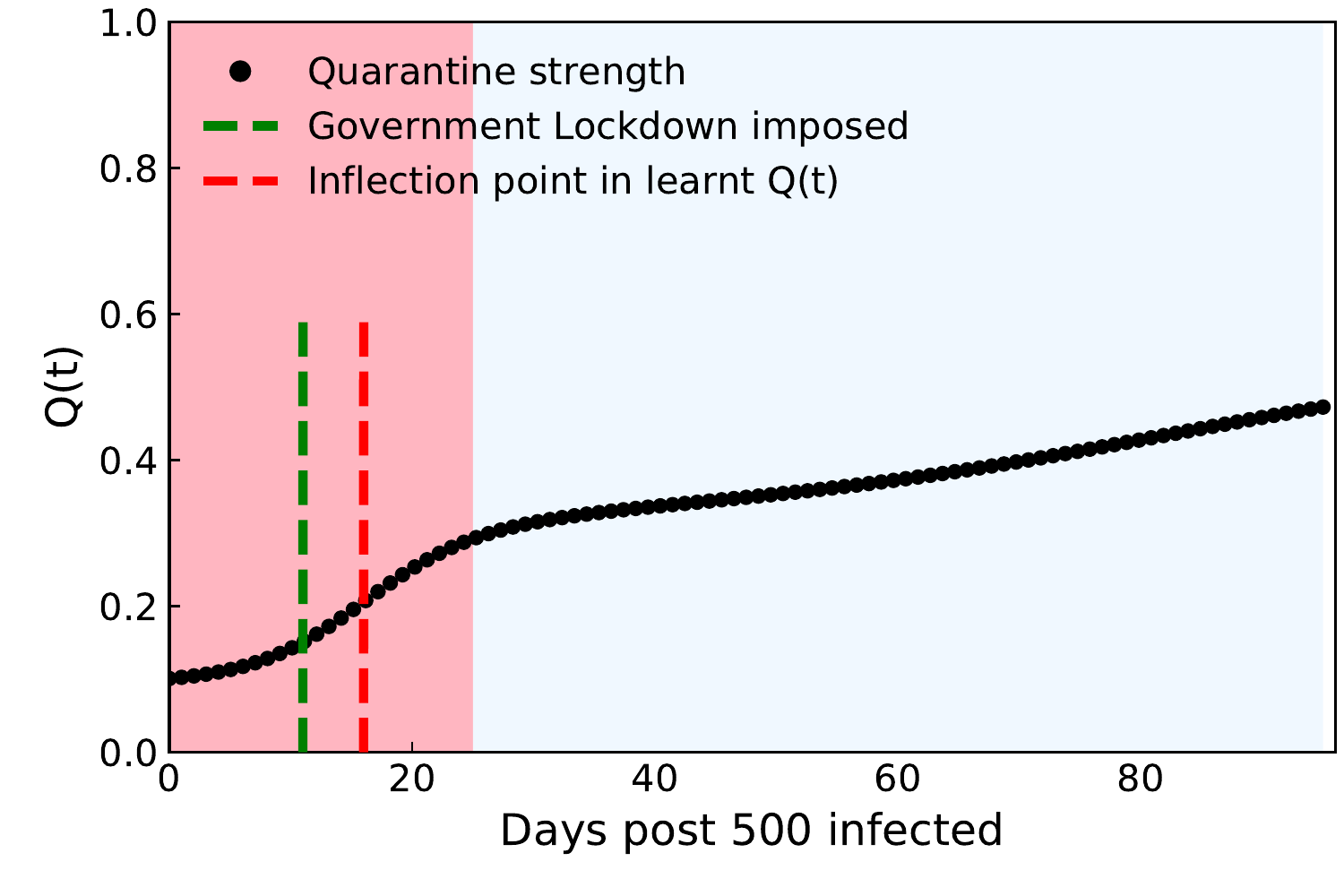}}\\
\end{tabular}
\caption{Quarantine strength $Q(t)$ learned by the neural network in the highest affected European countries as of June 1, 2020. The transition from the red to blue shaded region indicates the Covid spread parameter of value $C_{p} <1$ leading to halting of the infection spread. The green dashed line indicates the time when quarantine measures were implemented in the region under consideration, which generally matches well with an inflection point seen in the $Q(t)$ plot denoted by the red dashed line. For regions in which a clear inflection or ramp up point is not seen (Russia), the red dashed line is not shown.}\label{Europe2}
\end{figure}

\begin{figure}
\centering
\begin{tabular}{cc}
\subfloat[][Russia]{\includegraphics[width=0.33\textwidth]{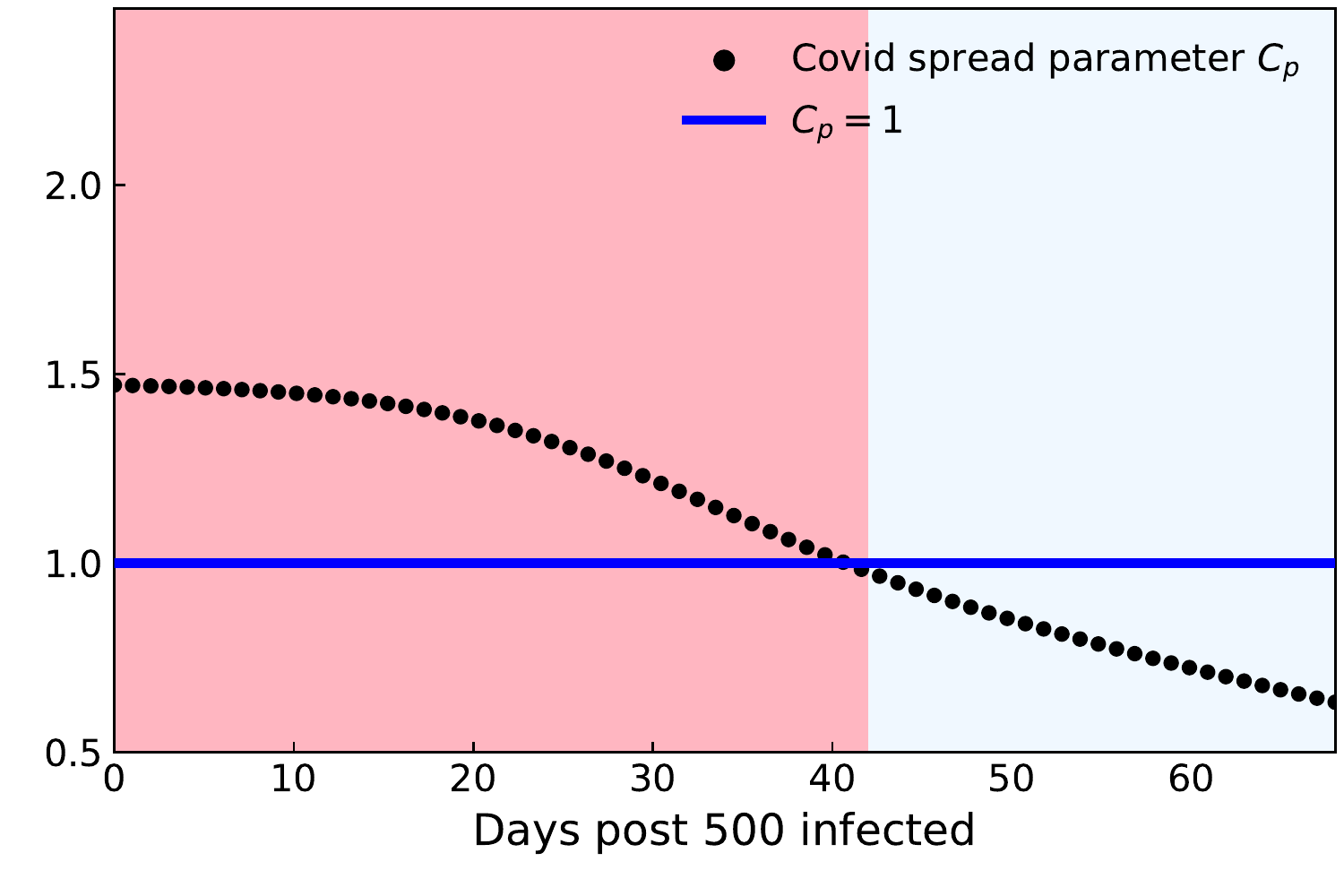}}
\subfloat[][UK]{\includegraphics[width=0.33\textwidth]{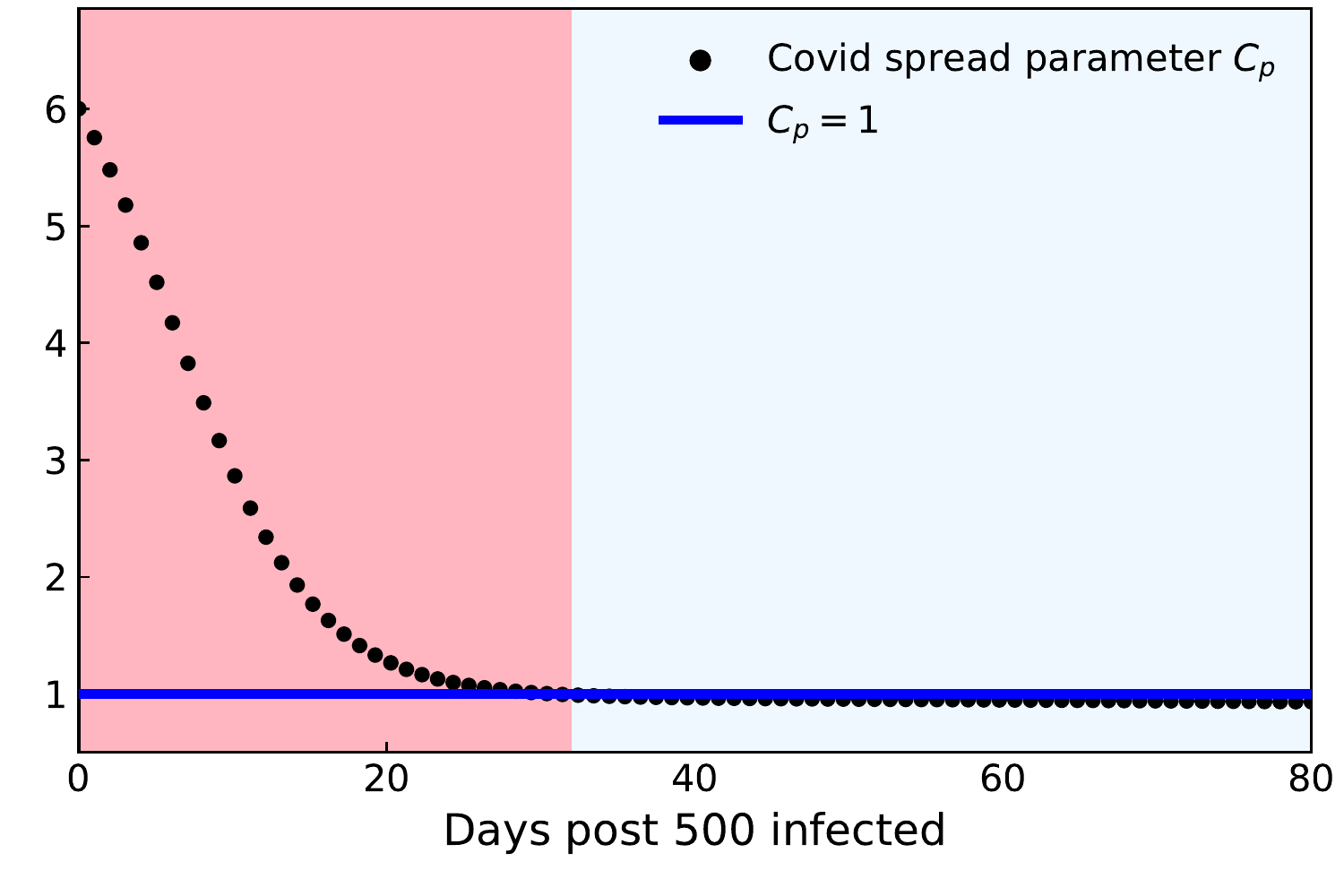}}\\
\subfloat[][Spain]{\includegraphics[width=0.33\textwidth]{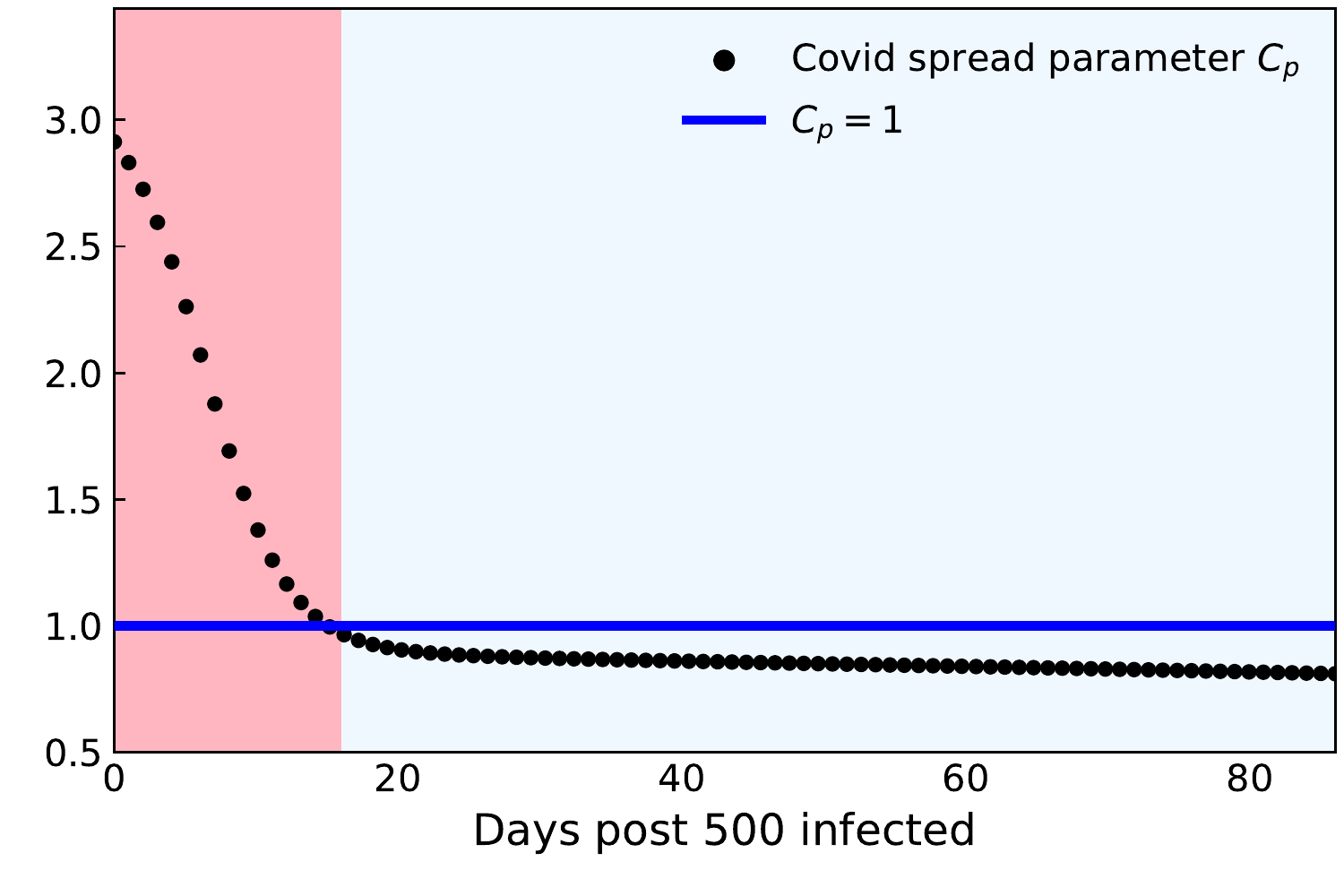}}
\subfloat[][Italy]{\includegraphics[width=0.33\textwidth]{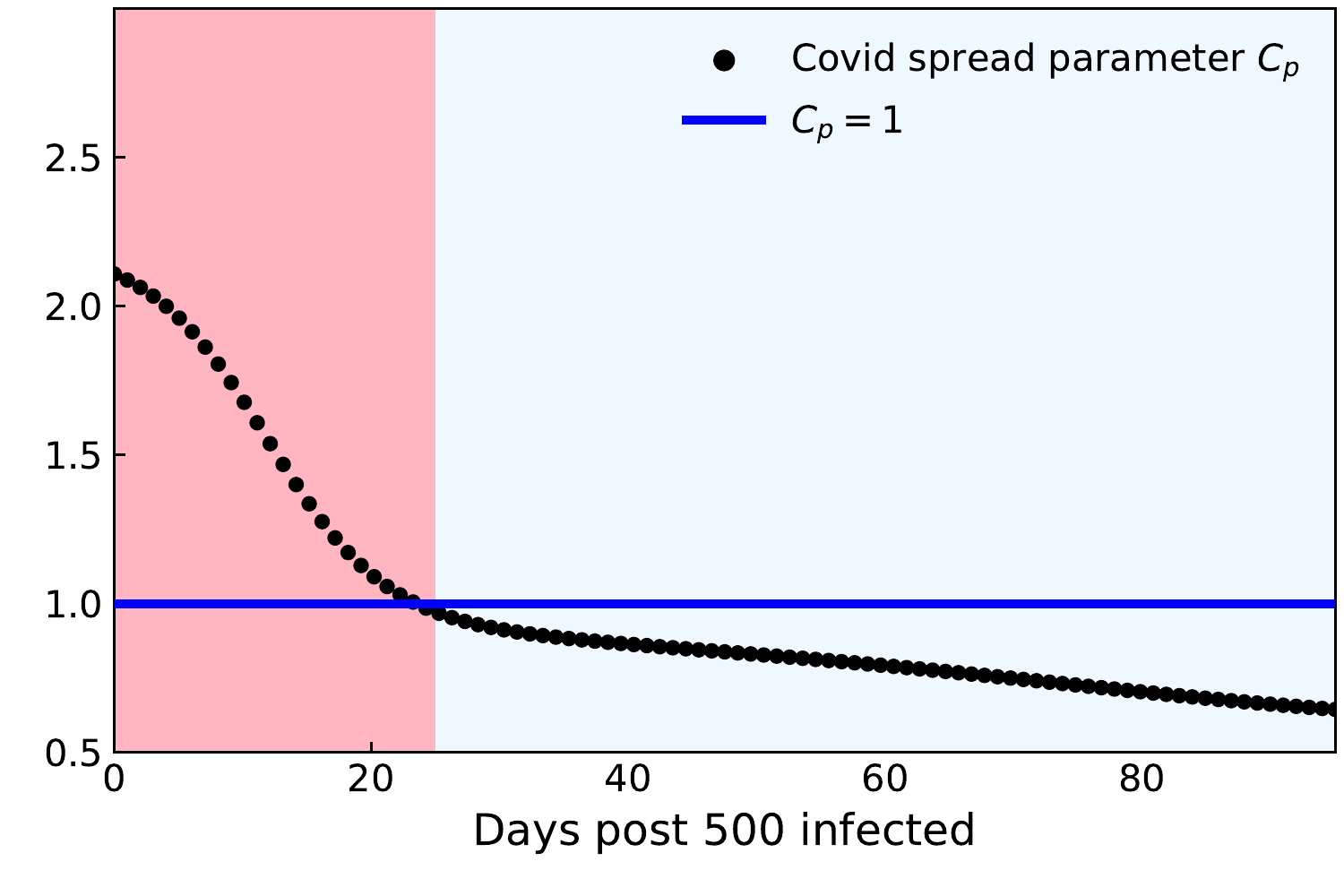}}\\
\end{tabular}
\caption{Control of COVID-19 quantified by the Covid spread parameter evolution in the highest affected European countries as of June 1, 2020. The transition from the red to blue shaded region indicates $C_{p} <1$ leading to halting of the infection spread.}\label{Europe3}
\end{figure}

Figure \ref{Europe1} shows the comparison of the model-estimated infected and recovered case counts with actual Covid-19 data for the highest affected European countries as of $1$ June 2020, namely: Russia, UK, Spain and Italy, in that order. We find that irrespective of a small set of optimized parameters (note that the contact rate $\beta$ and the recovery rate $\gamma$ are fixed, and not functions of time), a reasonably good match is seen in all four cases.\newline

Figure \ref{Europe2} shows the evolution of the neural network learnt quarantine strength function $Q(t)$ for the considered European nations. Inflection points in $Q(t)$ are seen for UK, Spain and Italy at $14$, $10$ and $16$ days, respectively, post detection of the $500^{\textrm{th}}$ case i.e on $23^{\text{th}}$ March, $15^{\text{th}}$ March and $14^{\text{th}}$ March, respectively. This is in good agreement with nationwide quarantine imposed on $25^{\text{th}}$ March, $14^{\text{th}}$ March and $9^{\text{th}}$ March in UK, Spain and Italy, respectively \cite{NatureNews, Spainquar, Italyquar}. \newline

Figure \ref{All}a shows the comparison of the contact rate $\beta$, quarantine efficiency as defined in the beginning of this subsection and  the recovery rate $\gamma$. It should be noted that the contact and recovery rates are assumed to be constant in our model, in the duration spanning the detection of the $500^{\textrm{th}}$ infected case and June $1^{\text{st}}$, 2020. The average contact rate in Spain and Italy is seen to be higher than Russia and UK over the considered duration of $2-3$ months, possibly because Russia and UK were affected relatively late by the virus, which gave sufficient time for the enforcement strict social distancing protocols prior to widespread outbreak. For Spain and Italy, the quarantine efficiency and also the recovery rate are generally higher than for Russia and UK, possibly indicating  more efficient testing, isolation and quarantine; and hospital practices in Spain and Italy. This agrees well with the ineffectiveness of testing, contact tracing and quarantine practices seen in UK \cite{Guardian}. Although the social distancing strength also varied with time, we do not focus on that aspect in the present study, and will be the subject of future studies. A higher quarantine efficiency combined with a higher recovery rate led Spain and Italy to bring down the Covid spread parameter (defined in (\ref{Cp})), $C_{p}$ from $>1$ to $<1$ in $16, 25$ days. respectively, as compared to $32$ days for UK and $42$ days for Russia (figure \ref{Europe3}).

\subsubsection{Quarantine efficiency map for Europe} 
\begin{figure}
\centering
\begin{tabular}{c}
\subfloat[]{\includegraphics[width=0.65\textwidth]{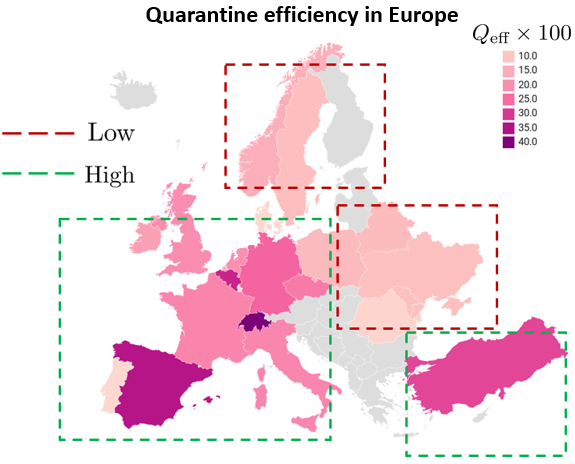}}\\

\end{tabular}
\caption{(a) Quarantine efficiency, $Q_{\textrm{eff}}$ defined in (\ref{Qeff}) for the $23$ highest affected European countries. Note that $Q_{\textrm{eff}}$ contains composite information about the quarantine and lockdown strength; and the testing and tracing protocols to identify and isolate infected individuals. Map also shows the demarcation between countries with a high $Q_{\textrm{eff}}$ shown by a green dotted line and those with a low $Q_{\textrm{eff}}$ shown by a red dotted line.}\label{Qeff_schem_europe}
\end{figure}

Figure \ref{Qeff_schem_europe} shows $Q_{\textrm{eff}}$ for the $23$ highest affected European countries. We can see that $Q_{\textrm{eff}}$ in the western European regions is generally higher than eastern Europe. This can be attributed to the strong lockdown measures implemented  in western countries like Spain, Italy, Germany, France after the rise of infections seen first in Italy and Spain \cite{Europequar}. Although countries like Switzerland and Turkey didn't enforce a strict lockdown as compared to their west European counterparts, they were generally successful in halting the infection count before reaching catastrophic proportions, due to strong testing and tracing protocols \cite{Swissquar, Turkeyquar}. Subsequently, these countries also managed to identify potentially infected individuals and prevented them from coming into contact with susceptibles, giving them a high $Q_{\textrm{eff}}$ score as seen in figure \ref{Qeff_schem_europe}. In contrast, our study also manages to identify countries like Sweden which had very limited lockdown measures \cite{Swedenquar}; with a low $Q_{\textrm{eff}}$ score as seen in figure \ref{Qeff_schem_europe}. This strengthens the validity of our model in diagnosing information about the effectiveness of quarantine and isolation protocols in different countries; which agree well with the actual protocols seen in these countries.

\subsection{USA}
\begin{figure}
\centering
\begin{tabular}{cc}
\subfloat[][New York]{\includegraphics[width=0.5\textwidth]{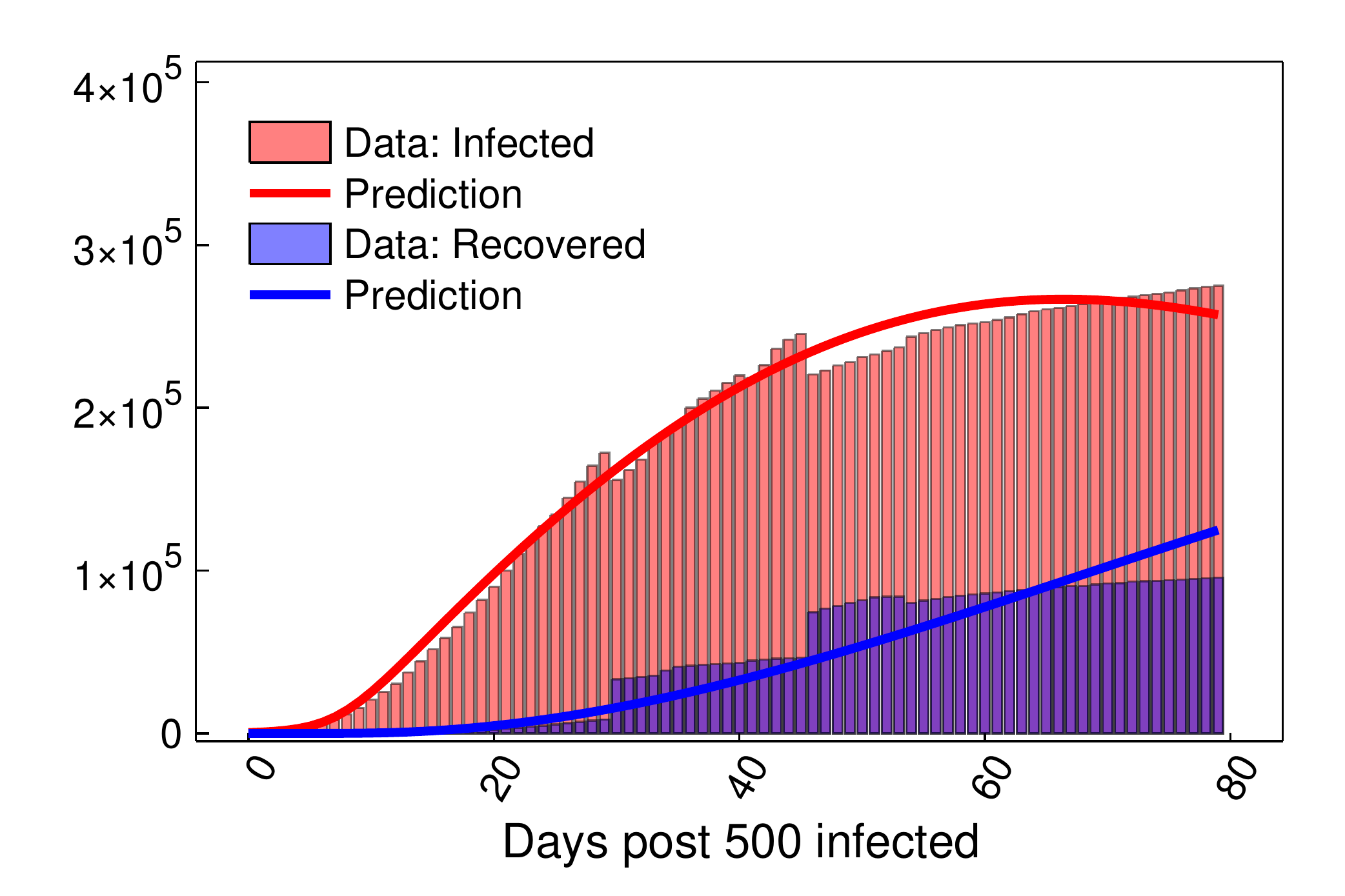}}
\subfloat[][New Jersey]{\includegraphics[width=0.5\textwidth]{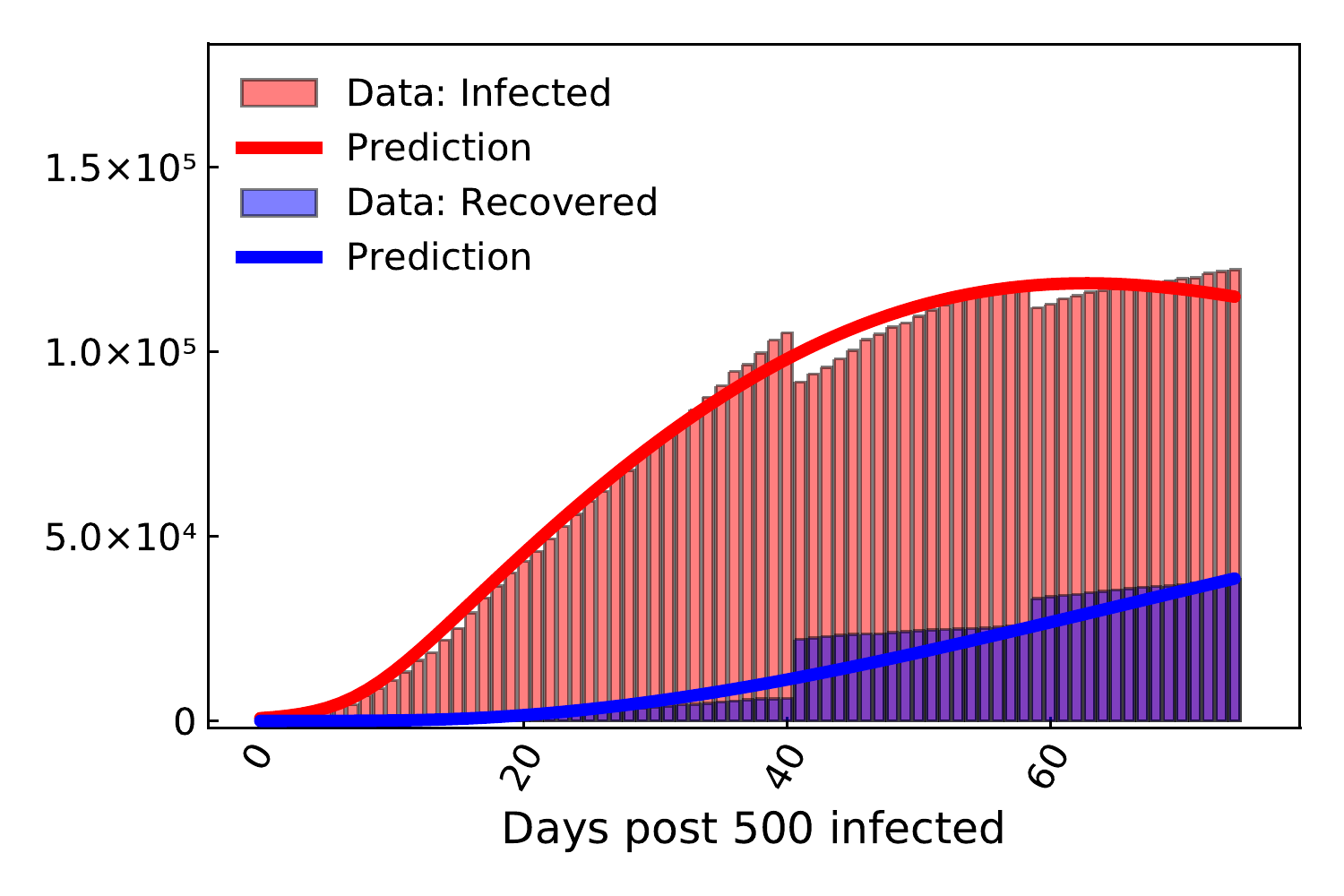}}\\
\subfloat[][Illinois]{\includegraphics[width=0.5\textwidth]{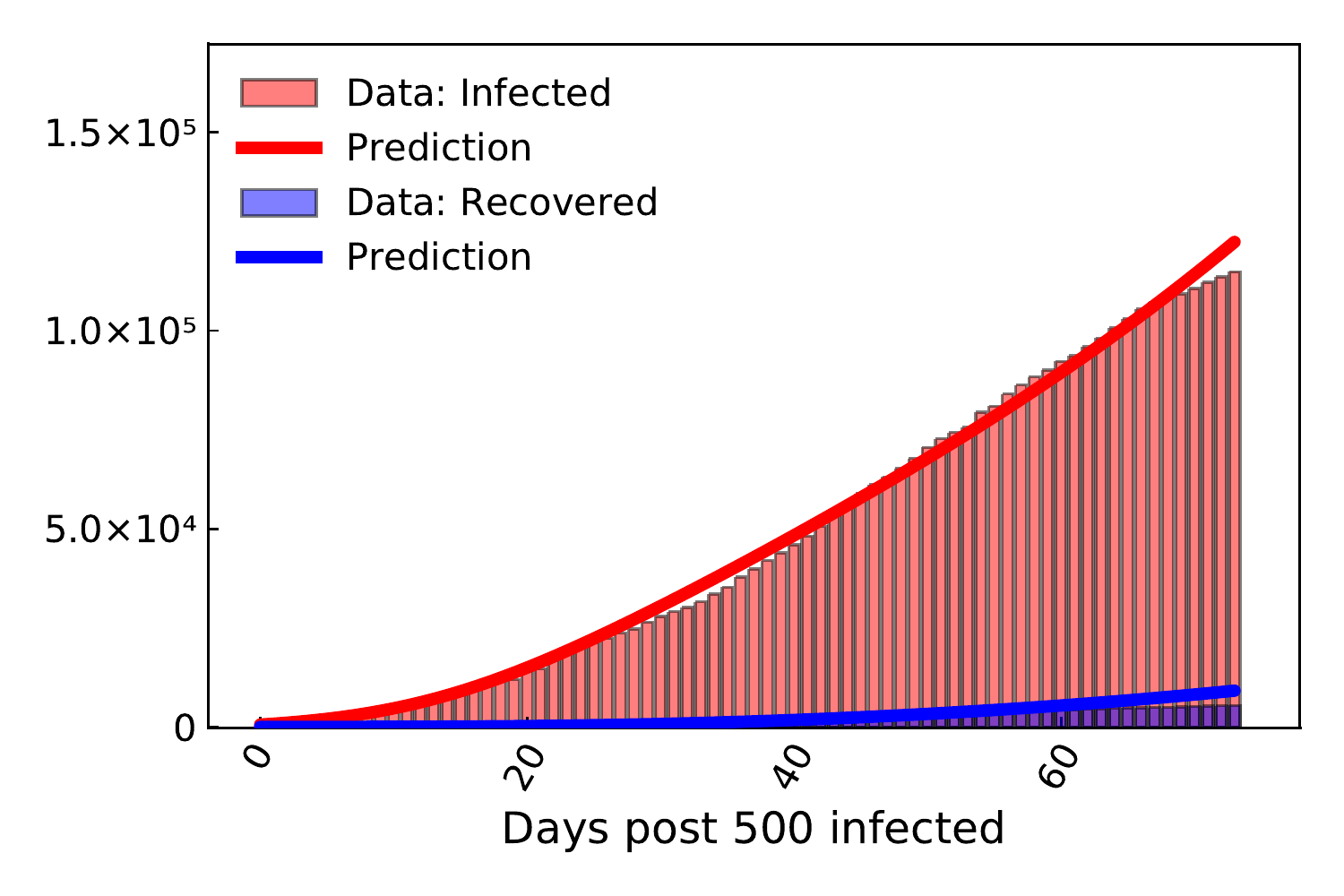}}
\subfloat[][California]{\includegraphics[width=0.5\textwidth]{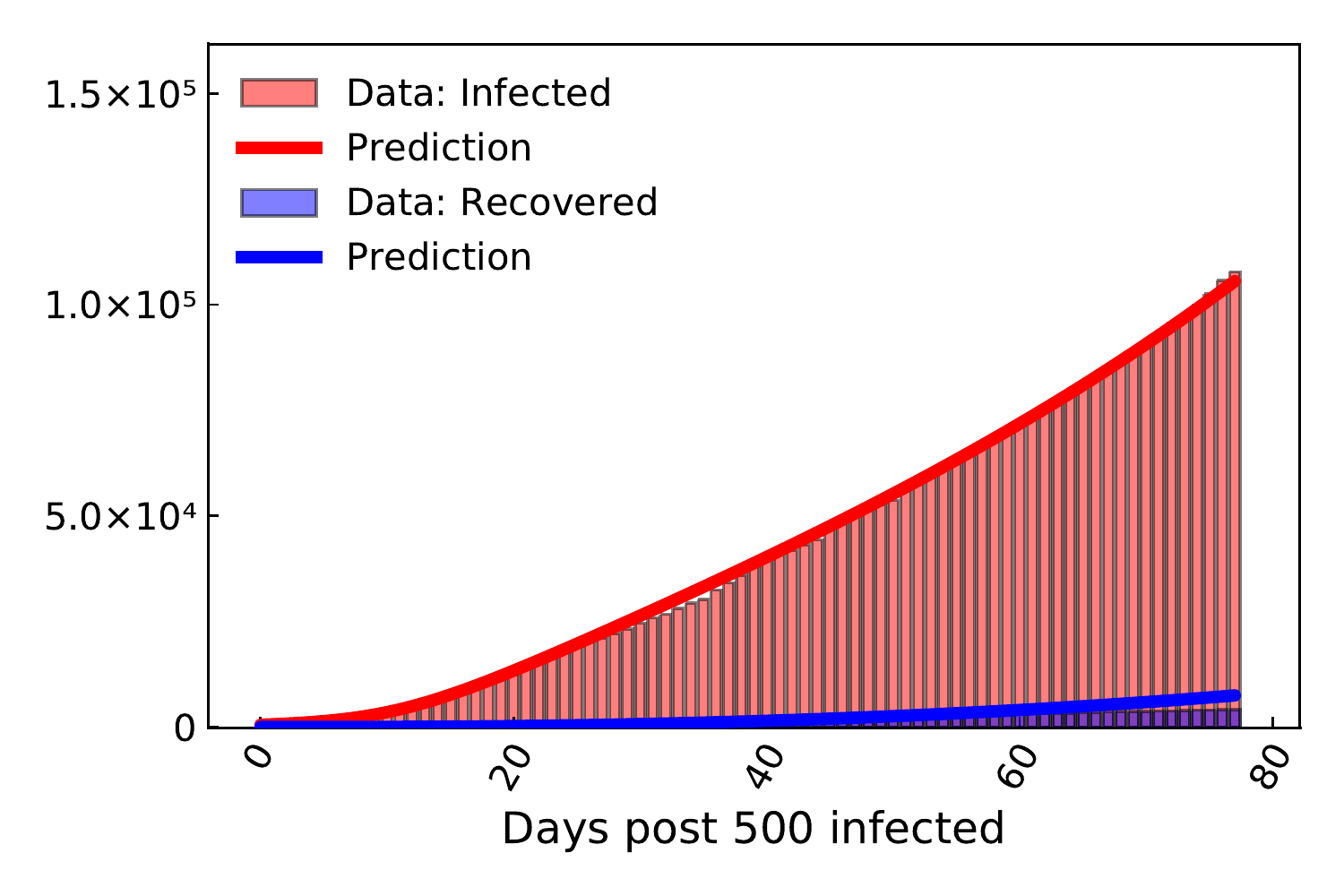}}\\
\end{tabular}
\caption{COVID-19 infected and recovered evolution compared with our neural network augmented model prediction in the highest affected USA states as of June 1, 2020.}\label{NAmerica1}
\end{figure}

\begin{figure}
\centering
\begin{tabular}{cc}
\subfloat[][New York]{\includegraphics[width=0.33\textwidth]{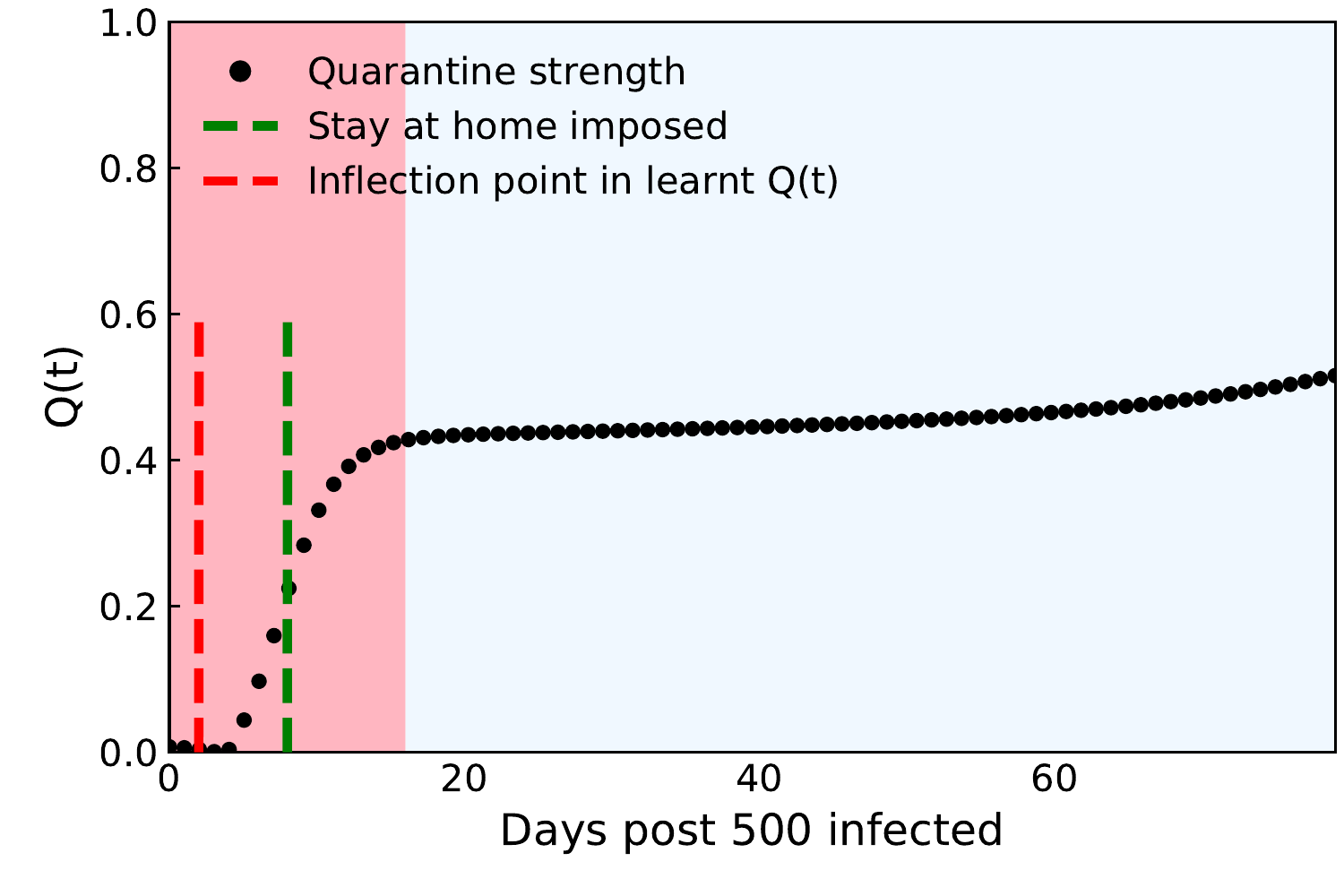}}
\subfloat[][New Jersey]{\includegraphics[width=0.33\textwidth]{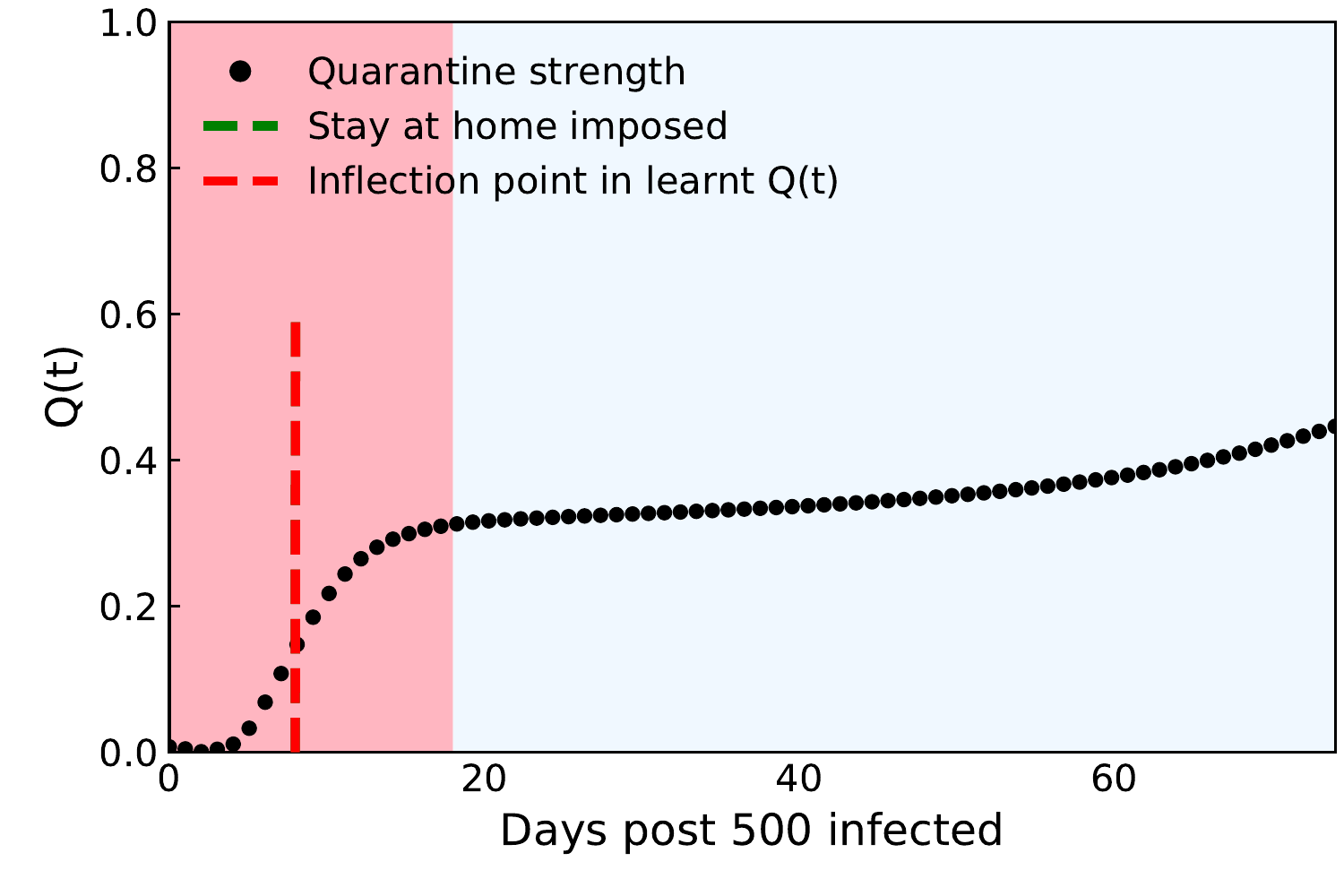}}\\
\subfloat[][Illinois]{\includegraphics[width=0.33\textwidth]{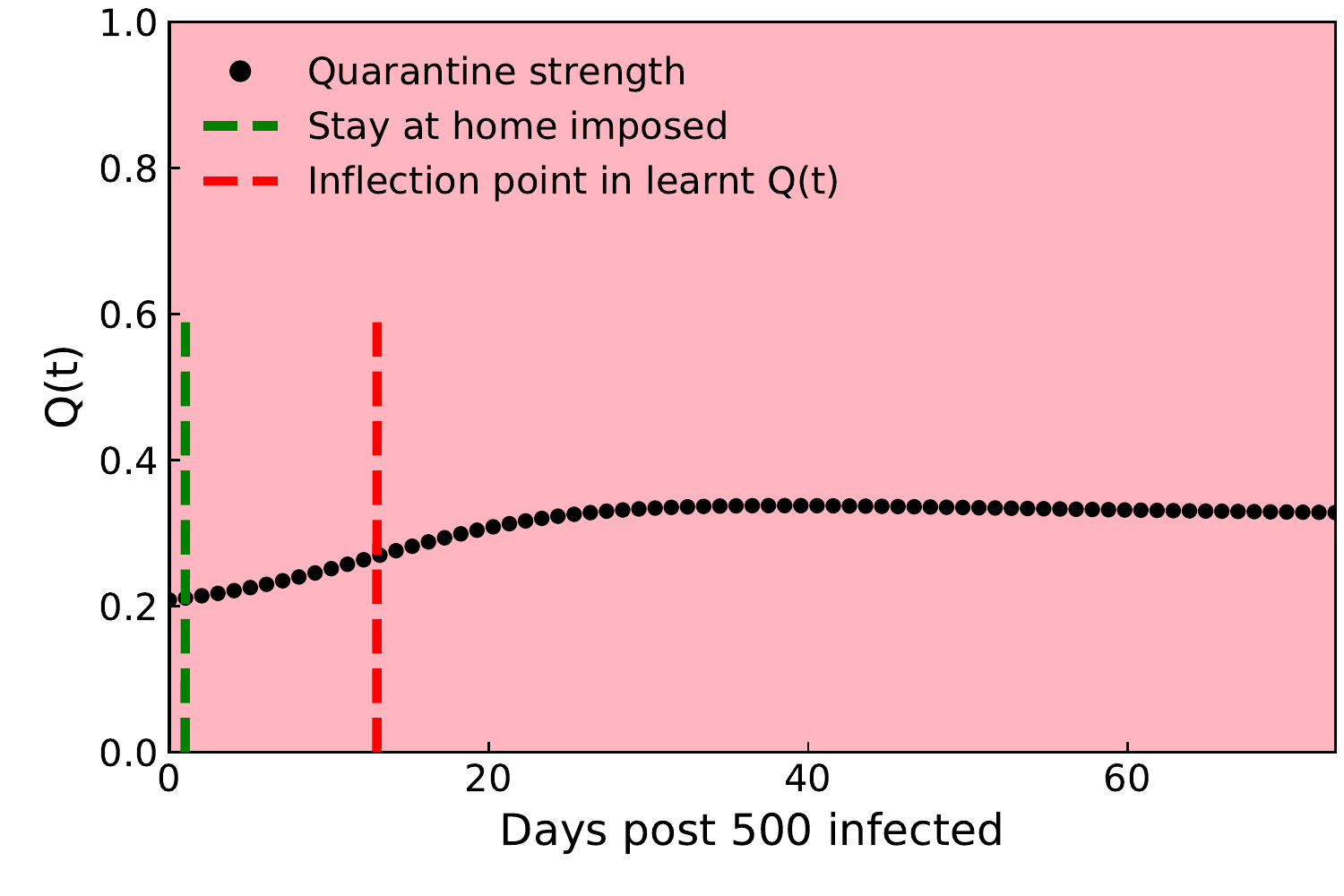}}
\subfloat[][California]{\includegraphics[width=0.33\textwidth]{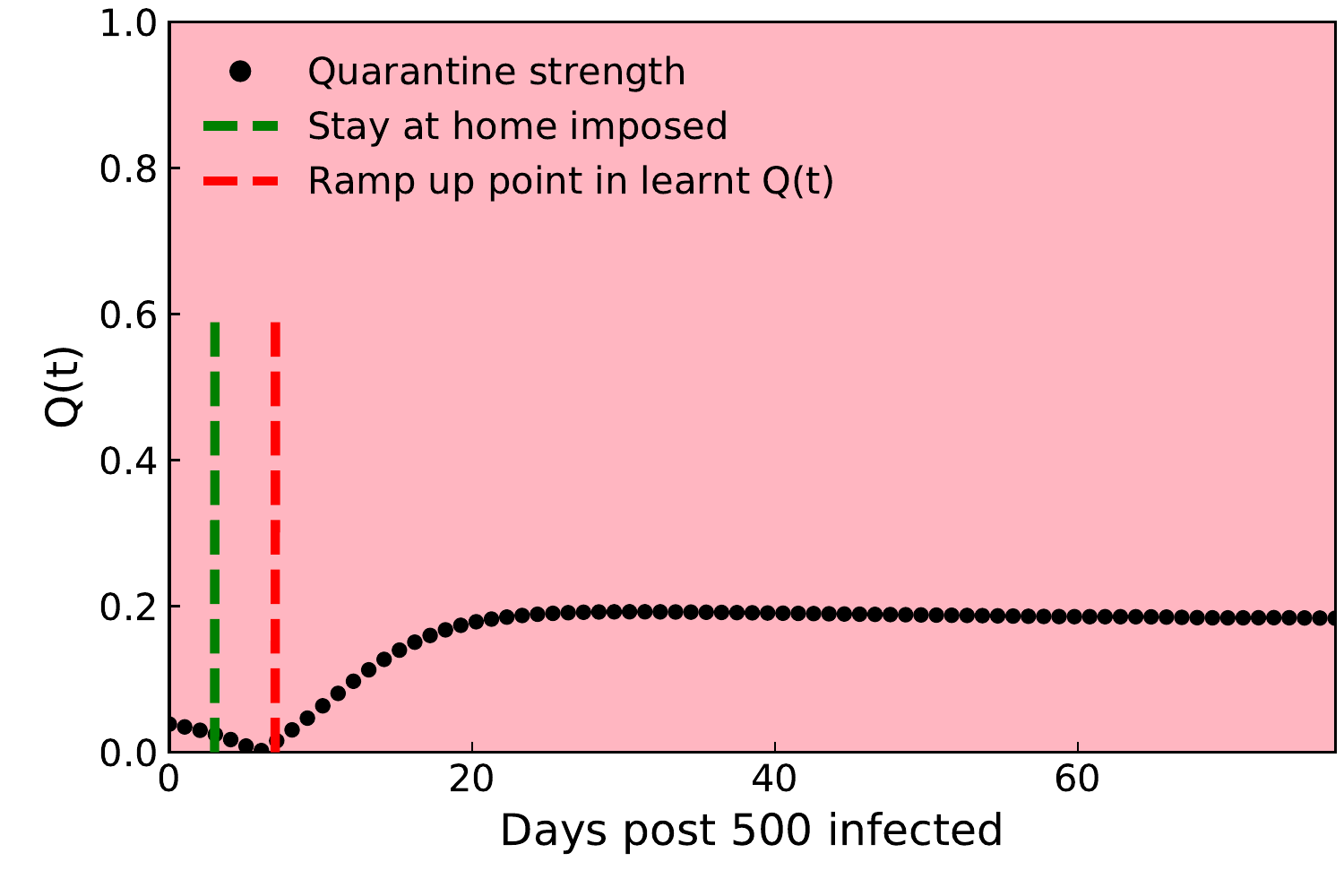}}\\
\end{tabular}
\caption{Quarantine strength $Q(t)$ learned by the neural network in the highest affected USA states as of June 1, 2020. The transition from the red to blue shaded region indicates the Covid spread parameter of value $C_{p} <1$ leading to halting of the infection spread. The green dashed line indicates the time when quarantine measures were implemented in the region under consideration, which generally matches well with an inflection point (for New York, New Jersey and Illinois) or a ramp up point (California) seen in the $Q(t)$ plot denoted by the red dashed line.}\label{NAmerica2}
\end{figure}

\begin{figure}
\centering
\begin{tabular}{cc}
\subfloat[][New York]{\includegraphics[width=0.33\textwidth]{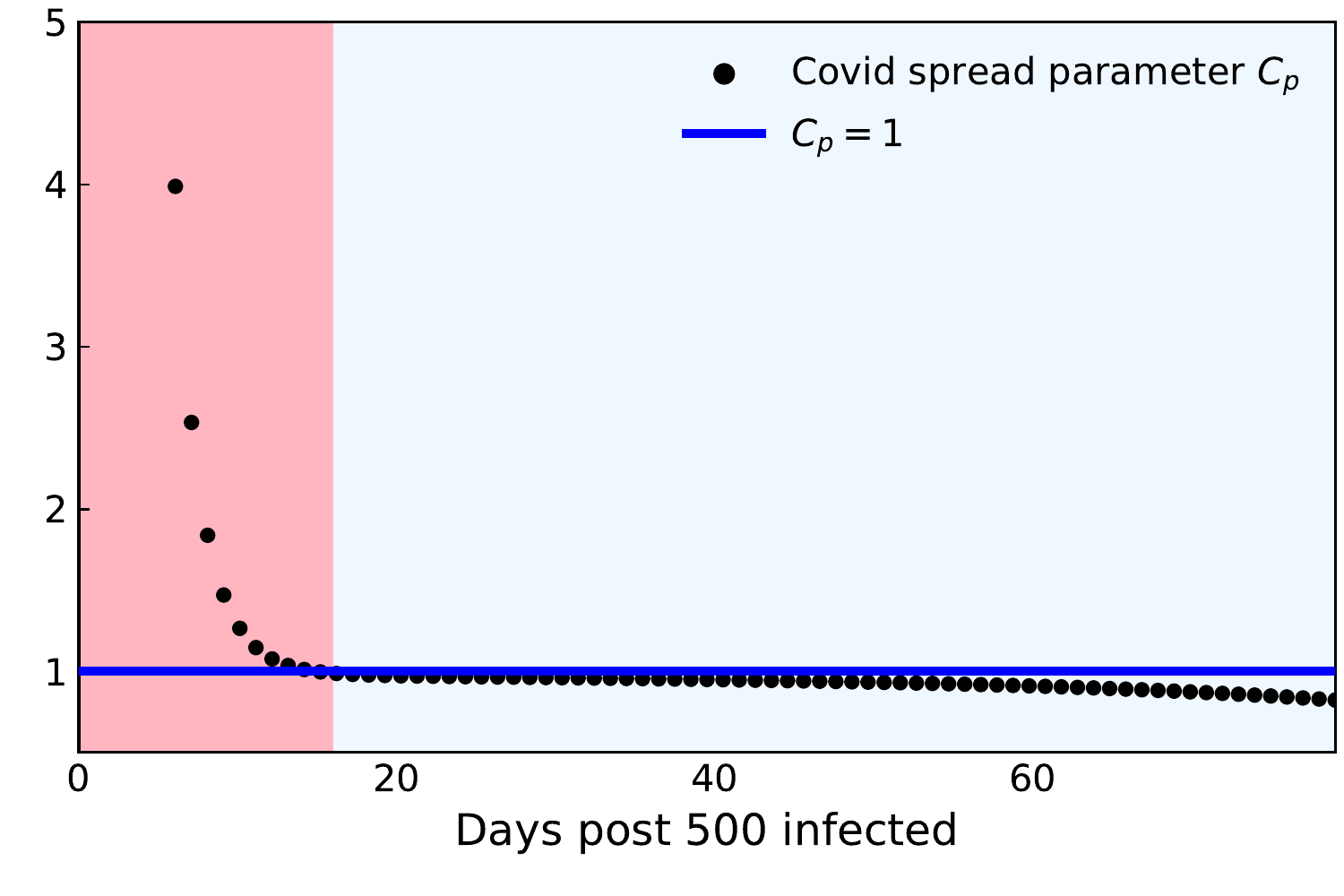}}
\subfloat[][New Jersey]{\includegraphics[width=0.33\textwidth]{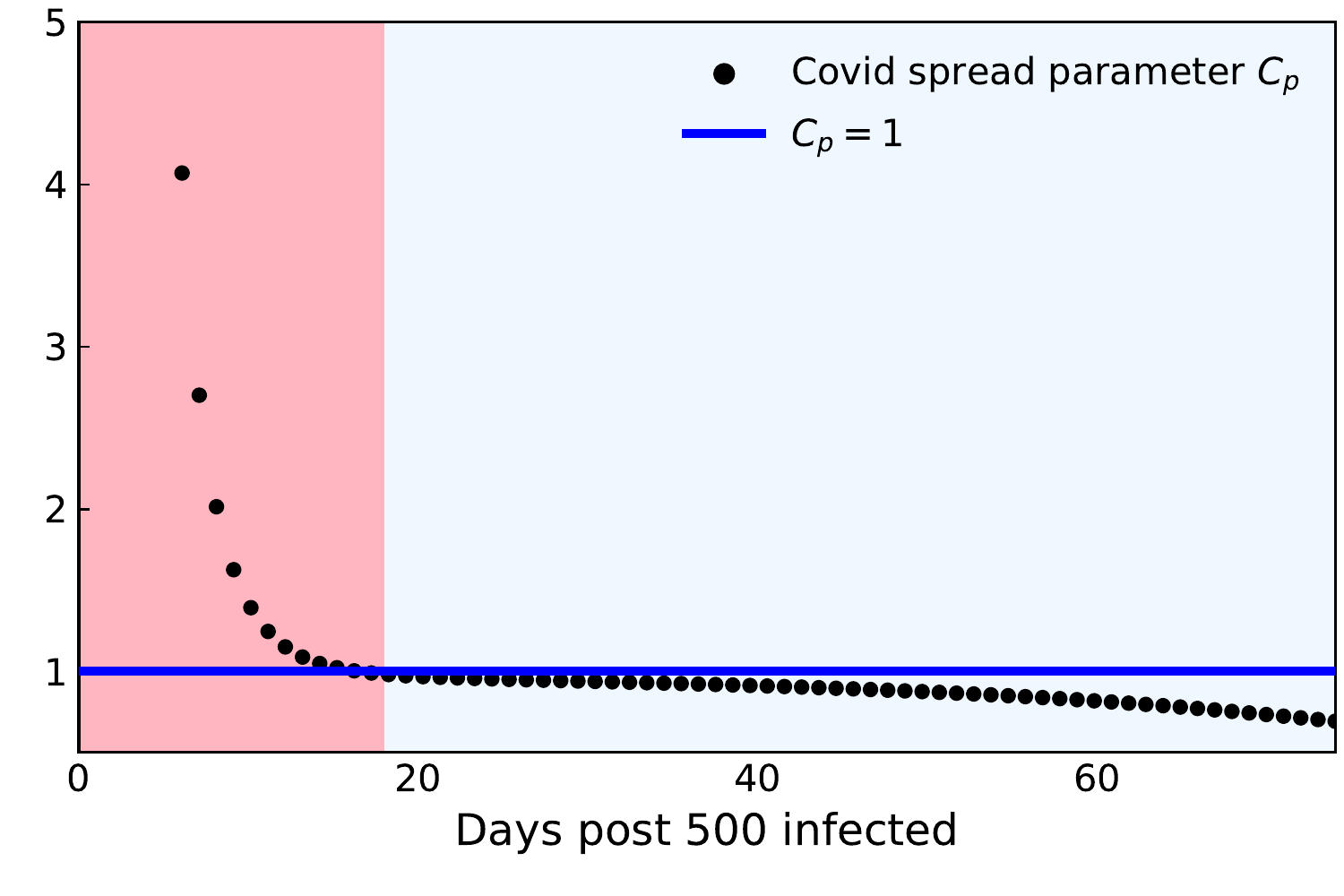}}\\
\subfloat[][Illinois]{\includegraphics[width=0.33\textwidth]{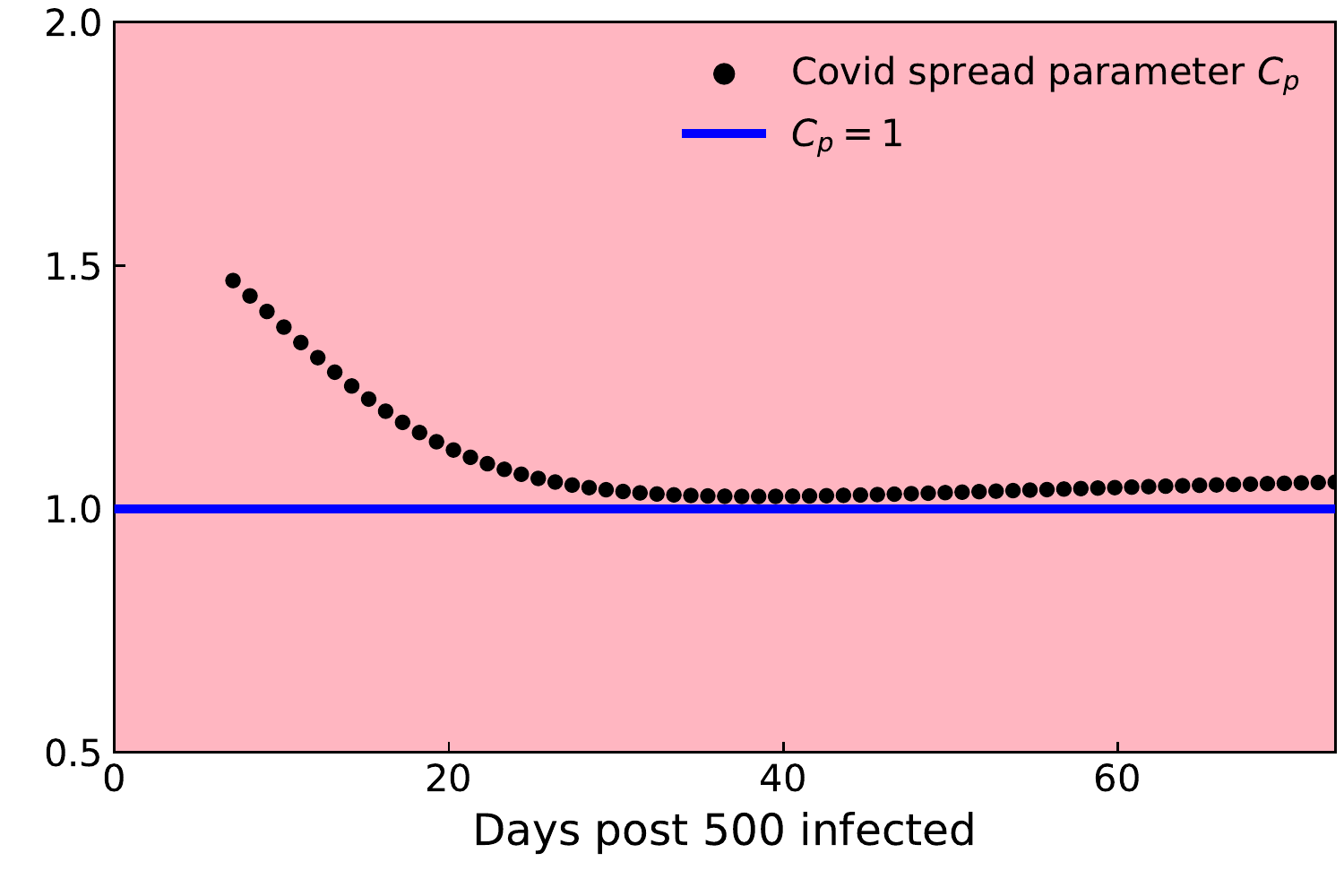}}
\subfloat[][California]{\includegraphics[width=0.33\textwidth]{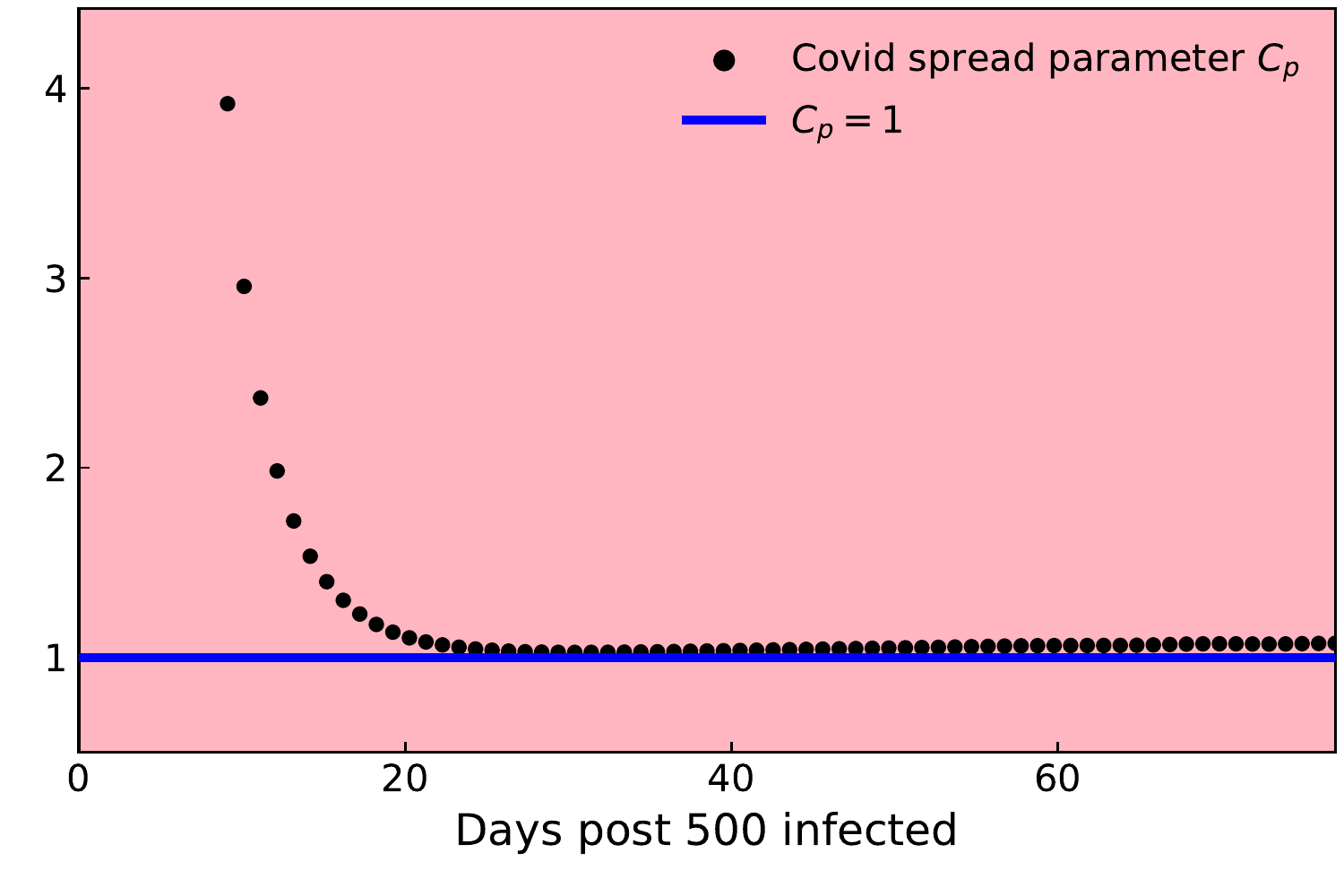}}\\
\end{tabular}
\caption{Control of COVID-19 quantified by the Covid spread parameter evolution in the highest affected USA states as of June 1, 2020. The transition from the red to blue shaded region indicates $C_{p} <1$ leading to halting of the infection spread.}\label{NAmerica3}
\end{figure}

Figure \ref{NAmerica1} shows  reasonably good match between the model-estimated infected and recovered case counts with actual Covid-19 data for the highest affected North American states (including states from Mexico, the United States, and Canada) as of $1$ June 2020, namely: New York, New Jersey, Illinois and California. $Q(t)$ for New York and New Jersey show a ramp up point immediately in the week following the detection of the $500^{\textrm{th}}$ case in these regions, {\it i.e.} on $19$ March for New York and on $24$ March for New Jersey (figure \ref{NAmerica2}). This matches well with the actual dates: $22$ March in New York and $21$ March in New Jersey when stay at home orders and isolation measures were enforced in these states. A relatively slower rise of $Q(t)$ is seen for Illinois while California showing a ramp up post a week after detection of the $500^{\textrm{th}}$ case. Although no significant difference is seen in the mean contact and recovery rates between the different US states, the quarantine efficiency in New York and New Jersey is seen to be significantly higher than that of Illinois and California (figure \ref{All}b), indicating the effectiveness of the rapidly deployed quarantine interventions in New York and New Jersey \cite{USDrastic}. Owing to the high quarantine efficiency in New York and New Jersey, these states were able to bring down the Covid spread parameter, $C_{p}$ to less than $1$ in $19$ days (figure \ref{NAmerica3}). On the other hand, although Illinois and California reached close to $C_{p} = 1$ after the $30$ day and $20$ day mark respectively, $C_{p}$ still remained greater than $1$ (figure \ref{NAmerica3}), indicating that these states were still in the danger zone as of June 1, 2020. An important caveat to this result is the reporting of the recovered data.\newline

Comparing with Europe, the recovery rates seen in North America are significantly lower (figures \ref{All}a,b). It should be noted that accurate reporting of recovery rates is likely to play a major role in this apparent difference. In our study, the recovered data include individuals who cannot further transmit infection; and thus includes treated patients who are currently in a healthy state and also individuals who died due to the virus. Since quantification of deaths can be done in a robust manner, the death data is generally reported more accurately. However, there is no clear definition for quantifying the number of people who transitioned from infected to healthy. As a result, accurate and timely reporting of recovered data is seen to have a significant variation between countries, under reporting of the recovered data being a common practice. Since the effective reproduction number calculation depends on the recovered case count, accurate data regarding the recovered count is vital to assess whether the infection has been curtailed in a particular region or not. Thus, our results strongly indicate the need for each country to follow a particular metric for estimating the recovered count robustly, which is vital for data driven assessment of the pandemic spread.\newline

\begin{figure}
\centering
\begin{tabular}{cc}
\subfloat[]{\includegraphics[width=0.45\textwidth]{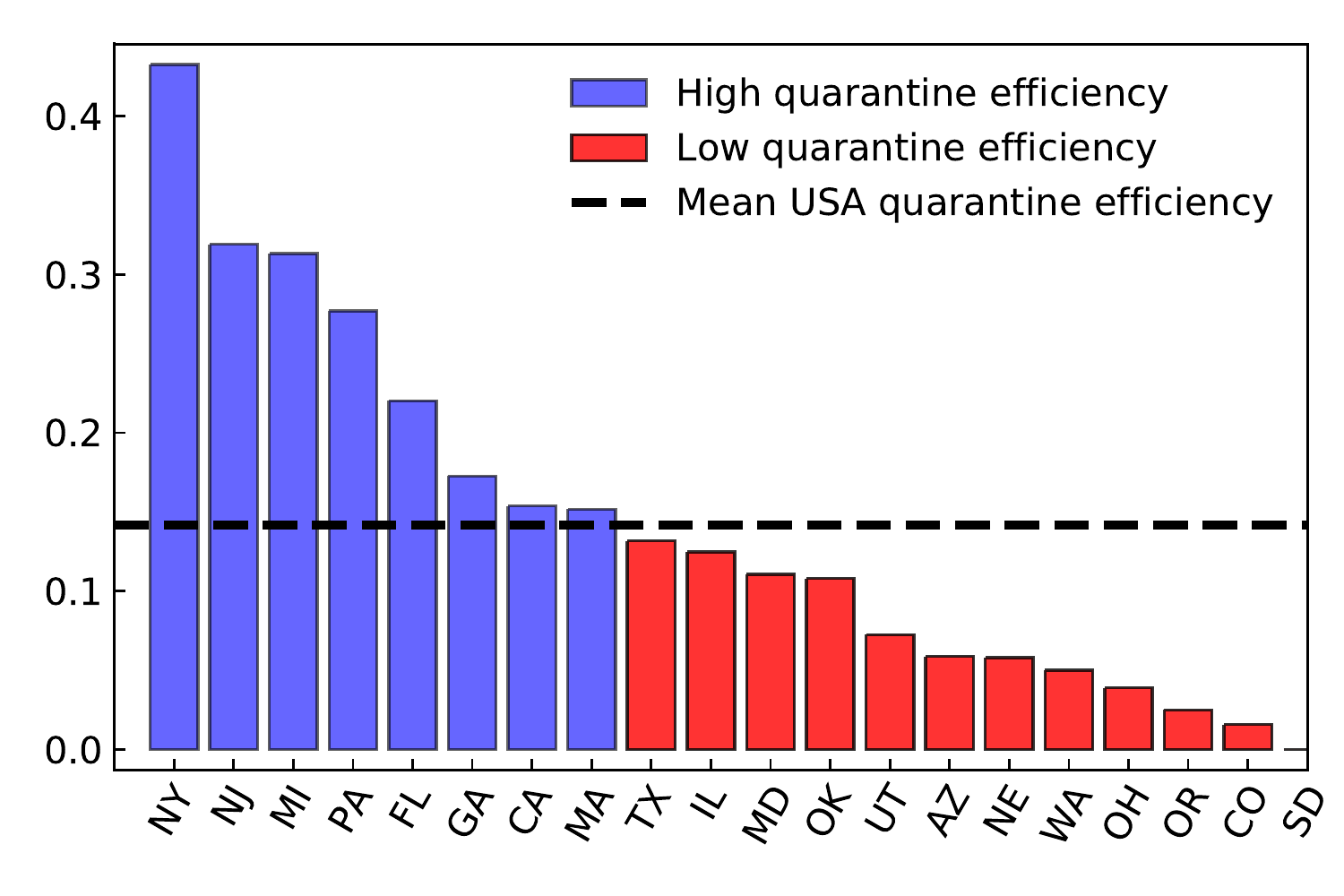}}\\
\subfloat[]{\includegraphics[width=\textwidth]{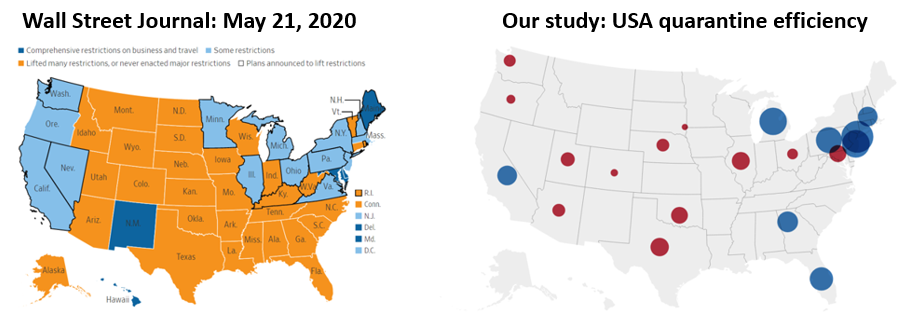}}

\end{tabular}
\caption{(a) Quarantine efficiency, $Q_{\textrm{eff}}$ defined in (\ref{Qeff}) for 20 major USA states. Note that $Q_{eff}$ contains composite information about the quarantine and lockdown strength; and the testing and tracing protocols to identify and isolate infected individuals. (b) Comparison between a report published in the Wall Street Journal on May 21 \cite{WSJMap} and the quarantine efficiency magnitude in our study. A generally strong correlation is seen between the magnitude of quarantine efficiency in our study and the level of restrictions actually imposed in different USA states.}\label{Qeff_schem}
\end{figure}

\subsubsection{Quarantine efficiency map for USA} 

Figure \ref{Qeff_schem}a shows the quarantine efficiency for 20 major US states spanning the whole country. Figure \ref{Qeff_schem}b shows the comparison between a report published in the Wall Street Journal on May 21 highlighting USA states based on their lockdown conditions \cite{WSJMap}, and the quarantine efficiency magnitude in our study. The size of the circles represent the magnitude of the quarantine efficiency. The blue color indicate the states for which the quarantine efficiency was greater than the mean quarantine efficiency across all US states, while those in red indicate the opposite. Our results indicate that the north-eastern and western states were much more responsive in implementing rapid quarantine measures in the month following early detection; as compared to the southern and central states. This matches the on-ground situation as indicated by a generally strong correlation is seen between the red circles in our study (states with lower quarantine efficiency) and the yellow regions seen in in the Wall Street Journal report \cite{WSJMap} (states with reduced imposition of restrictions) and between the blue circles in our study (states with higher quarantine efficiency) and the blue regions seen in the Wall Street Journal report \cite{WSJMap} (states with generally higher level of restrictions). This strengthens the validity of our approach in which the quarantine efficiency is recovered through a trained neural network rooted in fundamental epidemiological equations.

\subsection{Asia}
\begin{figure}
\centering
\begin{tabular}{cc}
\subfloat[][India]{\includegraphics[width=0.5\textwidth]{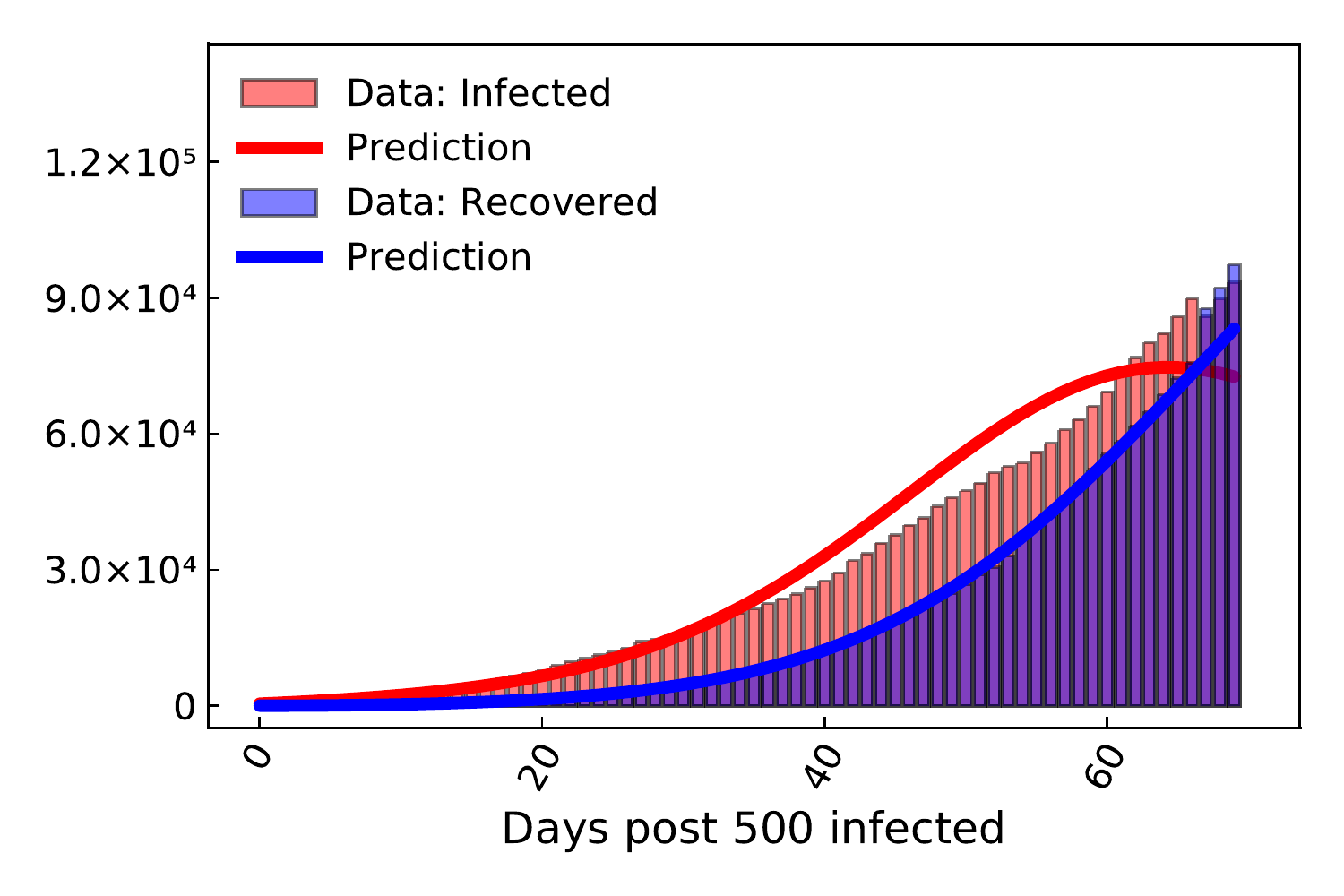}}
\subfloat[][China]{\includegraphics[width=0.5\textwidth]{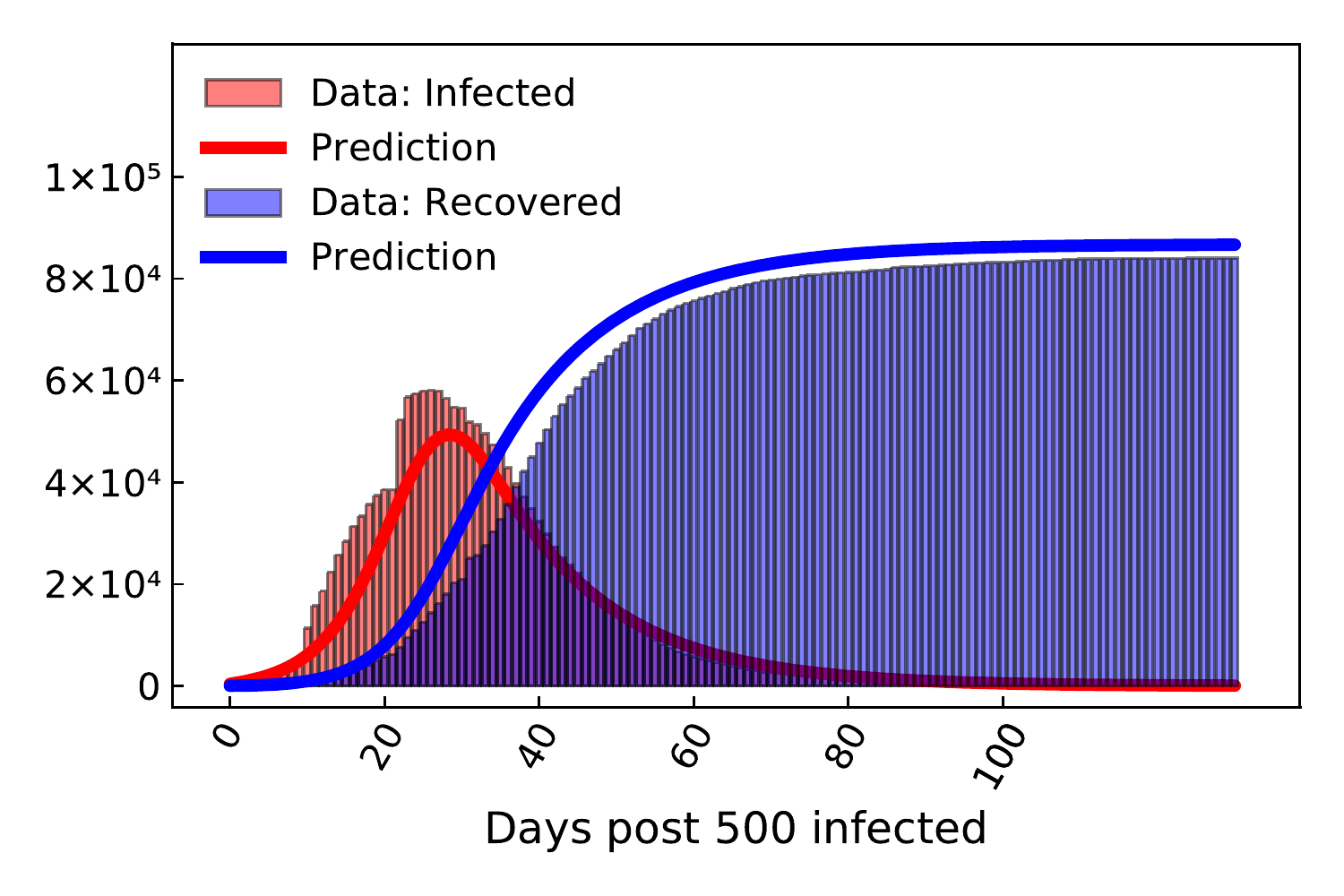}}\\
\subfloat[][South Korea]{\includegraphics[width=0.5\textwidth]{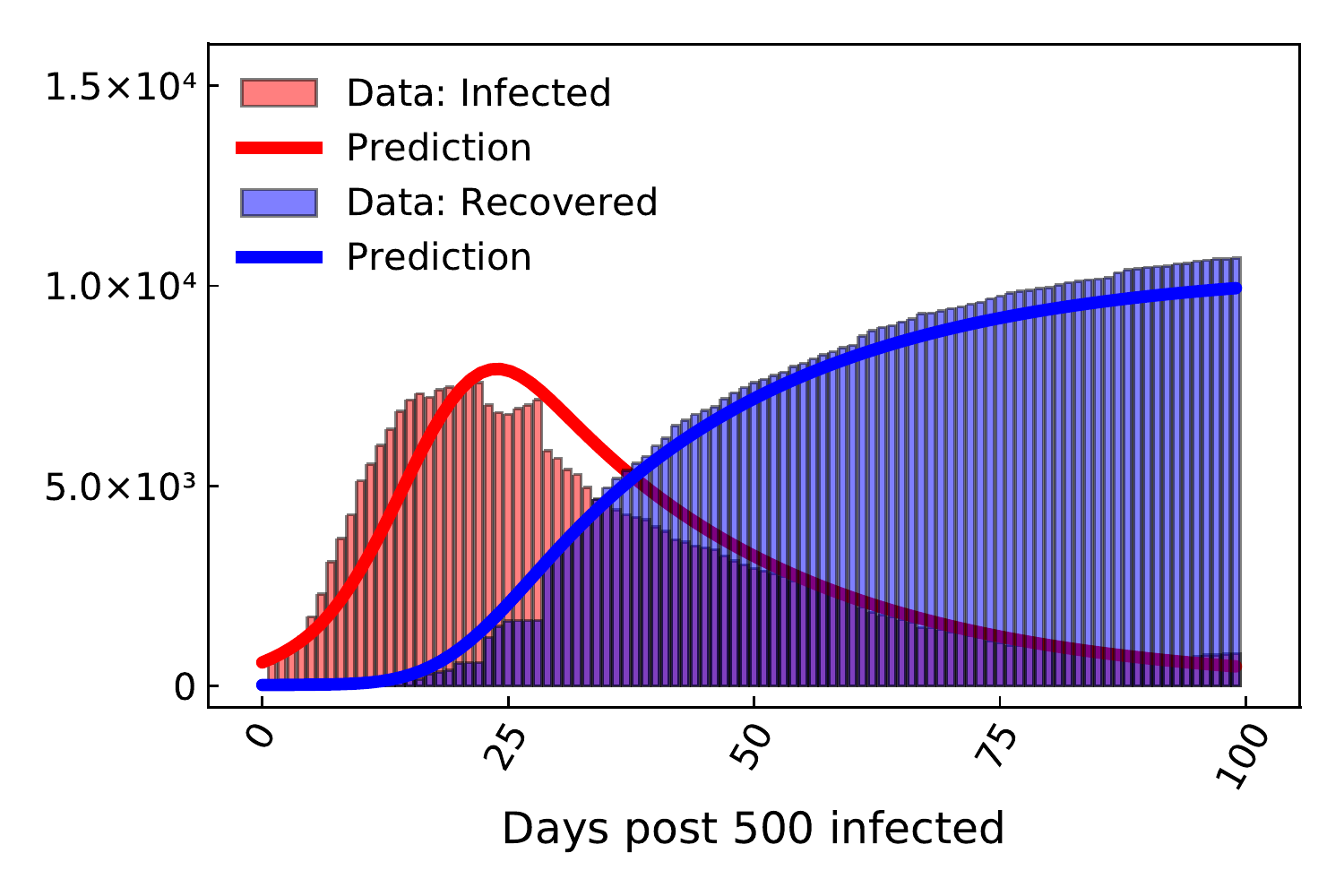}}
\end{tabular}
\caption{COVID-19 infected and recovered evolution compared with our neural network augmented model prediction in the highest affected Asian countries as of June 1, 2020.}\label{Asia1}
\end{figure}

\begin{figure}
\centering
\begin{tabular}{cc}
\subfloat[][India]{\includegraphics[width=0.33\textwidth]{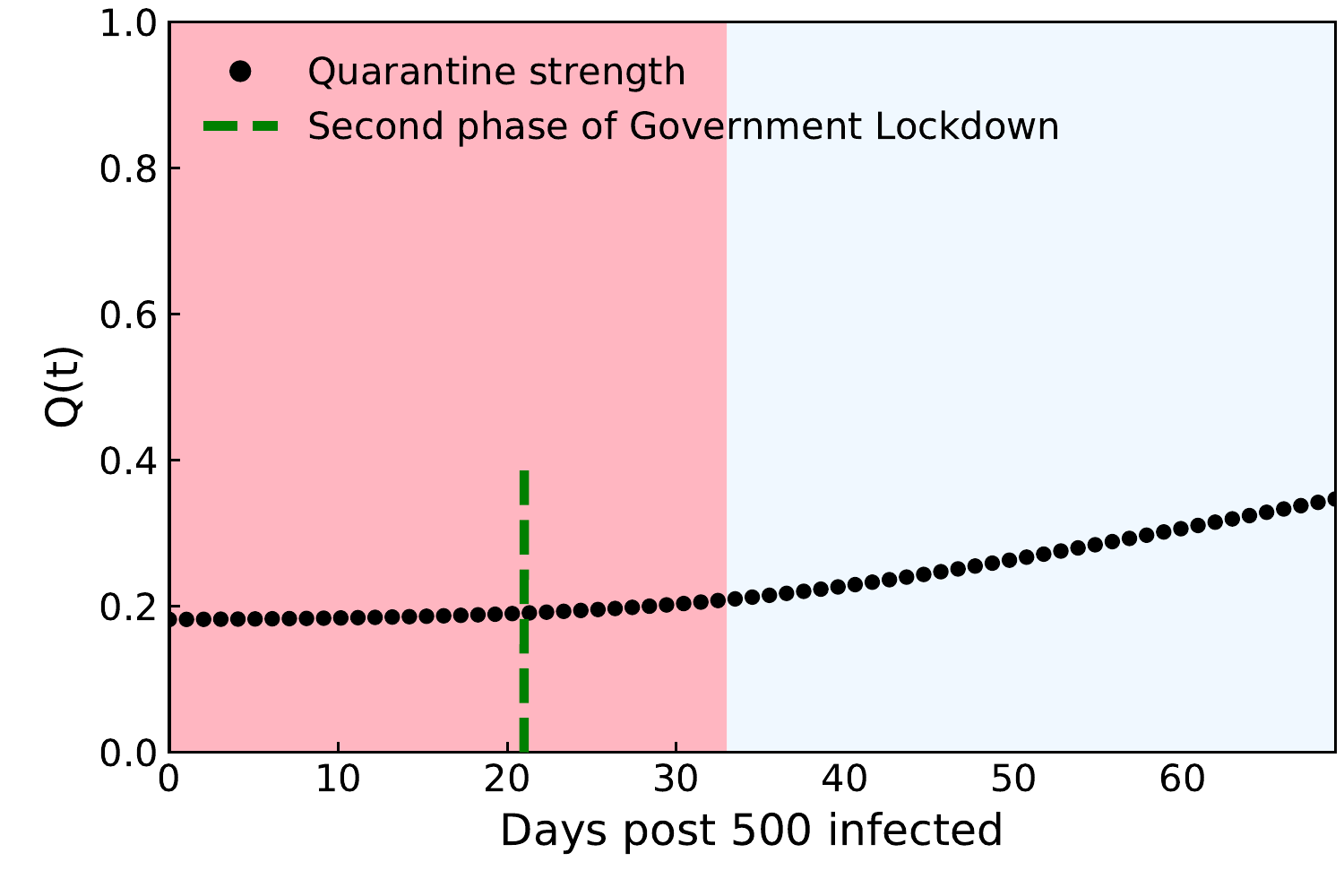}}
\subfloat[][China]{\includegraphics[width=0.33\textwidth]{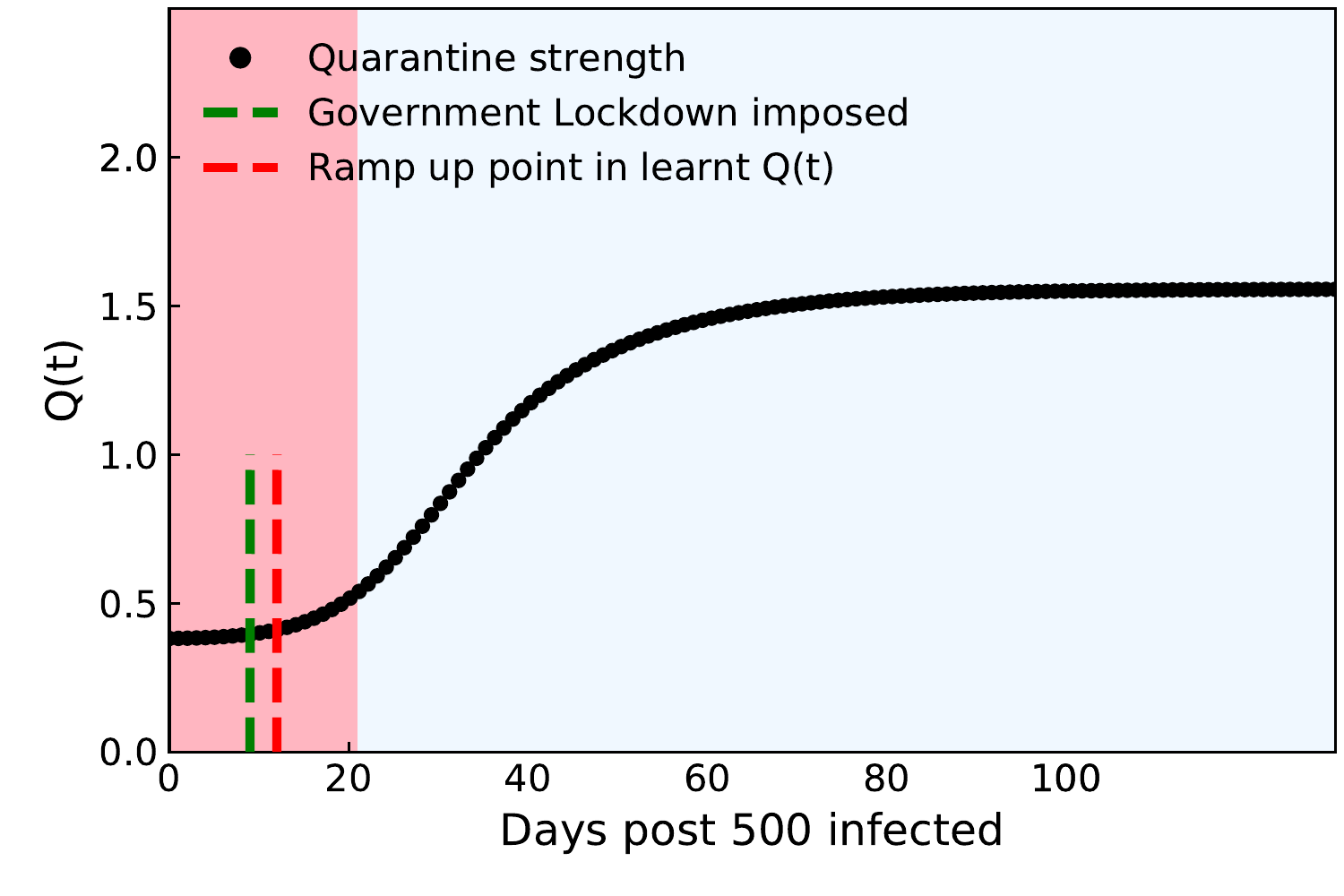}}\\
\subfloat[][South Korea]{\includegraphics[width=0.33\textwidth]{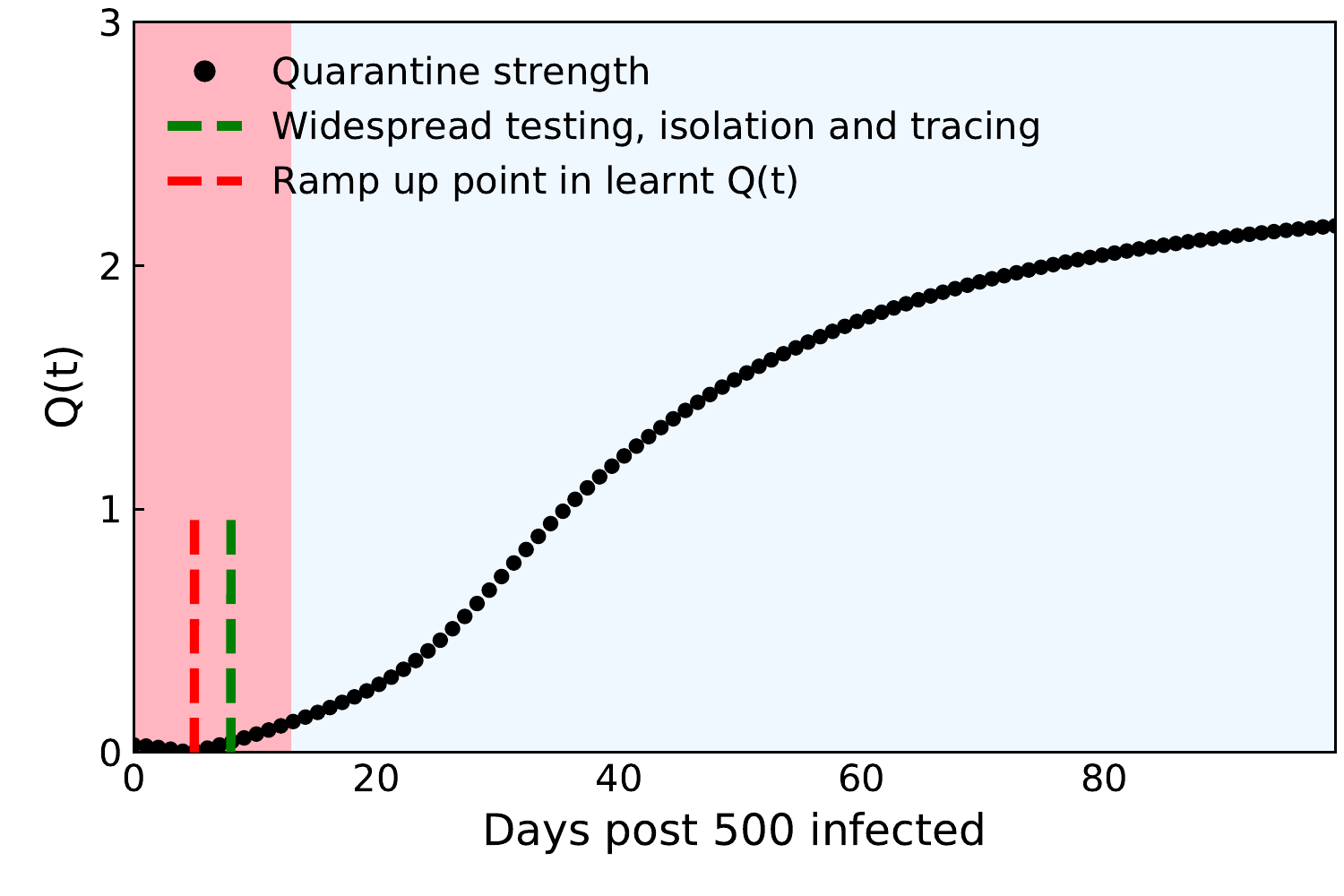}}
\end{tabular}
\caption{Quarantine strength $Q(t)$ learnt by the neural network in the highest affected Asian countries as of June 1, 2020. The transition from the red to blue shaded region indicates the Covid spread parameter of value $C_{p} <1$ leading to halting of the infection spread. The green dashed line indicates the time when quarantine measures were implemented in the region under consideration, which generally matches well with a ramp up point seen in the $Q(t)$ plot denoted by the red dashed line. For regions in which a clear inflection or ramp up point is not seen (India), the red dashed line is not shown.}\label{Asia2}
\end{figure}

\begin{figure}
\centering
\begin{tabular}{cc}
\subfloat[][India]{\includegraphics[width=0.33\textwidth]{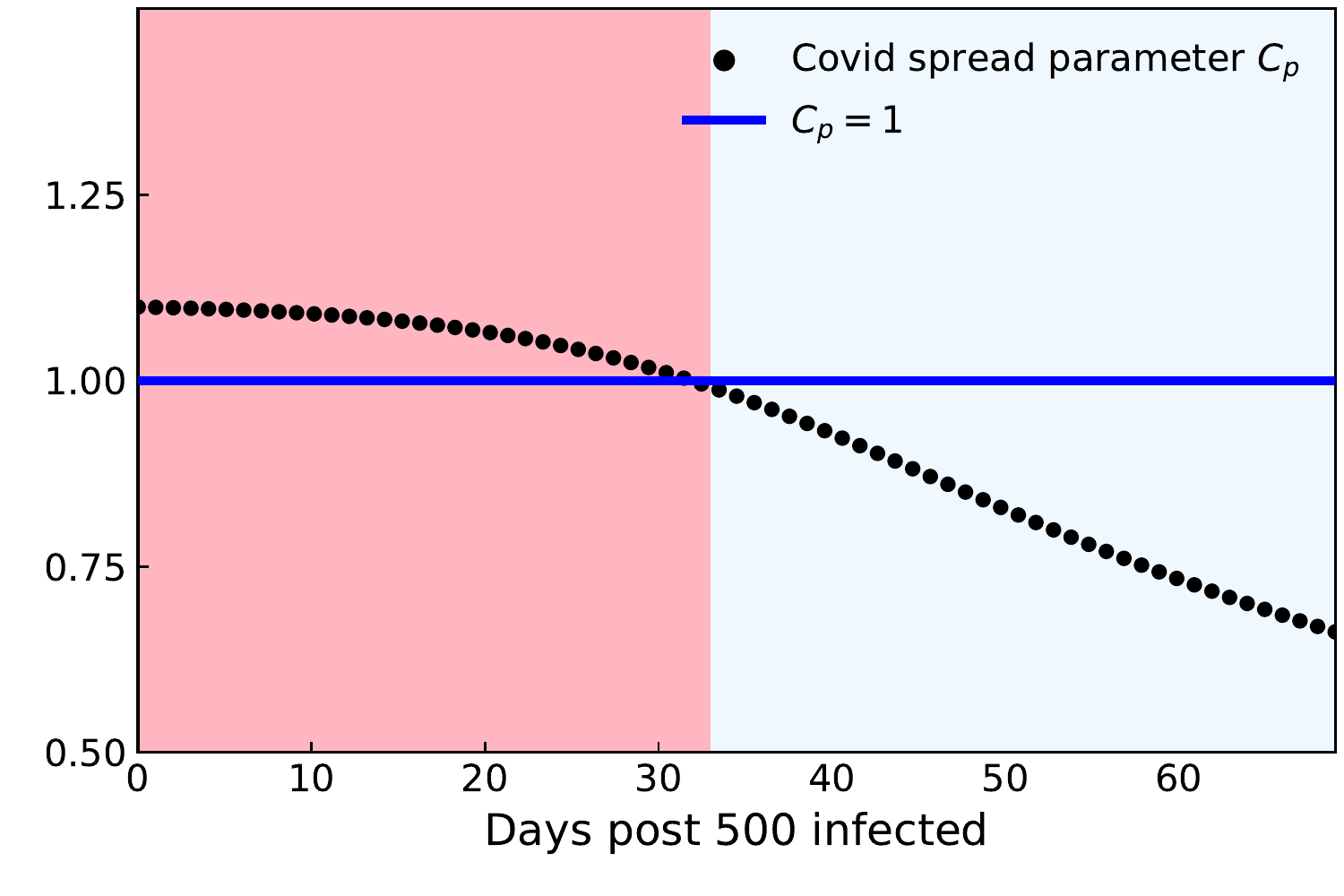}}
\subfloat[][China]{\includegraphics[width=0.33\textwidth]{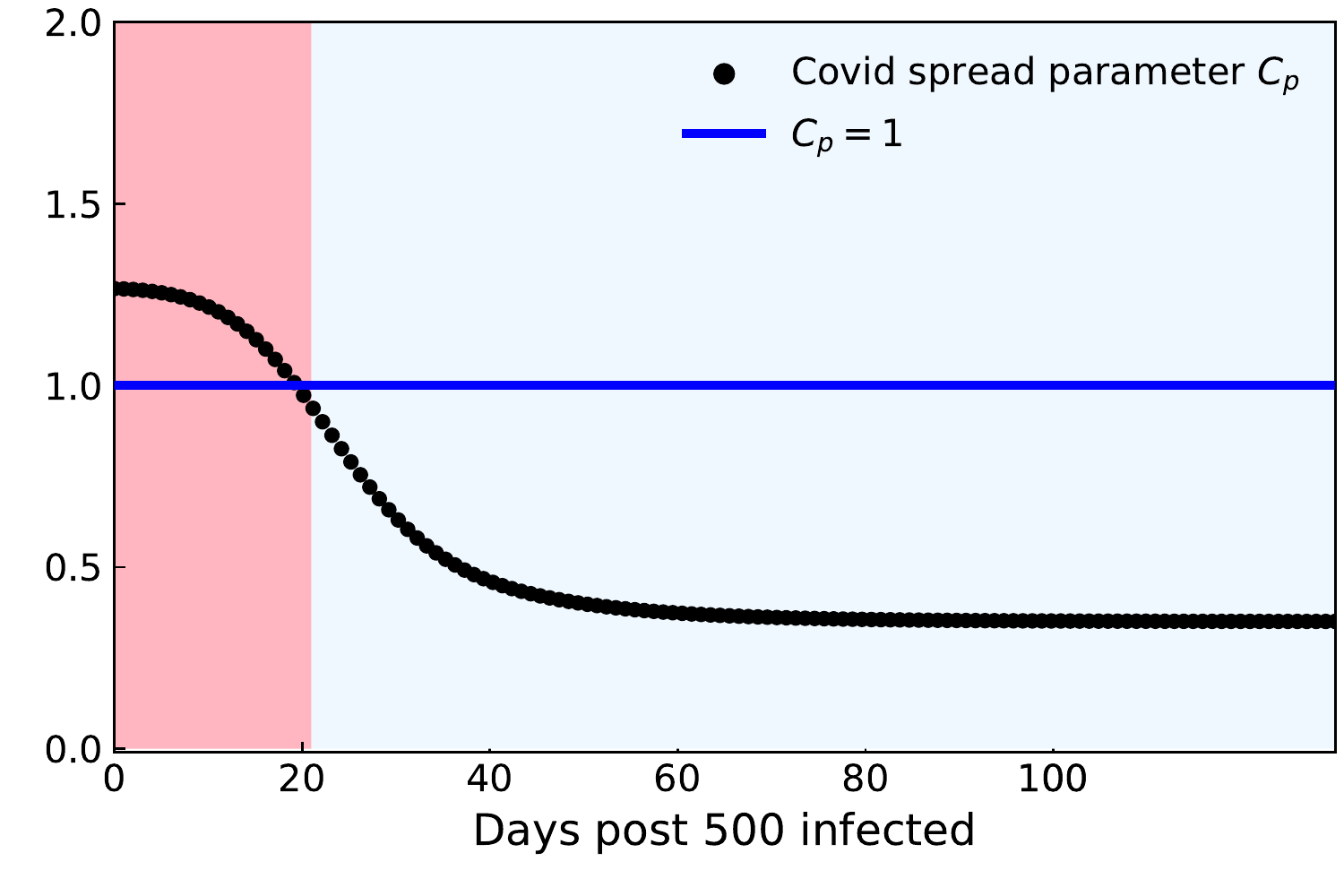}}\\
\subfloat[][South Korea]{\includegraphics[width=0.33\textwidth]{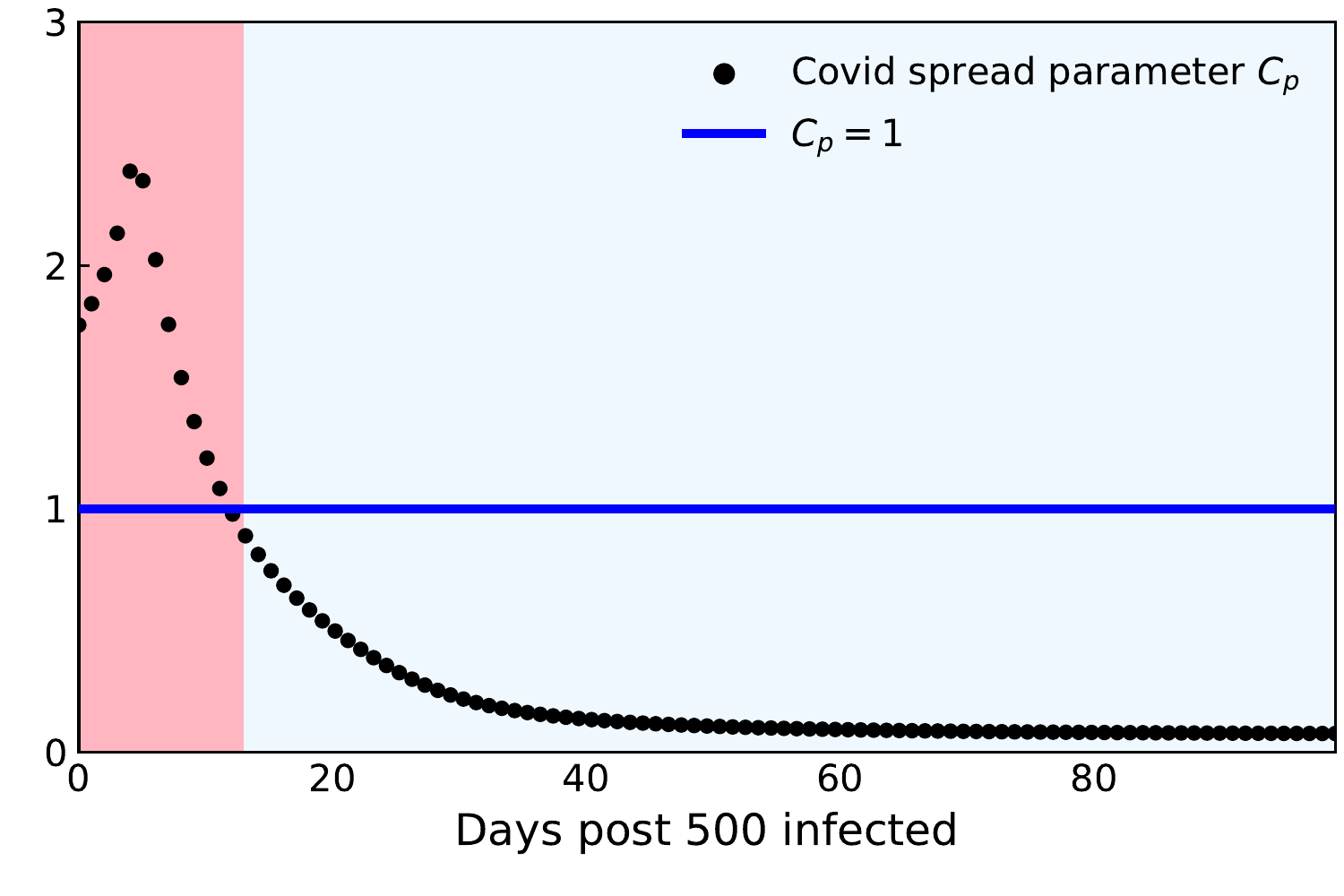}}
\end{tabular}
\caption{Control of COVID-19 quantified by the Covid spread parameter evolution in the highest affected Asian countries as of June 1, 2020. The transition from the red to blue shaded region indicates $C_{p} <1$ leading to halting of the infection spread.}\label{Asia3}
\end{figure}

Figure \ref{Asia1} shows reasonably good match between the model-estimated infected and recovered case count with actual Covid-19 data for the highest affected Asian countries as of $1$ June 2020, namely: India, China and South Korea. $Q(t)$ shows a rapid ramp up in China and South Korea (figure \ref{Asia2}) which agrees well with cusps in government interventions which took place in the weeks leading to and after the end of January \cite{cyranoski2020china} and February \cite{Korea_article} for China and South Korea respectively. On the other hand, a slow build up of $Q(t)$ is seen for India, with no significant ramp up. This is reflected in the quarantine efficiency comparison (figure \ref{All}c), which is much higher for China and South Korea compared to India. South Korea shows a significantly lower contact rate than its Asian counterparts, indicating strongly enforced and followed social distancing protocols \cite{SKEffective}. No significant difference in the recovery rate is observed between the Asian countries. Owing to the high quarantine efficiency in China and a high quarantine efficiency coupled with strongly enforced social distancing in South Korea, these countries were able to bring down the Covid spread parameter $C_{p}$ from $>1$ to $<1$ in $21$ and $13$ days respectively, while it took $33$ days in India (figure \ref{Asia3}).
\subsection{South America}
\begin{figure}
\centering
\begin{tabular}{cc}
\subfloat[][Brazil]{\includegraphics[width=0.5\textwidth]{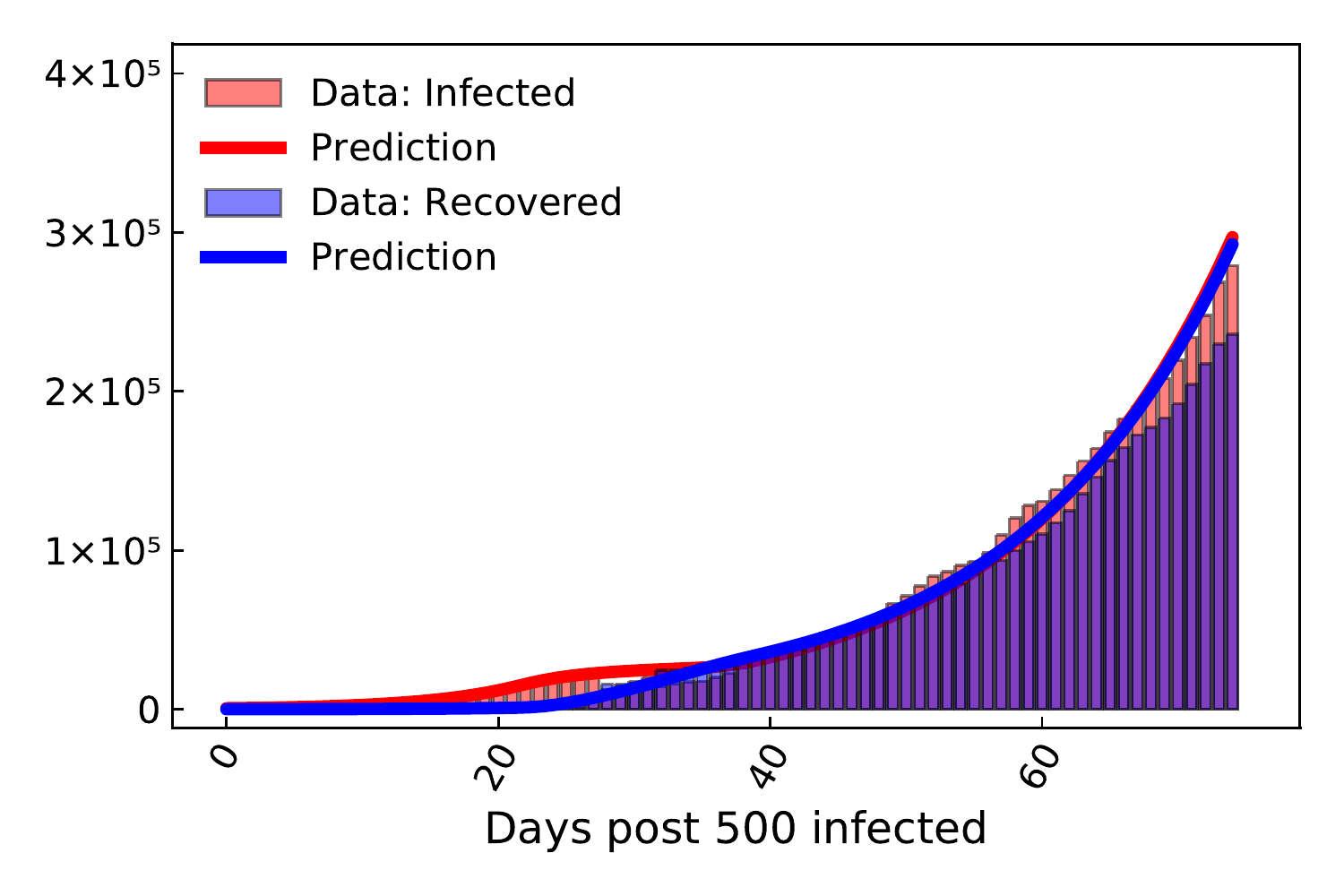}}
\subfloat[][Chile]{\includegraphics[width=0.5\textwidth]{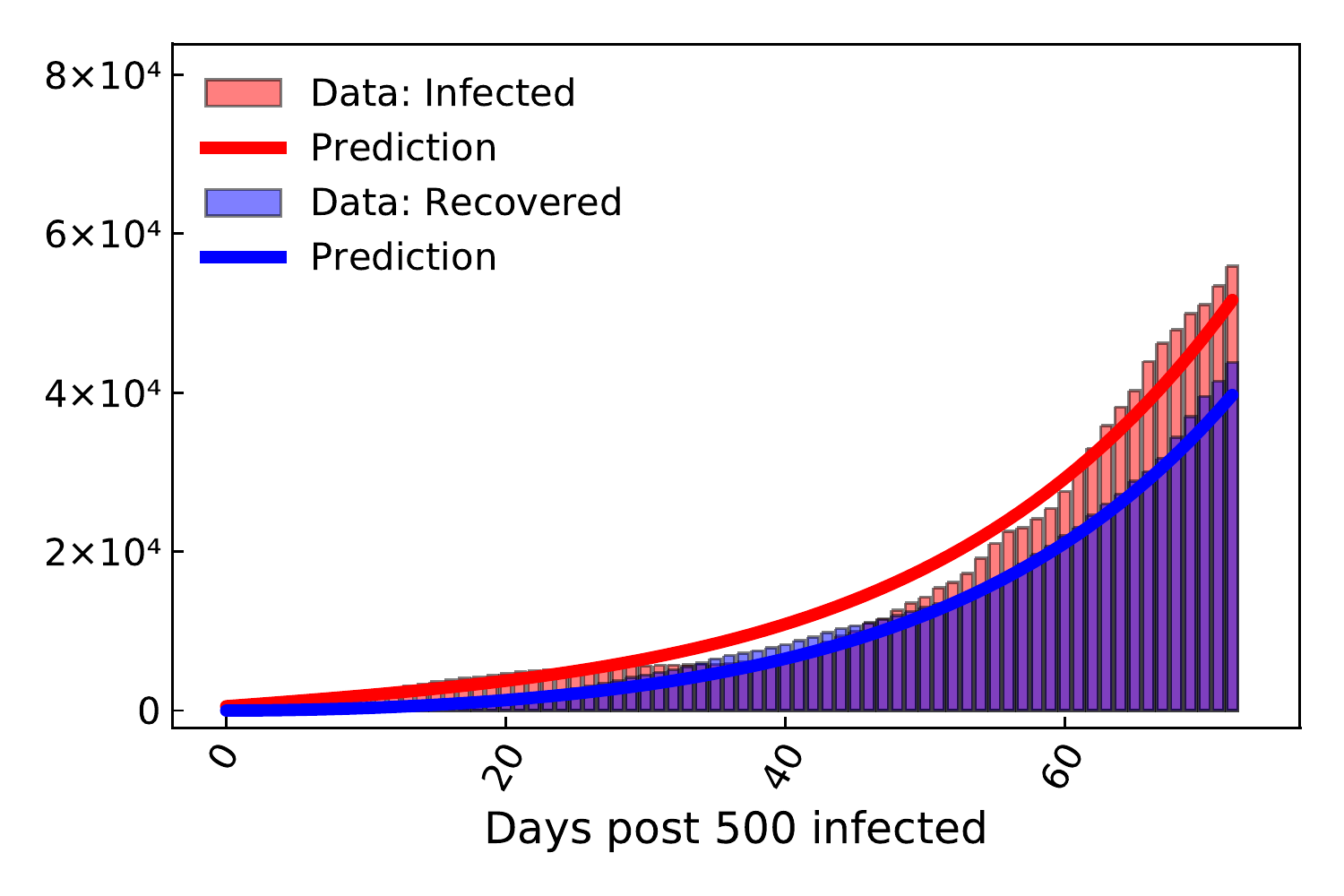}}\\
\subfloat[][Peru]{\includegraphics[width=0.5\textwidth]{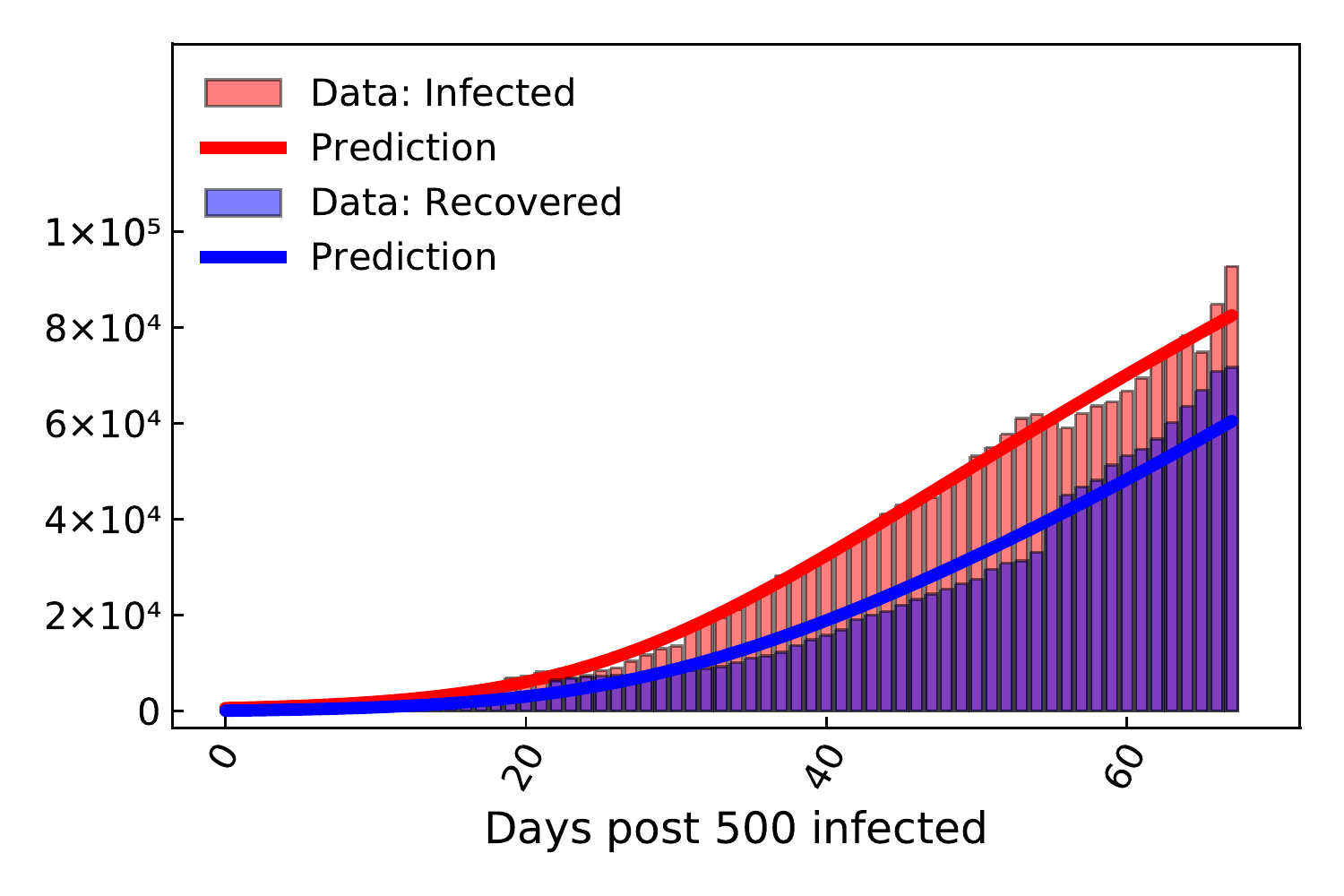}}
\end{tabular}
\caption{COVID-19 infected and recovered evolution compared with our neural network augmented model prediction in the highest affected South American countries as of June 1, 2020.}\label{SAmerica1}
\end{figure}

\begin{figure}
\centering
\begin{tabular}{cc}
\subfloat[][Brazil]{\includegraphics[width=0.33\textwidth]{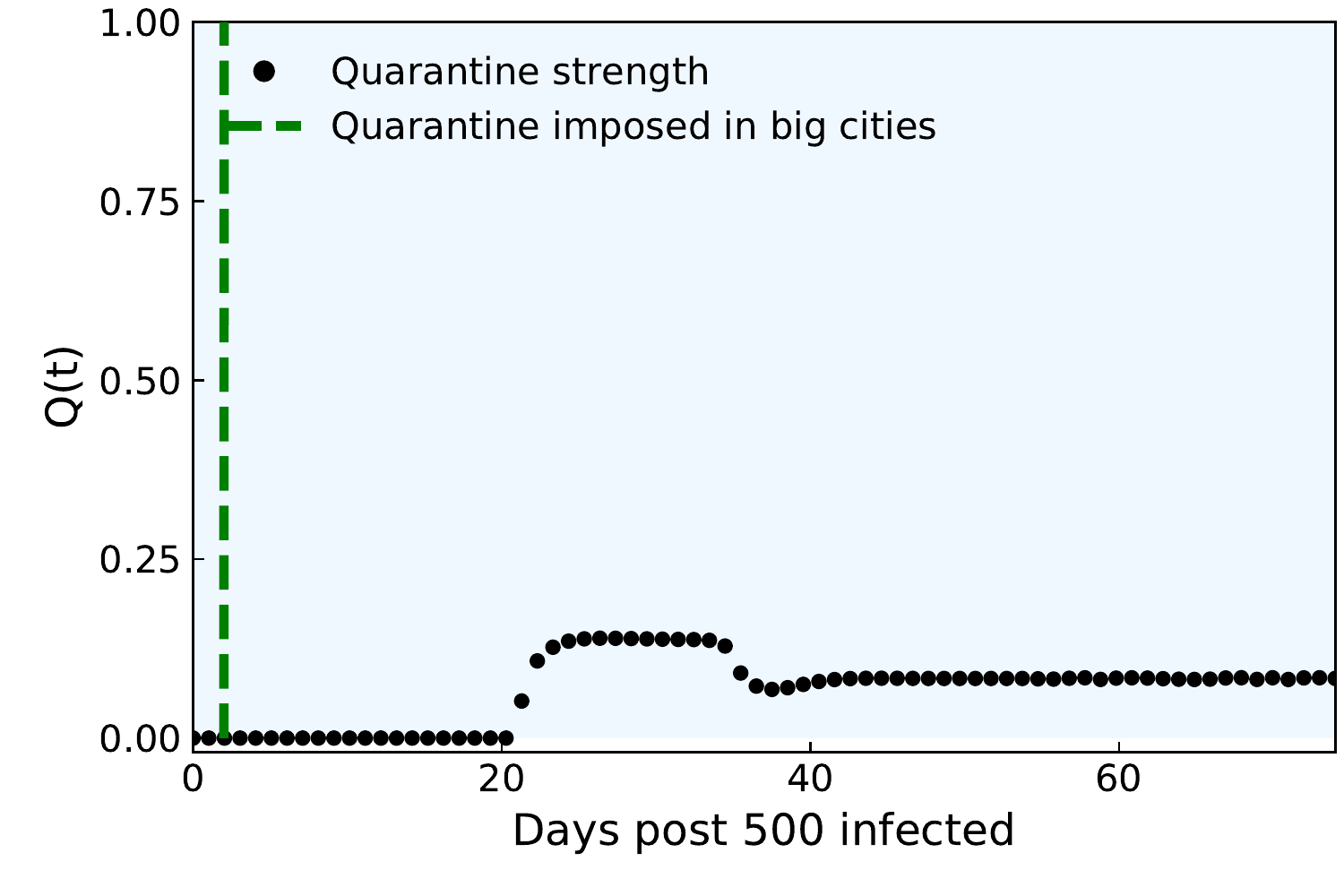}}
\subfloat[][Chile]{\includegraphics[width=0.33\textwidth]{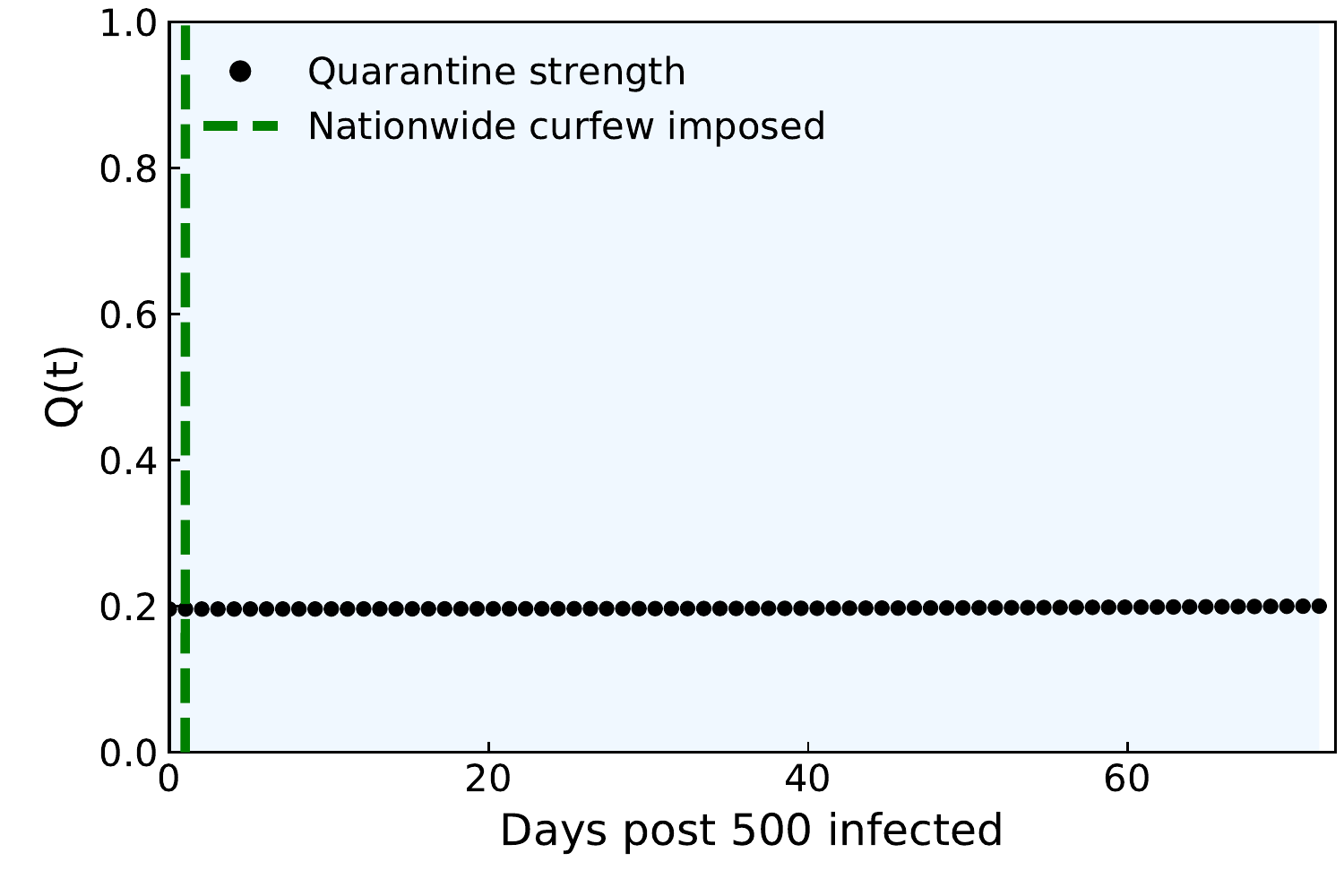}}\\
\subfloat[][Peru]{\includegraphics[width=0.33\textwidth]{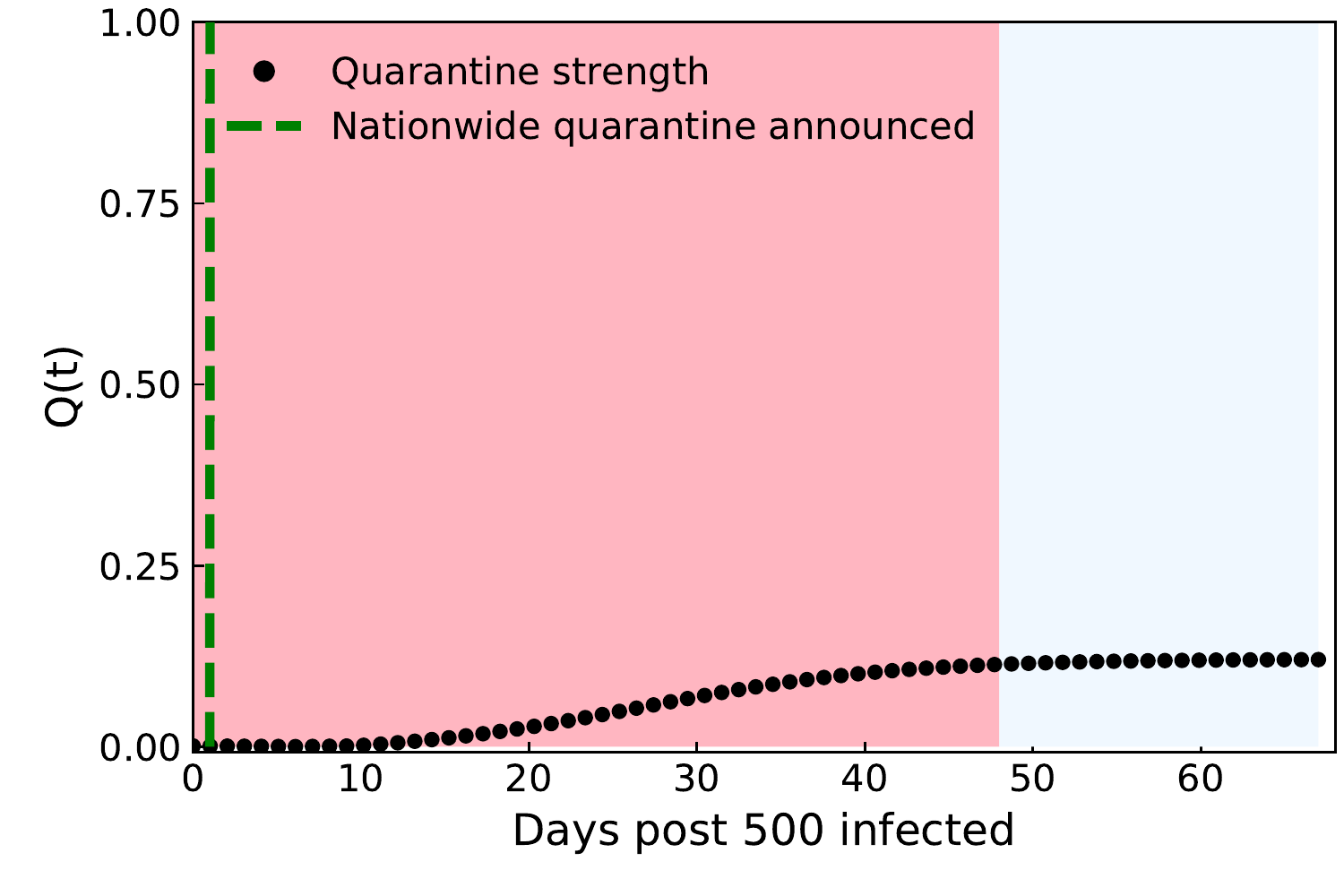}}
\end{tabular}
\caption{Quarantine strength $Q(t)$ learnt by the neural network in the highest affected South American countries as of June 1, 2020. The transition from the red to blue shaded region indicates the Covid spread parameter of value $C_{p} <1$ leading to halting of the infection spread. The green dotted line indicates the time when quarantine measures were implemented in the region under consideration. }\label{SAmerica2}
\end{figure}

\begin{figure}
\centering
\begin{tabular}{cc}
\subfloat[][Brazil]{\includegraphics[width=0.33\textwidth]{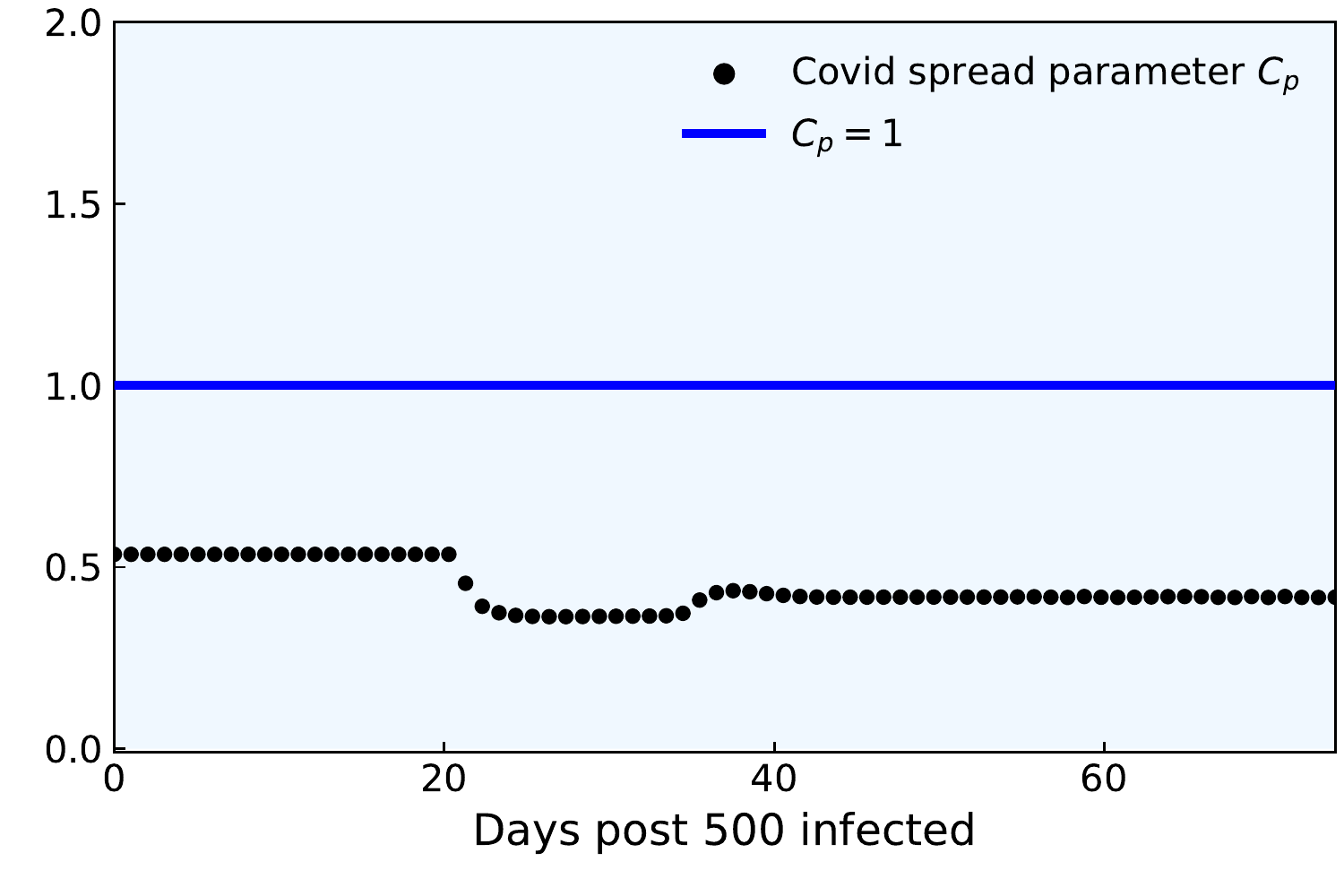}}
\subfloat[][Chile]{\includegraphics[width=0.33\textwidth]{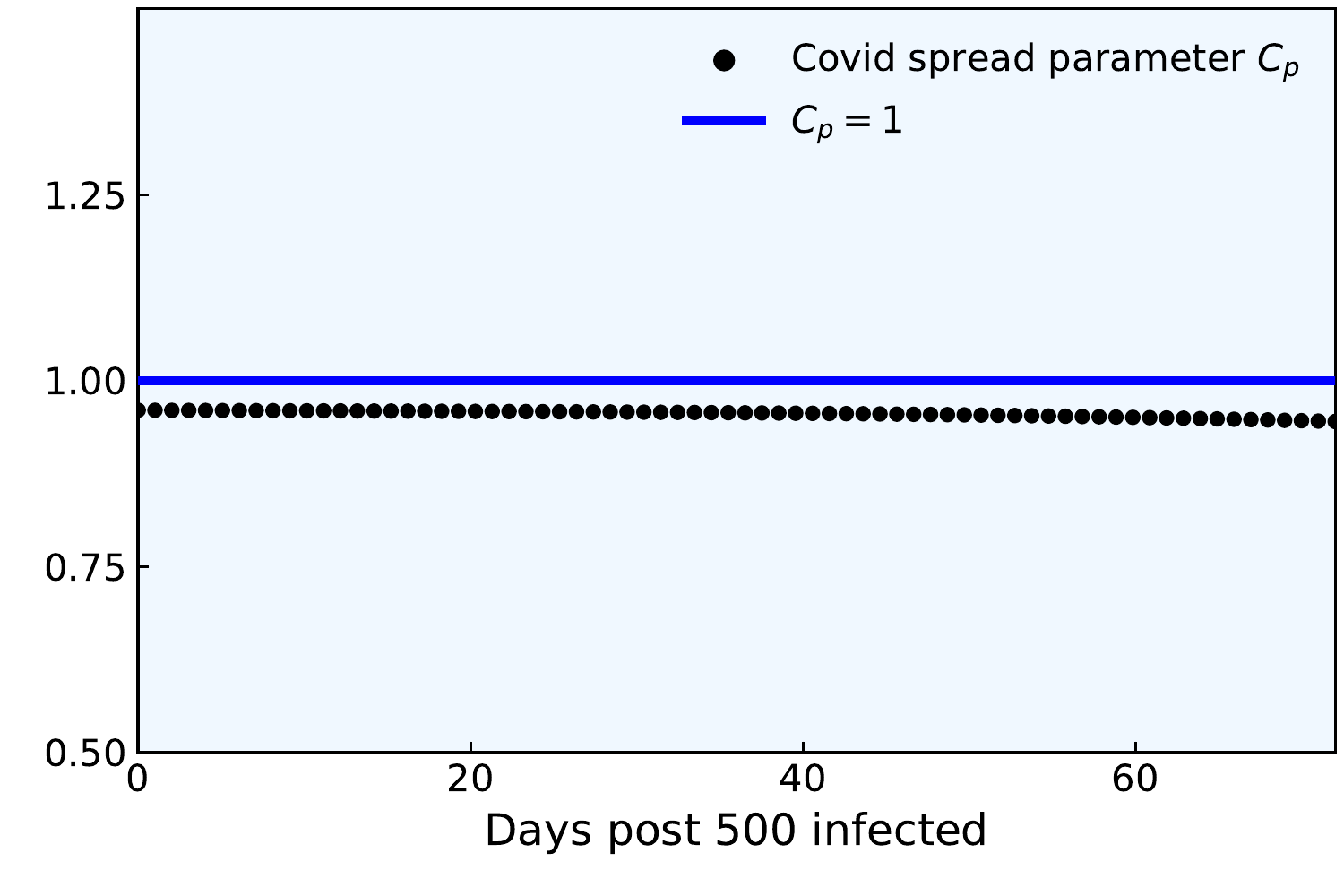}}\\
\subfloat[][Peru]{\includegraphics[width=0.33\textwidth]{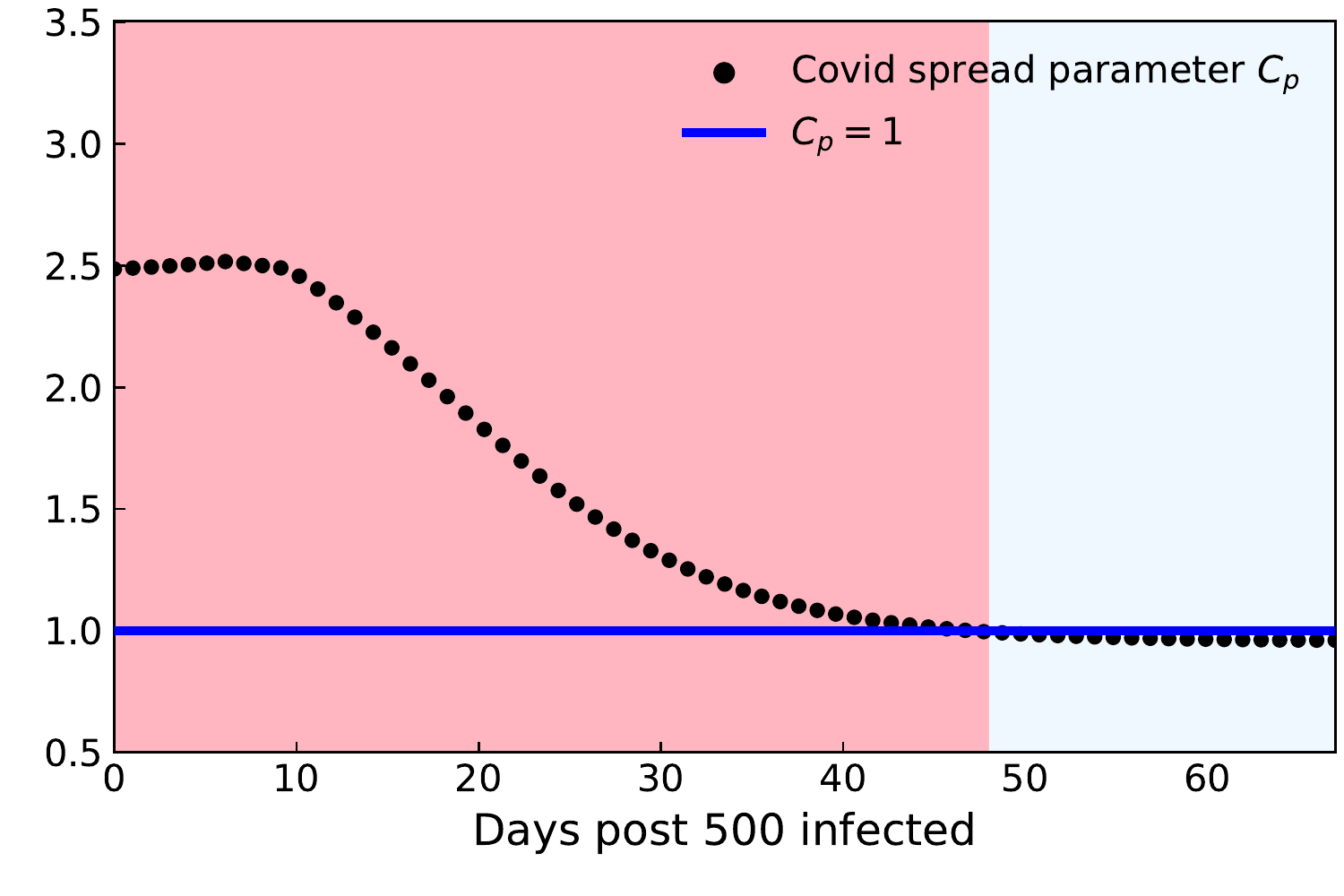}}
\end{tabular}
\caption{Control of COVID-19 quantified by the Covid spread parameter evolution in the highest affected South American countries as of June 1, 2020. The transition from the red to blue shaded region indicates $C_{p} <1$ leading to halting of the infection spread.}\label{SAmerica3}
\end{figure}

\begin{figure}
\centering
\begin{tabular}{cc}
\subfloat[][Europe]{\includegraphics[width=0.5\textwidth]{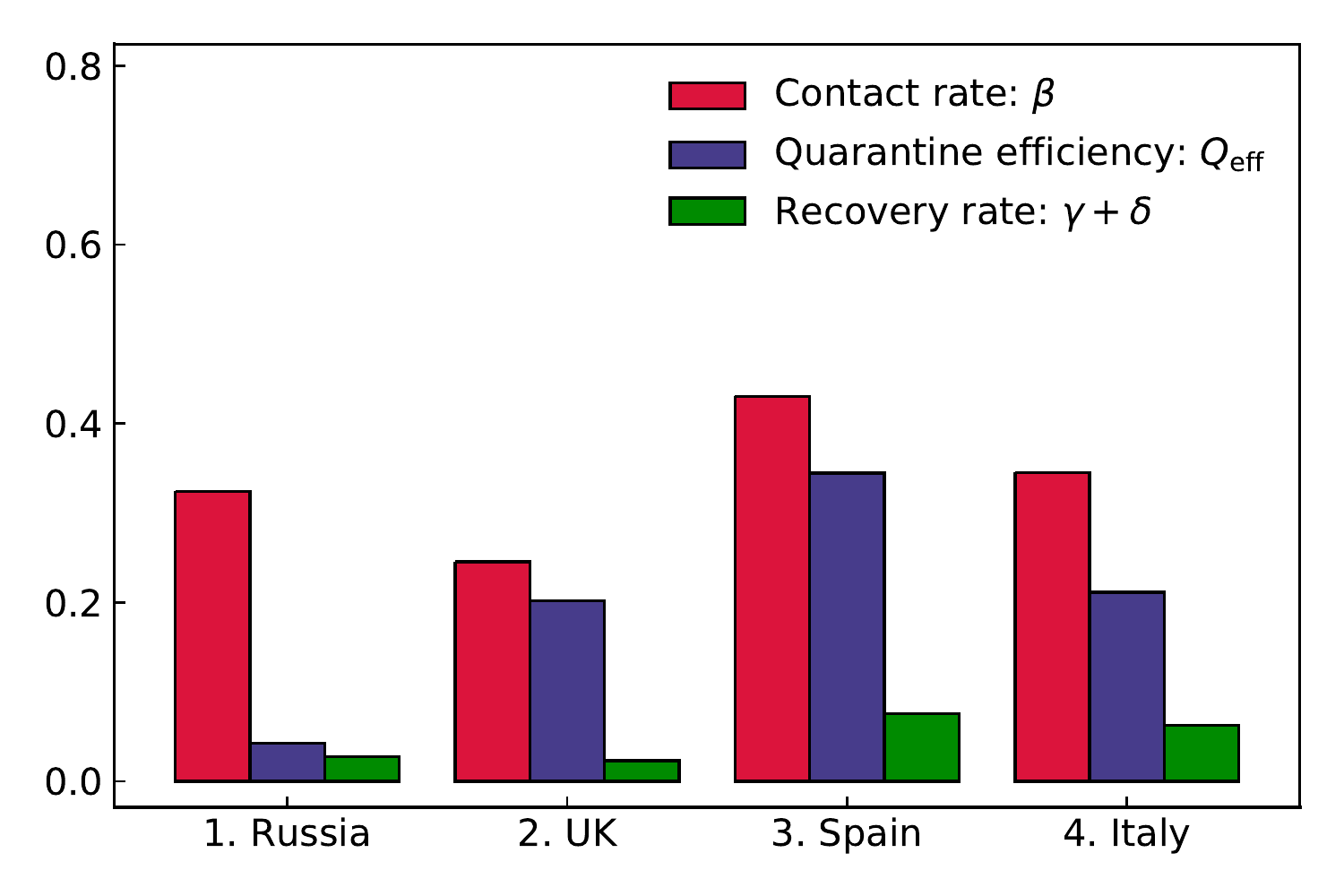}}
\subfloat[][USA]{\includegraphics[width=0.5\textwidth]{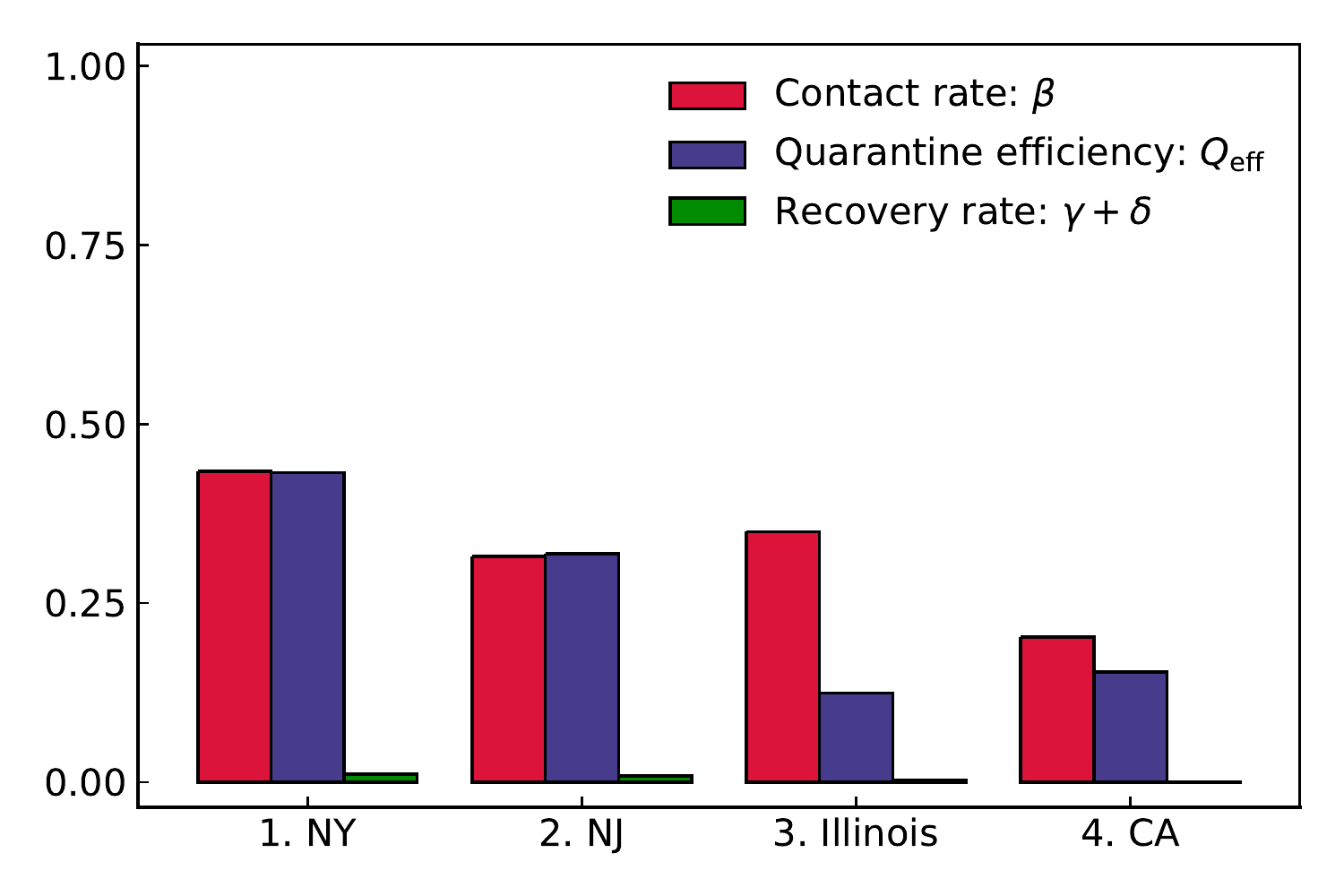}}\\
\subfloat[][Asia]{\includegraphics[width=0.5\textwidth]{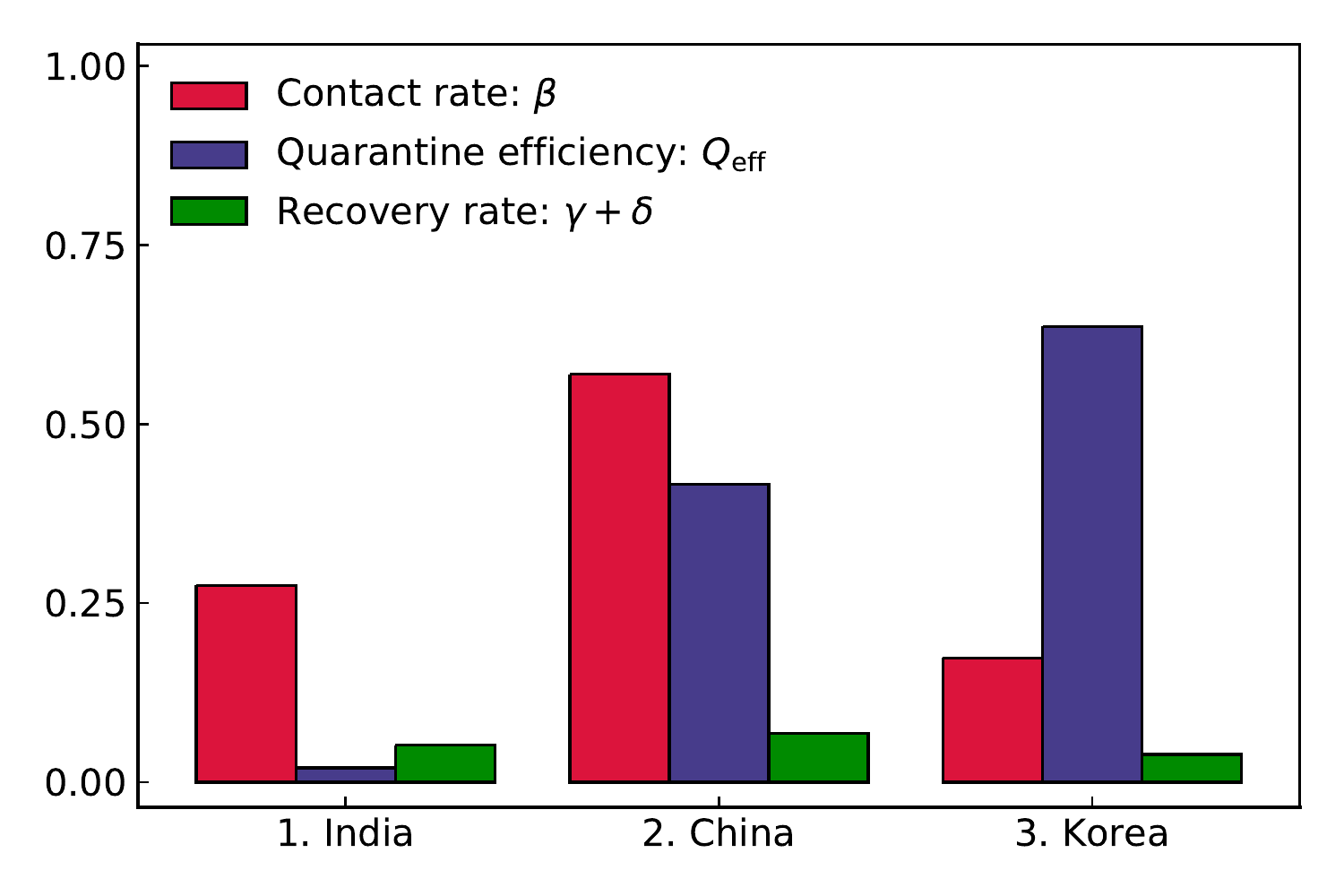}}
\subfloat[][South America]{\includegraphics[width=0.5\textwidth]{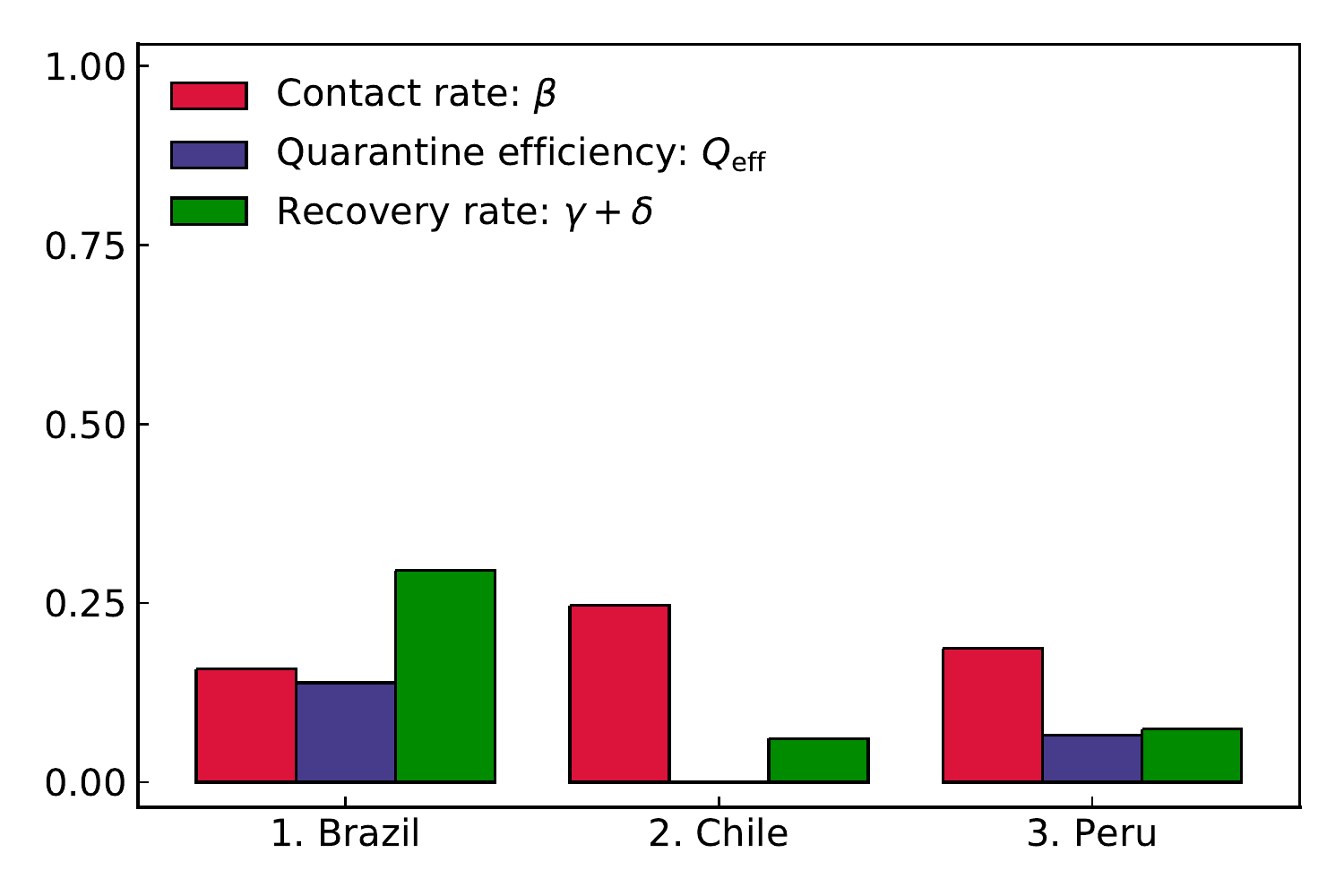}}\\
\end{tabular}
\caption{Global comparison of infection, recovery rates and quarantine efficiency.}\label{All}
\end{figure}

Figure \ref{SAmerica1} shows  reasonably good match between the model-estimated infected and recovered case count with actual Covid-19 data for the highest affected South American countries as of $1$ June 2020, namely: Brazil, Chile and Peru. For Brazil, $Q(t)$ is seen to be approximately constant $\approx 0$ initially with a ramp up around the $20$ day mark; after which $Q(t)$ is seen to stagnate (figure \ref{SAmerica2}a). The key difference between the Covid progression in Brazil compared to other nations is that the infected and the recovered (recovered healthy + dead in our study) count is very close to one another as the disease progressed (figure \ref{SAmerica1}). Owing to this, as the disease progressed, the new infected people introduced in the population were balanced by the infected people removed from the population, either by being healthy or deceased. This higher recovery rate combined with a generally low quarantine efficiency and contact rate (figure \ref{All}d) manifests itself in the Covid spread parameter for Brazil to be $<1$ for almost the entire duration of the disease progression (figure \ref{SAmerica3}a). 
For Chile, $Q(t)$ is almost constant for the entire duration considered (figure \ref{SAmerica2}b). Thus, although government regulations were imposed swiftly following the initial detection of the virus, leading to a high initial magnitude of $Q(t)$, the government imposition became subsequently relaxed. This maybe attributed to several political and social factors outside the scope of the present study \cite{LatinAmericaCOVID}. Even for Chile, the infected and recovered count remain close to each other compared to other nations. A generally high quarantine magnitude coupled with a moderate recovery rate (figure \ref{All}d) leads to $C_{p}$ being $<1$ for  the entire duration of disease progression (figure \ref{SAmerica3}b). In Peru, $Q(t)$ shows a very slow build up (figure \ref{SAmerica2}c) with a very low magnitude. Also, the recovered count is lower than the infected count compared to its South American counterparts (figure \ref{SAmerica1}c). A low quarantine efficiency coupled with a low recovery rate (figure \ref{All}d) leads Peru to be in the danger zone ($C_{p} >1$) for $48$ days post detection of the $500^{\textrm{th}}$ case (figure \ref{SAmerica3}c).

\section{DISCUSSION}
Our model captures the infected and recovered counts for highly affected countries in Europe, North America, Asia and South America reasonably well, and is thus globally applicable. Along with capturing the  evolution of infected and recovered data, the novel machine learning aided epidemiological approach allows us to extract valuable information regarding the quarantine policies, the  evolution of Covid spread parameter $C_{p}$, the mean contact rate (social distancing effectiveness), and the recovery rate. Thus, it becomes possible to compare  across different countries, with the model serving as an important diagnostic tool.\newline

Our results show a generally strong correlation between strengthening of the quarantine controls, {\it i.e.} increasing $Q(t)$ as learnt by the neural network model; actions taken by the regions' respective governments; and decrease of the Covid spread parameter $C_{p}$ for all continents considered in the present study.\newline

Based on the Covid-19 data collected (details in the Materials and Methods section), we note  that accurate and timely reporting of recovered data is seen to have a significant variation between countries; with under reporting of the recovered data being a common practice. In the North American countries, for example, the recovered data are significantly lower than its European and Asian counterparts. Thus, our results strongly indicate the need for each country to follow a particular metric for estimating the recovered count robustly, which is vital for data driven assessment of the pandemic spread. \newline

The key highlights of our model are: (a) it is highly interpretable with few free parameters rooted in an epidemiological model, (b) its reliance on only Covid-19 data and not on previous epidemics  and (c) it is highly flexible and adaptable to different compartmental modelling assumptions. In particular, our method can be readily extended to more complex compartmental models including hospitalization rates, testing rate and distinction between symptomatic and asymptomatic individuals. Thus, the methodology presented in the study can be readily adapted to any province, state or country globally; making it a potentially useful tool for policy makers in event of future outbreaks or a relapse in the current one.\newline

Finally, we have hosted our quarantine diagnosis results for the top $70$ affected countries worldwide on a public platform (https://covid19ml.org/ or https://rajdandekar.github.io/COVID-QuarantineStrength/), which can be used for informed decision making by public health officials and researchers alike. We believe that such a publicly available global tool will be of significant value for researchers who want to study the correlation between the quarantine strength evolution in a particular region with a wide range of metrics spanning from mortality rate to socio-economic landscape impact of Covid-19 in that region.\newline

Currently, our model lacks forecasting abilities. In order to do robust forecasting based on prior data available, the model needs to further augmented through coupling to with real-time metrics parameterizing social distancing, e.g. the publicly available Apple mobility data \cite{Apple}. This could be the subject of future studies.

\section{EXPERIMENTAL PROCEDURES}
\subsection{Augmented QSIR Model: Initial Conditions}
The starting point $t  =0$ for each simulation was the day at which $500$ infected cases were crossed, {\it i.e.} $I_{0} \approx 500$. The number of susceptible individuals was assumed to be equal to the population of the considered region.  Also, in all simulations, the number of recovered individuals was initialized from data at $t = 0$ as defined above. The quarantined population $T(t)$ is initialized to a small number $T(t=0) \approx 10$.  

\subsection{Augmented QSIR Model: Parameter estimation}
The time resolved data for the infected, $I_{\textrm{data}}$ and recovered, $R_{\textrm{data}}$ for each locale considered is obtained from the Center for Systems Science and Engineering (CSSE) at Johns Hopkins University. The neural network-augmented SIR ODE system was trained by minimizing the mean square error loss function 
\begin{equation}\label{loss2}
    L_{\text{NN}} (W, \beta, \gamma, \delta) = ||\textrm{log}(I(t) + T(t)) - \textrm{log}(I_{\textrm{data}}(t))||^{2} + ||\textrm{log}(R(t)) - \textrm{log}(R_{\textrm{data}}(t))||^{2} 
\end{equation}
that includes the neural network's weights $W$. For most of the regions under consideration, $W, \beta, \gamma, \delta$ were optimized by minimizing the loss function given in (\ref{loss2}). Minimization was employed using local adjoint sensitivity analysis \cite{cao2003adjoint, Rack2} following a similar procedure outlined in a recent study \cite{Rackauckas20} with the ADAM optimizer \cite{kingma2014adam} with a learning rate of $0.01$. The iterations required for convergence varied based on the region considered and generally ranged from $40,000 - 100, 000$. For regions with a low recovered count: all US states and UK, we employed a two stage optimization procedure to find the optimal $W, \beta, \gamma, \delta$. In the first stage, (\ref{loss2}) was minimized. For the second stage, we fix the optimal $\gamma, \delta$ found in the first stage to optimize for the remaining parameters: $W, \beta$ based on the loss function defined just on the infected count as $L(W, \beta) = ||\textrm{log}(I(t) + T(t))- \textrm{log}(I_{\textrm{data}}(t))||^{2}$. In the second stage, we don't include the recovered count $R(t)$ in the loss function, since $R(t)$ depends on $\gamma, \delta$ which have already been optimized in the first stage. By placing more emphasis on minimizing the infected count, such a two stage procedure leads to much more accurate model estimates; when the recovered data count is low. The iterations required for convergence in both stages varied based on the region considered and generally ranged from $30,000 - 100, 000$.\newline

The mean error resulting from different model seeding conditions, was seen to be $< 1\%$ for all regions considered.

Preliminary versions of this work can be found at \newline
\texttt{medRxiv 2020.04.03.20052084} and \texttt{arXiv:2004.02752}.

\section{DATA AND CODE AVAILABILITY}
Data for the infected and recovered case count in all regions was obtained from the Center for Systems Science and Engineering (CSSE) at Johns Hopkins University. All code files are available at https://github.com/RajDandekar/MIT-Global-COVID-Modelling-Project-1.
All results are publicly hosted at https://covid19ml.org/ or https://rajdandekar.github.io/COVID-QuarantineStrength/.

\section{ACKNOWLEDGEMENTS}
This effort was partially funded by the Intelligence Advanced Reseach Projects Activity (IARPA.) We are grateful to Emma Wang for help with some of the simulations, and to Haluk Akay, Hyungseok Kim and Wujie Wang for helpful discussions and suggestions. 

\section{DECLARATION OF INTERESTS}
The authors declare no conflicts of interest.

\bibliography{Paper_Draft1}

\bibliographystyle{pnasn}

\end{document}